\documentclass[reprint,amsmath, amssymb, aps, prl, superscriptaddress, nobibnotes, longbibliography, twocolumn, nobalancelastpage]{revtex4-2}

\usepackage{multirow}
\usepackage{minitoc}
\usepackage{appendix}

\usepackage{soul}
\usepackage{siunitx}
\usepackage{graphicx}
\usepackage{braket}
\usepackage[colorlinks]{hyperref}
\usepackage[capitalize]{cleveref}
\usepackage[normalem]{ulem}
\usepackage{comment}

\hypersetup{citecolor={black}}  
\usepackage{float}

\newcommand*{\citen}[1]{%
  \begingroup
    \romannumeral-`\x %
    \setcitestyle{numbers}%
    \cite{#1}%
  \endgroup   
}

\makeatletter
\newcommand*{\balancecolsandclearpage}{%
  \close@column@grid
  \newpage
  \twocolumngrid
}
\makeatother

\makeatletter
\newcommand*{\balancecolsandclearpagesingle}{%
  \close@column@grid
  \newpage
}
\makeatother

\crefname{EDfig}{Extended Data Fig.}{Extended Data Fig.}

\newcounter{SIfig}
\crefname{SIfig}{Fig.}{Fig.}
\Crefname{SIfig}{Fig.}{Fig.}

\newcounter{SItable}
\crefname{SItable}{Table}{Table}
\Crefname{SItable}{Table}{Table}

\begin{document}

\title{Universal distributed blind quantum computing with
solid-state qubits} 

\author
{
Y.-C. Wei$^{1\ast}$,
P.-J. Stas$^{1\ast}$,
A. Suleymanzade$^{1\ast\dagger}$,
G. Baranes$^{1,2\ast}$,
F. Machado$^{1,3}$,  
Y. Q. Huan$^{1}$, 
C. M. Knaut$^{1}$, 
Sophie W. Ding$^{4}$, 
M. Merz$^{5}$, 
E. N. Knall$^{4}$, 
U. Yazlar$^{1,6}$, 
M. Sirotin$^{1,2}$, 
I. W. Wang$^{1}$, 
B. Machielse$^{4, 7}$, 
S. F. Yelin$^{1}$, 
J. Borregaard$^{1, 7}$,  
H. Park$^{1, 8}$, 
M. Lon\v{c}ar$^{4}$, 
and M. D. Lukin$^{1\ddagger}$  \\
	\emph{ \small$^{1}$ Department of Physics, Harvard University, Cambridge, Massachusetts
02138, USA.\\
	\small$^{2}$Department of Physics and Research Laboratory of Electronics,\\
Massachusetts Institute of Technology, Cambridge, MA, USA.\\
	\small$^{3}$ITAMP, Harvard-Smithsonian Center for Astrophysics, Cambridge, MA, USA.\\
	\small$^{4}$John A. Paulson School of Engineering and Applied Sciences,\\ Harvard University, Cambridge, Massachusetts 02138, USA.\\
	\small$^{5}$Department of Physics, ETH Zurich, 8093 Zurich, Switzerland.\\
	\small$^{6}$Division of Materials Science \& Engineering, Boston University, Boston
Massachusetts 02215, USA.\\
    \small$^{7}$Lightsynq Technologies Inc., Brighton, Massachusetts 02135, USA.\\
	\small$^{8}$Department of Chemistry and Chemical Biology,\\ Harvard University, Cambridge, Massachusetts 02138, USA.\\ 
	\small$^\dagger$Corresponding author. Email: azizasuleymanzade@g.harvard.edu\\
\small$^{\ddagger}$Corresponding author. Email:  lukin@physics.harvard.edu\\
	\small$^\ast$These authors contributed equally to this work.
    }
}

\begin{abstract}
Blind quantum computing (BQC) is a promising application of distributed quantum systems, where a client can perform computations on a remote server without revealing any 
details of the applied circuit. While the most promising realizations of quantum computers 
are  based 
on various matter qubit platforms, implementing BQC on matter qubits remains an outstanding challenge. 
Using silicon-vacancy (SiV) centers in nanophotonic diamond cavities with an efficient optical interface, we experimentally demonstrate a universal quantum gate set consisting of single- and two-qubit blind gates over a distributed two-node network. Using these ingredients,  we perform a distributed algorithm with blind operations across our two-node network, paving the way towards blind quantum computation with matter qubits in distributed, modular architectures.
\end{abstract}

\maketitle


Quantum computers are expected to outperform their classical counterparts for certain problems~\cite{Preskill2018},
while the no-cloning theorem of quantum mechanics offers routes to information-theoretic security~\cite{PhysRevLett.67.661}, as harnessed in quantum key distribution~\cite{Kimble2008, PhysRevLett.85.441}.
Blind quantum computing (BQC) has the potential to combine these two concepts, allowing a client with limited quantum capabilities to utilize remote, powerful servers with information-theoretic privacy~\cite{childs2001secure, broadbent2009universal, arrighi2006blind, fitzsimons2017unconditionally, fitzsimons2017private}. 
The realization of BQC requires the exchange of quantum information between a remote client and servers through flying qubits, which is  effectively achieved with photons due to their ability to travel over long distances. In fact, BQC was first experimentally explored with all-photonic platforms, including generating cluster-state resources for blind protocols and  algorithms~\cite{barz2012demonstration, barz2013experimental, huang2017experimental}.

The ability to locally manipulate and store quantum information via matter qubits offers scalable ways for generating larger entangled resources~\cite{Baranes2024efficient, drmota2024verifiable, Thomas2022, oneway_repeater, qd_mete}. Moreover, recent progress in quantum computing with matter-based platforms, including neutral atoms~\cite{Bluvstein2023, radnaev2024}, superconducting qubits~\cite{google2023, acharya2024quantumerrorcorrectionsurface}, trapped ions~\cite{Pino2021, ryananderson2022implementingfaulttolerantentanglinggates}, semiconductor quantum dots~\cite{philips2022}, and solid-state defects~\cite{Abobeih2022, Tin_mete, Jenela_micro_tin} makes it timely to explore if BQC can be realized on such matter-based quantum systems. 
%
%
%
%
Matter-based BQC requires high-fidelity matter-photon entanglement, where photons travel to clients to mediate quantum gates applying
on the matter qubits~\cite{anders2010ancilla, sueki2013ancilla}.
This approach critically depends  on the availability of efficient optical interfaces — a challenge for matter-based quantum  platforms, which are typically not optimized for this purpose. 
Other challenges include further scaling up the number of qubits in BQC. The distributed quantum computing architecture provides a natural solution for systems with matter-photon interfaces
~\cite{Kimble2008, main2024distributed, Daiss2021, liang_pra}. Although matter-photon implementation of blind rotations has been recently demonstrated~\cite{drmota2024verifiable}, several key ingredients for matter-based BQC, including a universal blind gate set  
and blind operations on distributed matter resources, have not yet been experimentally explored.


Silicon-vacancy centers (SiV) integrated in diamond nanophotonic cavities~\cite{twofridge2024,robustnode2022, Bhaskar2020, raman_siv} provide an efficient matter-photon interface, making them well suited for BQC applications.
Here, we report the first experimental implementation of a universal gate set for matter-based BQC, including distributed blind operations, using a two-node SiV-based quantum network.

\begin{figure*}[htp]
    \centering
    \includegraphics[width=1\linewidth]{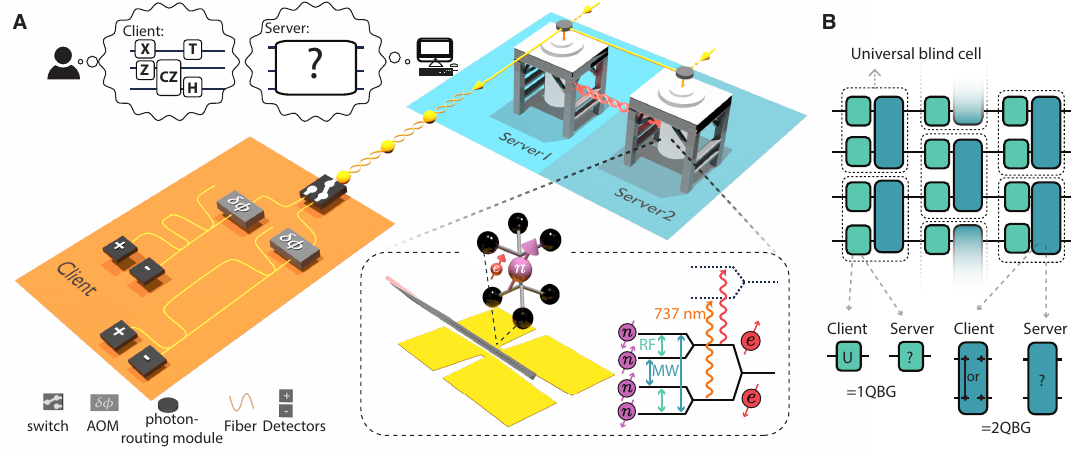}
    \caption{\textbf{Distributed BQC with a two-node quantum network with SiVs.} (\textbf{A}) Top left, an illustration of
    universal BQC protocol. The client can run a quantum algorithm on the server, while the server has no
    knowledge of the applied circuits. Middle, experimental layout of our implementation of distributed BQC. The client receives photons from the servers and performs measurements. Each server operates an $^{29}\text{Si}$V inside a nanophotonic cavity within a dilution refrigerator. A gold coplanar waveguide is used to deliver microwave (MW) and radio-frequency (RF) pulses. The $^{29}\text{Si}$V energy level diagram shows the MW and RF transitions in the two-qubit manifold (straight arrows) and spin-conserving optical transitions (wavy arrows). 
    (\textbf{B}) BQC circuit structure. Circuits are constructed from repeated universal blind cells. A universal blind cell allows the client to perform operations, including a universal single-qubit gate set ($U$) and $CZ$ gate, while the servers always observe the same quantum decoherence channel and, therefore, have no knowledge of the client's operations. 1QBG allows the client to perform an arbitrary gate within the set $U$, while 2QBG allows the client to apply either a maximally entangling or a non-entangling gate, such as $CZ$ or Identities ($I$).}
    \label{fig:fig1}
\end{figure*}

\subsection{BQC architecture with SiV spin qubits}

The goal of BQC is for a client to perform a computation on a server without revealing the applied quantum circuit (Fig.~\ref{fig:fig1}A top). 
A way to fulfill this requirement for arbitrary algorithms is to compose the entire circuit from universal blind cells with a periodic pattern as shown in Fig.~\ref{fig:fig1}B~\cite{broadbent2009universal, fitzsimons2017unconditionally, sueki2013ancilla, Baranes2024efficient, SI}.
%
%
The client can use such a universal blind cell to apply any desired quantum operations in the universal gate set, including arbitrary single-qubit rotations and a two-qubit gate.
%
%
However, the servers cannot distinguish between different choices of operations by the client, and therefore the quantum logic is hidden from the servers. 
A universal blind cell can be implemented from two sub-units: one-qubit blind gates (1QBG) and two-qubit blind gates (2QBG), explained in \cite{SI} section I and \cite{Baranes2024efficient}. The 1QBG allows the client to execute any arbitrary single-qubit operations from the universal gate set,  whereas 2QBG allows the client to choose between applying a maximally entangling two-qubit gate and applying a non-entangling gate. For both 1QBG and 2QBG, the client's choice of operation is unknown to the servers (Fig.~\ref{fig:fig1}B). The implementation of 1QBG and 2QBG completes a universal gate set for BQC.


%

We use a SiV-based two-node quantum network~\cite{twofridge2024} as a backbone for our BQC implementation. The network consists of two physically separated nodes and a photonic measurement apparatus, acting as two distributed servers and a client, respectively (Fig.~\ref{fig:fig1}A). 
The servers use spin qubits as computational resources,
and blind quantum gates are applied through high-fidelity matter-photon entanglement generated locally by servers.
The photons, which are part of the resulting entangled resource, are sent to the client for measurements. These measurements define the gate logic applied on the spin qubits while preserving secrecy. 

Each server node consists of a SiV center coupled to a nanophotonic cavity. Selectively implanting the $^{29}\text{Si}$ 
isotope creates a two-qubit register: one electron spin qubit with an optical transition at \SI{737}{\nano \meter} as a communication qubit and one nuclear spin qubit as a memory~\cite{robustnode2022}. In our implementation, we use both the electron ($e_1$) and nuclear ($n_1$) spins at server 1 and the electron spin ($e_2$) at server 2 as matter qubit resources. The nanophotonic cavity enables a high contrast electron spin-dependent reflectance of the SiV-cavity system, and we send microwave (MW) or radio-frequency (RF) pulses to a gold coplanar waveguide to apply high-fidelity single and two-qubit gates on the electron and nuclear spin qubits (Fig.~\ref{fig:fig1}A). The above features allow us to generate spin-photon entanglement~\cite{robustnode2022} and photon-mediated entanglement between two spins at separate nodes~\cite{twofridge2024}. 
%
%
Universal BQC requires controllable photon measurements in the $X-Y$ plane of the Bloch sphere. In our work, we use the presence of a photon in an early ($\ket{0}$) or late time-bins ($\ket{1}$) as the photonic qubit. The client uses an acousto-optical modulator (AOM) to apply a phase ($\phi$) between early and late time-bins, followed by a time-delay interferometer (TDI) to interfere both time-bin photons. This allows the photon measurement in an arbitrary basis in the $X-Y$ plane~\cite{SI}.


%

\subsection{Intra-node blind gates}
\begin{figure}[hbt]
    \centering
    \includegraphics[width=0.5\textwidth]{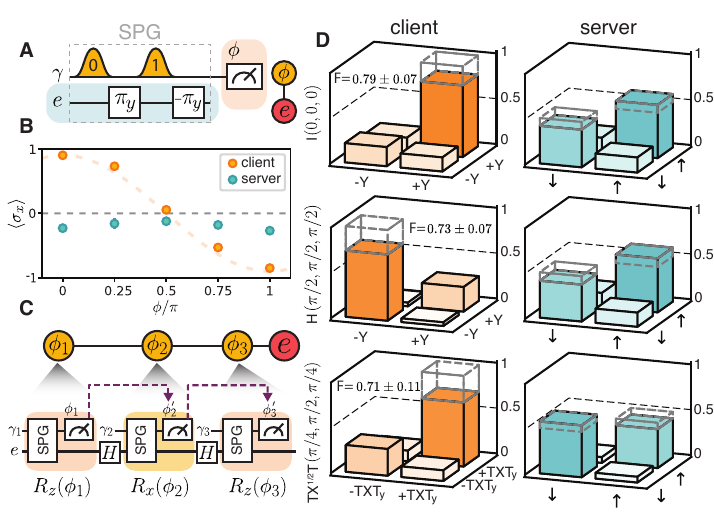}
    \caption{\textbf{Single qubit blind gates.} (\textbf{A}) $R_z(\phi)$ sequence. First the servers  perform a spin-photon gate (SPG), after which the client measures the photon in the basis $\ket{\pm_\phi}_\gamma$. (\textbf{B}) The expectation value of the state of the matter qubit along the positive $x$-axis as a function of the single qubit rotation angle as measured by the client and the server. By generating a photon-electron Bell pair and measuring the photon in different bases, $\ket{0} + e^{i\phi} \ket{1}$, the client rotates the electron spin around the $z$-axis. Here, the electron is initialized in $\ket{+}$. (\textbf{C}) Universal 1QBG consisting of 3SPGs generates a three-photon linear cluster state with the electron as the fourth qubit. (\textbf{D}) Measured density matrices of the matter qubit after performing Identity, Hadamard, and $TX^{1/2}T$ gates to the electron initialized at $\ket{+i}_e$. We observe the corresponding fidelities of $0.79\pm0.07$, $0.73\pm0.07$, $0.71\pm0.11$. We measure $\chi$ for the three cases to be $0.04^{+0.10}_{-0.04}$~bits. }
    \label{fig:single_blind}
\end{figure}

We first focus on the 1QBG, which utilizes the $z$-axis rotation with a hidden angle $\phi$ ($R_z(\phi)$) as its basic building block.
We use the spin-photon gate (SPG) shown in Fig.~\ref{fig:single_blind}A to entangle an electron in an arbitrary initial state $\ket{\psi}_e = (\alpha \ket{\uparrow} + \beta \ket{\downarrow})_e$ with a photon ($\gamma$) through spin state-dependent reflectivity~\cite{robustnode2022, twofridge2024} as follows:
\begin{align*}
    (\ket{0} + \ket{1})_\gamma(\alpha \ket{\uparrow} + \beta \ket{\downarrow})_e
    \xrightarrow{\text{SPG} } (\alpha \ket{0\uparrow} + \beta \ket{1 \downarrow})_{\gamma, e}.
\end{align*}
Due to the inevitable photon loss, this entanglement generation occurs probabilistically with experimental success probability $\eta\sim10^{-3}$ but is heralded by the detection of the photon by the client. We implement $R_z(\phi)$ by first performing a SPG at the server, after which the client measures the photon in the $\ket{\pm_{\phi}}_\gamma = (\ket{0}\pm e^{i\phi}\ket{1})_\gamma/\sqrt{2}$ basis (Fig.~\ref{fig:single_blind}A). We use $s=0(1)$ to denote the measurement results $\ket{+_{\phi}}_\gamma (\ket{-_{\phi}}_\gamma)$, each occurring with equal probability.
By applying the feedback $Z^s$, the resulting final electron state is $\alpha \ket{\uparrow}_e + e^{i\phi} \beta \ket{\downarrow}_e = R_z(\phi)\ket{\psi}_e$, completing the desired $z$-axis rotation on the qubit. 
This feedback is not applied by the server; instead the client implements this by adaptively adjusting the measurement phase of subsequent blind gate(s) as explained in~\cite{SI} section I. No measurement information or feedback actions are needed at the server side. 

Since the client's measurement outcome $s$ is unknown to the servers, they observe a mixed state averaged over $s=0$ and $1$ without any information about the phase $\phi$, rendering the operation blind (Fig.~\ref{fig:single_blind}B).
We  measure an average gate fidelity of $94.8\pm0.3\%$ for the blind $R_z$ gate.
We further characterize the information leakage, as the Holevo information $\chi$ of the client's operations gained by the servers (see \cite{SI} for definition). This is measured by examining the resulting states at the server side over the different client's operations.
For an $n$-qubit operation, $\chi$ ranges from $0$ to $n$~bits, with $\chi=0$ for perfectly blind operations.
We find $\chi$ to be $0.0045^{+0.018}_{-0.0045}$~bits, far below $1$ bit to fully reveal $\phi$~\cite{SI}.

To implement a universal 1QBG, we apply 3 SPGs interleaved with Hadamard gates, realizing $R_z(\phi_3)R_x(\phi_2)R_z(\phi_1)$ on the spin qubit. The adaptive feedback must be carried out in real-time for non-Clifford gates, necessary for universal operations. This is done by applying the $i^{th}$-photon blind rotation $\phi_i'$, as a function of the previously run outcome $s_{i-1}$ and the desired rotation $\phi_i$ (Fig.~\ref{fig:single_blind}C)~\cite{SI}. This protocol generates a three-photon linear cluster state with the electron as the fourth qubit, and the client measures the first, second, and third photon in the $\ket{\pm_{\phi_1}}_{\gamma}$, $\ket{\pm_{\phi_2'}}_{\gamma}$, and $\ket{\pm_{\phi_3'}}_{\gamma}$ bases, respectively (Fig.~\ref{fig:single_blind}C). This implements an arbitrary rotation about an arbitrary axis on a Bloch sphere.
%
To demonstrate the universality of this gate, we implement three different 1QBGs: Identity ($I$), Hadamard ($H$), and $TX^{1/2}T$ - an arbitrarily designed non-Clifford gate. $T$ gate, or the $\pi/4$ phase shift gate, is a non-Clifford gate which is  essential for universal quantum computation. By applying this gate to $\ket{+i}_e = \frac{1}{\sqrt{2}}(\ket{0}_e + i\ket{1}_e)$, we observe the resulting state fidelities to be $0.79\pm0.07$, $0.73\pm0.07$, $0.71\pm0.11$, respectively,
%
%
and $\chi$ for these three gates is $0.04^{+0.10}_{-0.04}$~bits (Fig.~\ref{fig:single_blind}D). 

\begin{figure}[hbt]
    \centering
    \includegraphics[width=0.5\textwidth]{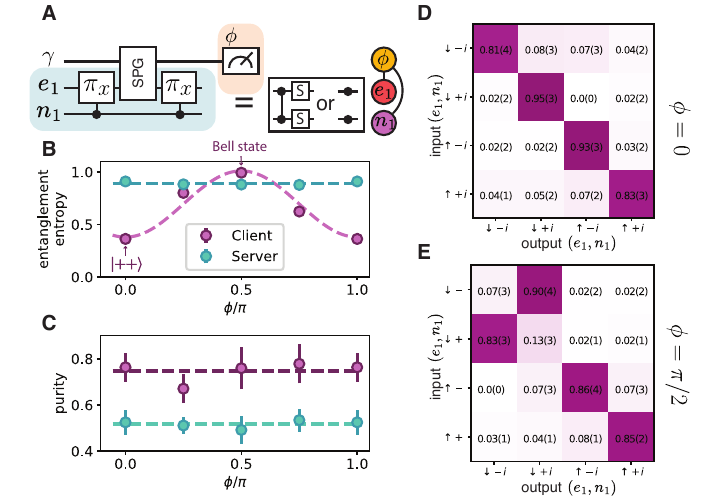}
    \caption{\textbf{Intra-node two-qubit blind gate.} (\textbf{A}) Left: gate sequence for the intra-node 2QBG. We use $e_1$ and $n_1$ as our computational qubits. Right: cluster state after applying the intra-node 2QBG to $\ket{++}$. By varying $\phi$, the entanglement degree between two matter qubits is tuned. (\textbf{B}) Entanglement entropy and (\textbf{C}) state purity as a function of the photon measurement angle set by the client, which ranges from a fully-entangling gate to Identity. At $\phi=0$, we observe a product-state fidelity of $0.85\pm0.01$. At $\phi=\pi/2$, we observe a Bell-state fidelity of $0.85\pm0.02$. We measure $\chi$ over $\phi=\{0, \pi/2\}$ to be $0.032^{+0.12}_{-0.032}$~bits. (\textbf{D}) and (\textbf{E}) show the gate truth table. This is done by initializing  different initial states as inputs and measuring the population distribution of the resulting output states when $\phi=0$ and $\phi=\pi/2$, respectively. $\phi=0 (\pi/2)$ corresponds to the $I$ gate ($S_{e_1} S_{n_1} CZ$ gate). From these truth tables, the client observes the averaged resulting state fidelity of $0.86\pm0.02$ and $0.88\pm0.02$ for the $I$ gate and $S_{e_1} S_{n_1} CZ$ gates, respectively.}
    \label{fig:intranode}
\end{figure}

We next implement an intra-node 2QBG using $e_1$ and $n_1$ spins in server 1. Crucially, the 2QBG can be either an entangling or non-entangling gate based on the client's measurement, while the servers should observe output states independent from this choice. We achieve this with the gate sequence shown in Fig.~\ref{fig:intranode}A consisting of local two-qubit gates and an SPG. 
An SPG with the client's measurement phase $\phi=0$ ($\pi/2$) results in the sequence being $I$ ($S_{e_1}S_{n_1}CZ$), a non-entangling (entangling) gate. Here $S$ is the $\pi/2$-phase shift gate. We emphasize, by utilizing local matter-qubit operations, we can implement this entanglement-hiding gate with fewer probabilistic photon operations compared with all-photonic platforms~\cite{SI, broadbent2009universal}. 

We first probe our intra-node 2QBG by measuring the gate truth table using all combinations of $\ket{\uparrow/\downarrow}_{e_1}$ and $\ket{+i/-i}_{n_1}$ or $\ket{+/-}_{n_1}$ as initial input states. Fig.~\ref{fig:intranode}D, E show the probability distribution of resulting output states when the client measures the photon in the $\ket{\pm_{0}}_\gamma$ and $\ket{\pm_{\pi/2}}_\gamma$ basis, corresponding to a $I$ and $S_{e_1}S_{n_1}CZ$ operation, respectively. 
We then apply the 2QBG to the initial state $\ket{++}_{e_1,n_1}$. 
By choosing $I$ or $S_{e_1}S_{n_1}CZ$ operations, the client observes a resulting product-state fidelity of $0.85\pm0.01$ or Bell-state fidelity of $0.85\pm0.02$, respectively.
We then vary $\phi$, continuously tuning the entangling degree of the gate. The client observes a corresponding change of entanglement entropy of the resulting state, whereas the server always observes an entanglement entropy close to $1$, consistent with the mixed state generated by our protocol (Fig.~\ref{fig:intranode}B). Fig.~\ref{fig:intranode}C shows that the purity of the two-qubit system remains high for the client but stays at $\sim0.5$ for the server, indicating that the system is in a coherent quantum state for the client but appears mixed to the server. We characterize $\chi$ over $\phi= \{ 0, \pi/2\}$ to be 
$0.032^{+0.12}_{-0.032}$~bits, far below $1$ bit required to fully decode the client's choice between entangling and non-entangling gates. 
The complete protocol, dataset, and error budgets are given in \cite{SI}.



\subsection{Distributed two-qubit blind gate}

\begin{figure*}[hbt]
    \centering
    \includegraphics[width=0.75\textwidth]{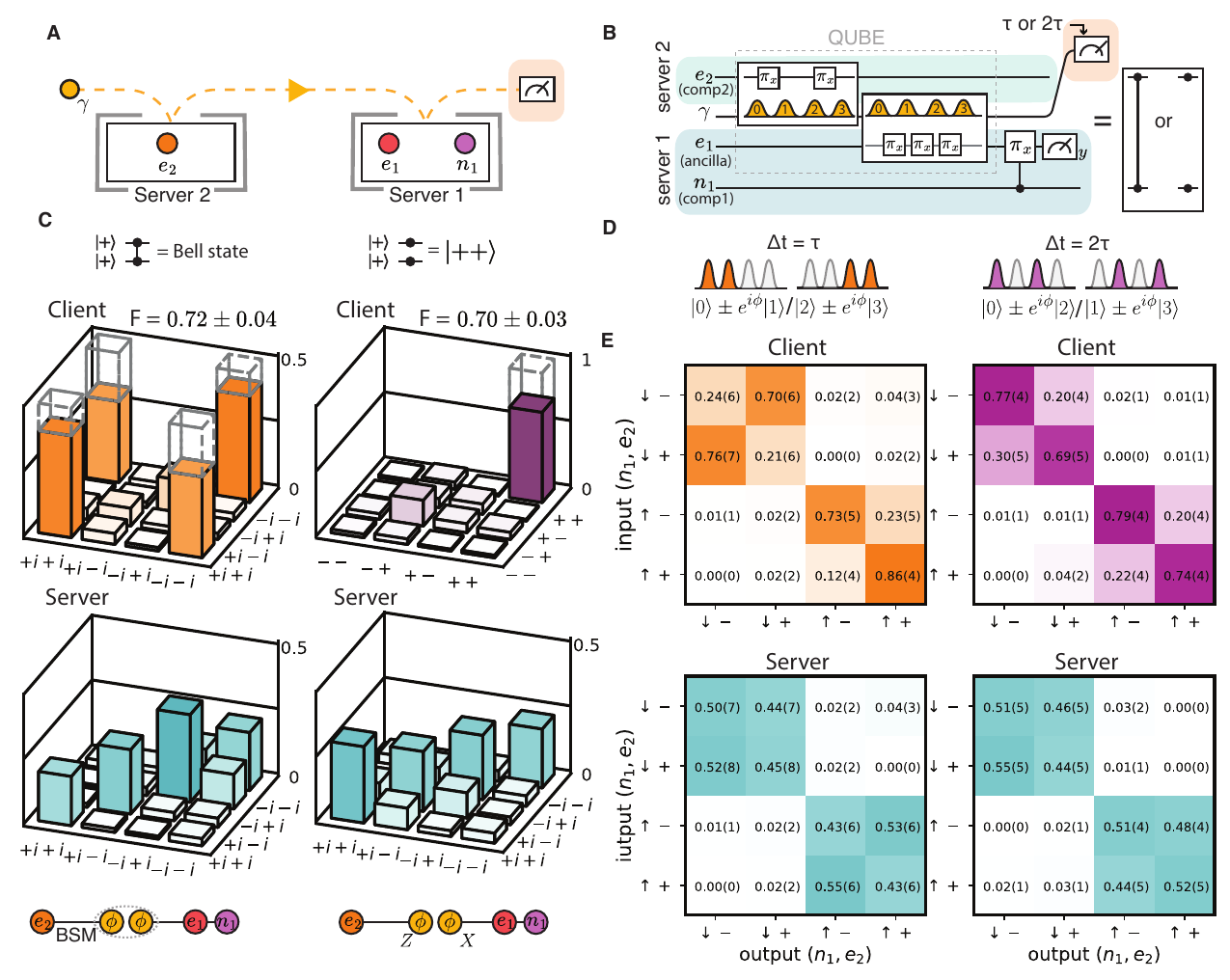}
    \caption{\textbf{Distributed two-qubit blind gate.} (\textbf{A}) Diagram of the experimental setup for the distributed 2QBG. (\textbf{B}) Gate sequence of the distributed 2QBG. $n_1, e_2$ are computational qubits, and $e_1$ acts as an ancilla. (\textbf{C}) Density matrix of $n_1, e_2$ after applying the distributed 2QBG to the initial state $\ket{++}_{n_1, e_2}$. Left (Right), the client observes a Bell-state (product-state) fidelity of $0.72\pm0.04$ ($0.70\pm0.03$) with $CZ$ ($I$) operation. 
    The servers observe indistinguishable density matrices over different client's operations $\{ CZ, I\}$, and $\chi$ over them is $0.12\pm0.06$~bits. Note that $1$ bit is necessary to reveal the client's choice.
    (\textbf{D}) Left: interfering $\ket{0}, \ket{1}$ or  $\ket{2}, \ket{3}$ with a short TDI establishes the entanglement between $e_1, e_2$. Right: interfering $\ket{0}, \ket{2}$ or  $\ket{1}, \ket{3}$ with a long TDI disentangles  $e_1, e_2$ without causing $e_2$ to decohere. (\textbf{E}) Left (Right) column: the gate truth tables for the $CZ (I)$ operations as observed by the client and server. The client observes an averaged resulting state fidelity of $0.76\pm0.03$ and $0.75\pm0.03$ with $CZ$ and $I$ operation, respectively.}
    \label{fig:qube}
\end{figure*}

To utilize computational resources across different servers, we now perform the second type of 2QBG --  a distributed 2QBG. 
We use  $n_1$ at server 1 and $e_2$ at server 2 as computational qubits, while $e_1$ at server 1 is an ancilla qubit that always starts at $\ket{+}_{e_1}$ and is measured out at the end of the sequence (Figs.\ref{fig:qube}A, B). The gate sequence starts with distributing entanglement between $e_1$ and $ e_2$ over the two-server network, such that this entanglement can be switched on or off secretly by the client. This switchable entanglement is achieved by using a 4-time-bin photonic qudit to carve out four combinations of $e_1, e_2$ states using a specially designed gate -- quantum universal blind entanglement gate (QUBE). For any initial state $(\alpha\ket{\uparrow} + \beta\ket{\downarrow})_{e_2}$, the QUBE gate results in the state:
\begin{align*}
    (\ket{0} + \ket{1} + \ket{2} + \ket{3})_{\gamma}(\ket{\uparrow} + \ket{\downarrow})_{e_1}(\alpha\ket{\uparrow} + \beta\ket{\downarrow})_{e_2}
        \\
        \xrightarrow{\text{QUBE}} (\alpha \ket{0}\ket{\uparrow \uparrow} + \beta \ket{1}\ket{\downarrow \downarrow} + \beta \ket{2}\ket{\uparrow\downarrow } + \alpha \ket{3}\ket{\downarrow\uparrow })_{\gamma, e_1, e_2}
\end{align*}
where $\ket{0}_{\gamma}, \ket{1}_{\gamma}, \ket{2}_{\gamma}, \ket{3}_{\gamma}$ are four photonic time bins separated by a time difference $\tau$. To turn on (off) the entanglement, the client secretly switches the photon to a TDI with delay time $\tau$ ($2\tau$) (Fig.~\ref{fig:qube}D). A short TDI interferes $\ket{0}_{\gamma}, \ket{1}_{\gamma}$ or $\ket{2}_{\gamma}, \ket{3}_{\gamma}$ photonic bins, which establishes entanglement between $e_1, e_2$. A long TDI interferes $\ket{0}_{\gamma}, \ket{2}_{\gamma}$ or $\ket{1}_{\gamma}, \ket{3}_{\gamma}$ and keeps them unentangled without decohering our computational qubits. A local $e_1-n_1$ gate is then applied to teleport this switchable entanglement to $n_1$ and $e_2$, completing a switchable gate, $CZ$ or $I$ (Fig.~\ref{fig:qube}B)~\cite{SI}. The gate is  generated with the experimental success probability $\eta \sim 10^{-5}$ and is heralded by single photon detection by the client.

We first characterize our distributed 2QBG by measuring the population transfer with all combinations of $\ket{\uparrow/\downarrow}_{n_1}$ and $\ket{+/-}_{e_2}$ as input states. The client observes an averaged output-state fidelity of $0.76\pm0.03$ and $0.75\pm0.02$ when choosing the $CZ$ and $I$ operation, respectively (Fig.~\ref{fig:qube}E top). 
The servers observe two indistinguishable truth tables since the client's choice is hidden from the servers (Fig.~\ref{fig:qube}E bottom). Through our gate design, the phase of the state observed from the servers is maximally scrambled, making $CZ$ and $I$ indistinguishable to the servers for any arbitrary input state~\cite{SI}. We further apply our distributed 2QBG to the initial state $\ket{++}_{n_1 e_2}$. This generates a Bell state with fidelity $0.72\pm0.04$ or a product state with fidelity $0.70\pm0.03$ when the client applies $CZ$ or $I$, respectively (Fig.~\ref{fig:qube}C top). Over different client's choices, the servers observe two indistinguishable output states (Fig.~\ref{fig:qube}C bottom). We find $\chi$ over $\{CZ, I\}$ to be $0.12\pm0.06$~bits, far below $1$ bit to fully reveal the client's choice. Fig.~\ref{fig:qube}C bottom shows their corresponding cluster states and explained in \cite{SI}.

\subsection{Distributed algorithm with blind operations}

The above demonstrations complete a universal gate set for matter-based BQC with a distributed quantum computing architecture. We now  utilize these ingredients to realize  a Deutsch-Jozsa-type algorithm with hidden oracles. We define a function with $n$ inputs $f: \{0,1\}^n \rightarrow \{ 0, 1\}$, with the constraint that the output over all inputs is either a "constant" value or "balanced" values, i.e., half $0$, half $1$. An unknown party (black box) implements an oracle based on $f$, while a user wants to know if this oracle is constant or balanced by making queries.
A classical algorithm takes maximally $2^{n-1} + 1$ queries, while the Deutsch-Jozsa quantum algorithm takes a single oracle query $O_f$~\cite{DJ1992}. Here, $O_f$ is a quantum operation applied to $n$ computational qubits and one ancilla qubit.

\begin{figure}[hbt]
    \centering
    \includegraphics[width=0.5\textwidth]{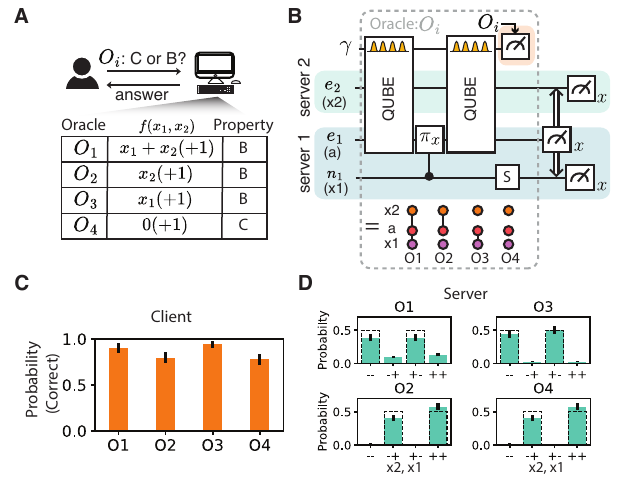}
    \caption{\textbf{Implementation of the distributed algorithm with blind oracles.} (\textbf{A}) Top, high-level schematic diagram. Four oracles are available from the server. The client implements an algorithm for a desired oracle $O_i$, in order to know if it is constant (C) or balanced (B).  
    Bottom, all four oracles, their corresponding functions, and their constant-or-balanced properties for two-input functions. Two functions with trivial bit flip  are grouped together.
    (\textbf{B})  Gate sequence for the Deutsch-Jozsa-type algorithm. $x_1 (x_2)$ denotes the computational qubit that encodes the first (second) input of the function $f(x)$, where $a$ denotes an ancilla qubit. This sequence generates four possible entangled states based on the client's measurement, which corresponds to four oracles. The oracle choice is hidden from the servers. (\textbf{C}) The output of the algorithm. We obtain a probability of $0.90\pm0.05$, $0.80\pm0.06$, $0.94\pm0.04$, $0.78\pm0.06$ of determining the correct outcome for four oracles listed in (\textbf{A}), respectively. (\textbf{D}) The server cannot distinguish the implementation of $O_1$ from $O_3$ and $O_2$ from $O_4$. This 1-bit oracle hiding is characterized by  the information leakage over $\{O_1, O_3 \}$ to be $0.05^{+0.19}_{-0.05}$~bits and  $\{O_2, O_4 \}$ to be $0.07^{+0.14}_{-0.07}$~bits, respectively. Dash bars indicate the theory expectation.}
    \label{fig:dj}
\end{figure}

We implement a Deutsch-Jozsa-type algorithm with four oracles ($O_1 - O_4$, see Fig.~\ref{fig:dj}A) available from the servers. The client chooses an index $i$ and intends to find out if $O_i$ is constant or balanced, without revealing the choice of $i$ to the servers.
We use $n_1, e_2$ as computational qubits and $e_1$ as an ancilla qubit. The servers apply the sequence shown in Fig.~\ref{fig:dj}B, entangling three matter qubits with an $8$-timebin photonic qudit. The client projects the photon in a measurement basis specific to the desired $O_i$, generating a corresponding $e_2-e_1-n_1$ entangled state (Fig.~\ref{fig:dj}B bottom). The server post-selects on the measurement result of the ancilla qubit, 
equivalently acting $O_i$ on the computational qubits
(see \cite{SI} for details). The $X$-basis measurement results of the computational qubits, together with the client's secret photon measurement outcome, determine if $O_i$ is constant or balanced.

Fig.~\ref{fig:dj}C shows the client's implementation of the algorithm with four oracles, with an averaged probability of $0.85\pm0.03$ of correctly determining the constant-or-balanced property. 
By the algorithm design, the inter-node entanglement is hidden from the servers. This makes the servers unable to recognize the implementation of $O_1$ from $O_3$ and $O_2$ from $O_4$, resulting in a $1$-bit oracle hiding. We characterize this oracle hiding by measuring the information leakage over $\{ O_1, O_3 \}$ to be $0.05 ^{+0.19}_{-0.05}$~bits and $\{ O_2, O_4 \}$ to be $0.07 ^{+0.14}_{-0.07}$~bits, both below $1$ bit necessary to fully distinguish between two oracles (Fig.~\ref{fig:dj}D).
Here, we measure the classical information leakage because the output of the algorithm only contains single-basis measurement results.
%
While in our implementation the traditional quantum advantage is not preserved due to the 
probabilistic nature of QUBE gate~\cite{SI}, we emphasize that the delegated implementation on the remote servers allows the client to execute oracle queries by simple instructions without knowing the explicit form of the oracle functions,   demonstrating the operation in which an oracle-based algorithm can be practically useful.


\subsection{Discussion and outlook}
Our experiments demonstrate key ingredients for universal BQC using matter qubits. We implement universal blind gates on a single qubit, and two-qubit blind gates within the same server and distributed over two servers, fulfilling a universal gate set and thereby extending the scope of BQC to matter-based quantum computing systems with a distributed architecture. 

In order to increase the scale and the complexity of the BQC algorithms, the key challenge is associated with the probabilistic gate implementation due to the low photon detection probability ($\eta$), corresponding to 
$\eta \sim 10^{-3}$ and $\eta \sim 10^{-5}$ for the intra-node and inter-node operations, respectively.
Deterministic local gates on matter qubits, however, enable efficient blind gate implementation. Notably, our 2QBG implementation requires only one photon, as opposed to the five photons needed in all-photonic platforms~\cite{SI, broadbent2009universal, fitzsimons2017unconditionally}.
%
%
%
For more complex circuits involving multi-photon detection, the running time of algorithms increases exponentially with gate depth ($D$) as $\sim (1/\eta)^D$. 
This can be circumvented by using additional ancillary qubit memories, which hold quantum information while communication qubits repeat entanglement attempts until success~\cite{Pompili2021, main2024distributed, drmota2024verifiable, Chou2018, Zhong2021} and enables deterministic entanglement generation and deterministic algorithm implementation with the running time scaling linearly with the depth as $\sim D/\eta$~\cite{SI}. 
Our demonstrated building blocks can be directly applied to such memory-based architectures~\cite{SI}.
In our platform, this can be achieved  by using weakly coupled $^{13}\text{C}$ spins~\cite{ PhysRevX.9.031045} as quantum memories. Such ancillary qubits also enable deterministic inter-node operations~\cite{Pompili2021}, which allows  scaling up the number of qubits by interconnecting multiple nodes without the exponential cost of the success rate.

Beyond this specific implementation,  
our methods can apply to other different matter-qubit platforms with matter-photon entangling capabilities. Platforms like neutral atoms or trapped ions can adopt the demonstrated  methods through applying local gates with tens of qubits and transporting them to an optical-access zone for matter-photon entanglement~\cite{main2024distributed, orevi2021}.
%
%
Superconducting qubits can be entangled with microwave photons~\cite{20-qubit-microwavephoton} and use cryogenic links~\cite{Storz2023} or microwave-to-optical transduction~\cite{Jiang2023, Kumar2023} for server-client communication.
%
%
%
%
Finally, we note that deep circuit BQC will require substantial reduction in gate error rates, achievable by integrating matter-photon entanglement with error correction, as proposed recently~\cite{sinclair2024faulttolerantopticalinterconnectsneutralatom, pattison2024fastquantuminterconnectsconstantrate} and explored in relation to fault-tolerant, matter-based BQC in \cite{Baranes2024efficient}.  In combination with recent experimental  advances in quantum error correction~\cite{Bluvstein2023, acharya2024quantumerrorcorrectionsurface, ryananderson2022implementingfaulttolerantentanglinggates, google2023} and error suppression in  matter-photon entanglement~\cite{robustnode2022, twofridge2024, grinkemeyer2024errordetectedquantumoperationsneutral}, our work may pave the way towards eventual realizaion of fault-tolerant BQC.

\balancecolsandclearpage

\section*{Acknowledgments}
We thank Mihir Bhaskar, David Levonian, Denis Sukachev, and Madison Sutula for useful
discussions and experimental help, Chawina De-Eknamkul for support with the tapered-fiber-optical interface, Daniel Riedel for the support with sample annealing, Eugene Knyazev and Francisca Abdo Arias for their insightful discussions and
feedback on the manuscript,
and Jim MacArthur for assistance
with electronics.

\textbf{Funding:}
This work was supported by the AWS Center for Quantum Networking, the National Science Foundation (Grant No. PHY-2012023), NSF Center for Ultracold Atoms, the NSF Engineering Research Center for Quantum Networks (Grant No. EEC-1941583), CQN
(EEC-1941583), and NSF QuSeC-TAQS OMA-2326787. Devices were fabricated at the Harvard
Center for Nanoscale Systems, NSF award no. 2025158.
G.B. acknowledges support from the MIT Patrons of Physics Fellows Society.
Y.Q.H acknowledges support from the A*STAR National Science Scholarship.

{\textbf{Author contributions:}}
Y.-C.W., P.-J.S., A.S., and G.B. planned
the experiment and analyzed the data. G.B., F.M., J. B., and I. W. formulated the theory. Y.-C.W., P.J.S., A.S., Y.Q.H, and C.M.K. built the setup and performed the experiment.  B.M. and E.N.K. fabricated the devices.
All work was supervised by S.F.Y., H.P., M.L., and M.D.L.
All authors discussed the results and contributed to the manuscript. Y.-C.W, P.J.S., A.S., and G.B. contributed equally to this work.

{\textbf{Data availability}}
All data related to the current study are available from the corresponding author upon
reasonable request.

\textbf{Code availability}
All analysis code related to the current study are available from the corresponding author
upon reasonable request.

\textbf{Competing Interests}
The authors declare no competing interests.


\bibliographystyle{science}
\bibliography{scibib}



\clearpage
\onecolumngrid



\newpage


\renewcommand{\thefigure}{S\arabic{figure}}
\renewcommand{\thetable}{S\arabic{table}}
\renewcommand{\theequation}{S\arabic{equation}}
\renewcommand{\thepage}{S\arabic{page}}
\setcounter{figure}{0}
\setcounter{table}{0}
\setcounter{equation}{0}
\setcounter{page}{1} 











\renewcommand{\thefigure}{S\arabic{figure}}
\renewcommand{\thetable}{S\arabic{table}}
\renewcommand{\theequation}{S\arabic{equation}}
\renewcommand{\thepage}{S\arabic{page}}
\setcounter{figure}{0}
\setcounter{table}{0}
\setcounter{equation}{0}
\setcounter{page}{1} 
\setcounter{section}{0}

\makeatletter

\begin{center}
\large \textbf{Supplementary Materials}
\end{center}


\setcounter{secnumdepth}{5}
\setcounter{tocdepth}{5}

\section{Protocol}

\subsection{Brickwork and universal blind unit cell}

In measurement-based quantum computing (MBQC), the implementation of universal blind quantum computing (UBQC) utilizes a resource state known as the brickwork state~\cite{broadbent2009universal, fitzsimons2017private}. This state can be interpreted as a circuit, where each row represents a single logical qubit, which propagates through the circuit via a sequence of unitary operations secretly controlled by the client, as illustrated in Fig.~\ref{fig:SI_unit_cells}B. The brickwork state, or its corresponding circuit which we hereafter call the brickwork circuit, is constructed from "universal blind cells". Each universal blind cell is capable of implementing a universal set of gates on two qubits. The flexibility of these cells allows for arbitrary single qubit rotations on each qubit as well as as a CZ or identity gate, depending on the measurement bases selected within the unit cell~\cite{fitzsimons2017private}. One example of a universal blind cell implementation is shown in Fig.~\ref{fig:SI_unit_cells}B.

More formally, the functionality of the universal blind cell is to allow the remote client to perform a universal set of gates over two qubits, including $\{U_{1q} (\text{for both qubits}), CZ\}$. Here $U_{1q}$ denotes a universal gate set of single qubit operations. The actual operation is unknown to the servers, as they always observe the same quantum channel, denoted as $\mathcal{N}(\rho)$, regardless what the actual gate is applied (Fig.~\ref{fig:SI_unit_cells}F). The circuit shown in Fig.~\ref{fig:SI_unit_cells}B is an example taken from \cite{broadbent2009universal} to fulfill the above requirement, but it is not the only way to realize a universal blind cell.

Because each unit cell is universal, when patterned periodically as shown in Fig.~\ref{fig:SI_unit_cells}A, the brickwork state or circuit can implement any arbitrary quantum circuit. This structure therefore is a universal resource for quantum computing~\cite{broadbent2009universal}. The blind property of the universal blind cell hides the actual gate and therefore the whole circuit from the server. The design of the brickwork state or circuit ensures that all gates have the same "shape" on the surface, thereby concealing the nature of the computation from the server, with only the topology of the brickwork state being revealed.

\subsubsection{Brickwork Decomposition}

\begin{figure}
    \centering
    \includegraphics[width=\linewidth]{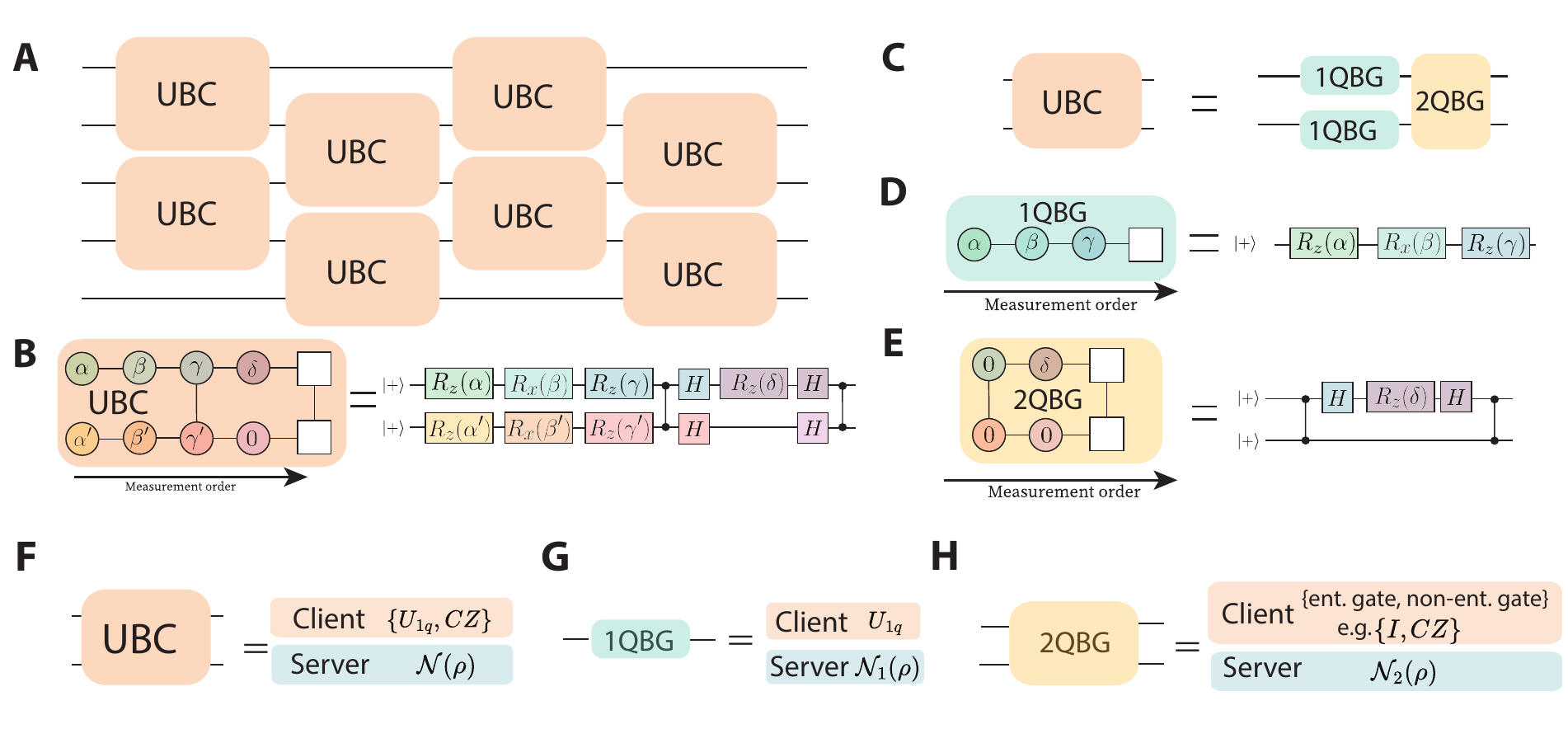}
    \caption{\textbf{Universal Blind Quantum Computing Decomposition.} (\textbf{A}) The structure of measurement-based quantum computing using the brickwork state repeating a periodic pattern of universal blind cells (UBC). (\textbf{B}) Illustration of the universal blind cell in a circuit-based context, showing the different gates that can be generated using different measurement angles. (\textbf{C}) The UBC can be decomposed into smaller sub-cells of 1-qubit blind gates (1QBG) and 2-qubit blind gates (2QBG). (\textbf{D,E}) Circuit-based illustration of the 1QBG and 2QBG, showing the universal set of gates that can be generated by choosing different measurement angles. (\textbf{F}) A universal blind cell allows the client to implement a universal quantum gate set, including universal single-qubit operations $U_{1q}$ and $CZ$, while the servers always observe a quantum channel $\mathcal{N}(\rho)$. (\textbf{G}) 1QBG: the client can implement $U_{1q}$, while the servers always observe a quantum channel $\mathcal{N}_1(\rho)$. (\textbf{H}) 2QBG: the client can choose to implement an entangling or non-entangling gate (e.g., $I$ or $CZ$), while the servers always observe a quantum channel $\mathcal{N}_2(\rho)$.}
    \label{fig:SI_unit_cells}
\end{figure}

We here consider two more basic logic gates: the one-qubit blind gate (1QBG) and two-qubit blind gate (2QBG)~\cite{Baranes2024efficient},
We here define the requirements for 1QBG or 2QBG:
\begin{itemize}
    \item 1QBG allows the client to apply a \textbf{universal set of single qubit gates}, denoted as $U_{1q}$, while the actual operation remains hidden from the server as they always see the same quantum channel, denoted as $\mathcal{N}_1(\rho)$, regardless of the real applied gate (Fig.~\ref{fig:SI_unit_cells}G).
    
    \item 2QBG allows the client to apply a \textbf{maximally entangling gate or non-entangling gate}, for example, $\{ I, CZ\}$ .This choice remains hidden from the server as they always see the same quantum channel, denoted as $\mathcal{N}_2(\rho)$ (Fig.~\ref{fig:SI_unit_cells}H).
\end{itemize}

Note that any maximally entangling gate can be decomposed into single-qubit operations and a $CZ$. Therefore, 2QBG provides an essential entangling element $CZ$ for the universal gate set. All of $U_{1q}$ is covered by 1QBG. Combining two 1QBGs (one on each qubit) and one 2QBG as shown in Fig.~\ref{fig:SI_unit_cells}C, the allowed operations therefore include $\{ U_{q} \text{ (for both qubits)},  CZ\}$, which fulfills the required gate set. The servers always observe the same quantum channel,
\begin{align*}
\mathcal{N}_2^{(q1 + q2)}( \mathcal{N}_1^{(q2)}( \mathcal{N}_1^{(q1)}(\rho) ) )
\end{align*}, which does not contain any information about the gate operation, therefore satisfying the blindness requirement. Thus, we can use 1QBG and 2QBG to construct a universal blind cell. 
We take the Fig.~\ref{fig:SI_unit_cells}B as an example, and its decomposition is shown in Fig.~\ref{fig:SI_unit_cells}D,E. 1QBG can implement a universal set of single qubit gates by decomposing the gate into 3 Euler angles as illustrated in Fig.~\ref{fig:SI_unit_cells}D, so that we can construct a rotation of any arbitrary angle around an arbitrary axis. Fig.~\ref{fig:SI_unit_cells}E shows that the 2QBG sub-cell can generate a maximally entangling gate or an identity gate by choosing $\delta = \pi/2$ or $\delta=0$, respectively.
The foundational papers on UBQC~\cite{broadbent2009universal, fitzsimons2017private}, as well as the work on matter-photon hybrid BQC~\cite{Baranes2024efficient} further explore these concepts in more detail.

We note that for any blind gate controlled by the client's measurement basis: 
\begin{itemize}
    \item 1. The blind gate allows the client to deterministically control the qubits at the servers without revealing classical information about the client's decisions.
    \item 2. The blind gate applies a scrambling channel (e.g., $\mathcal{N}_2$ for 2QBG) to the qubit(s) at the servers, which can be decoded with the client secret measurement outcomes. Only the client can decode the operation and recover the unitary quantum operation (e.g., $\{\text{CZ}, I\}$). The scrambling channel needs to be designed such that the servers always observe the same quantum channel (e.g., $\mathcal{N}_2$) independent of the client's operations (e.g., $\{\text{CZ}, I\}$) so that no information is leaked even if the server probes the quantum state before or after the operation.
\end{itemize}
For information-theoretic security, these two properties are necessary for any blind gates in the sequence. Property 1 allows the client to run the sequence deterministically without directly sending the server any classical information. Property 2 guarantees that the quantum state of the server's qubit(s) (without access to the client's measurement results) does not contain information about the client's decisions. This is so that even a dishonest server that might probe the quantum state before and after an operation to learn the nature of the operation still cannot gain any information. To make sure the server implements the desired gate sequences, the cheating behavior of the servers can be detected through 
verification protocols~\cite{fitzsimons2017unconditionally, barz2012demonstration, barz2013experimental, drmota2024verifiable} by interleaving computation rounds with test rounds.

Furthermore, if all blind gates satisfy both property 1 and property 2, they can be used as more flexible building blocks for blind circuits, for example, by arranging the sequence of blind gates in a non-uniform way or interleaving blind gates with deterministic non-blind gates in the circuit to boost algorithm performance~\cite{Baranes2024efficient}. In our work, the most fundamental building blocks include $z$-axis blind rotations (1QBG is composed of three $z$-axis blind rotations) and two different types of 2QBG. As shown in the main text and the later section, they all satisfy properties 1 and 2.

\begin{figure}
    \centering
    \includegraphics[width=1\linewidth]{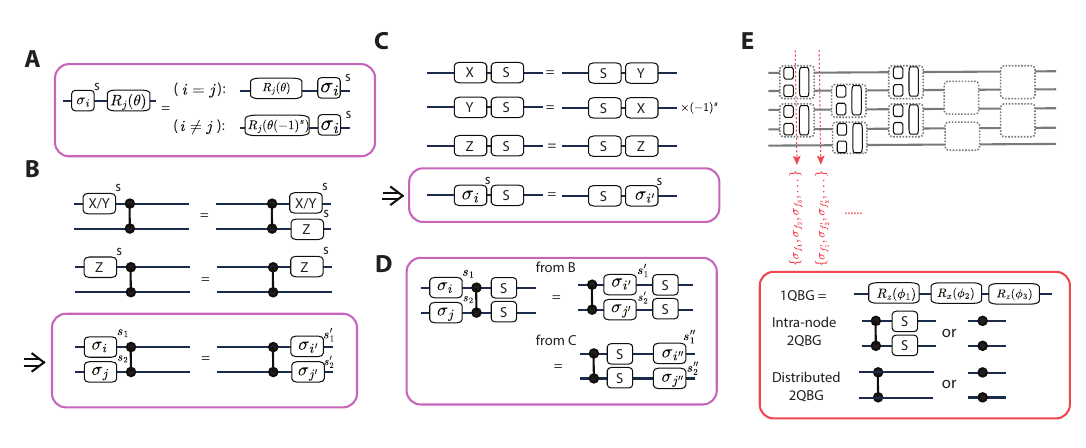}
    \caption{\textbf{Propagation of feedback operators.} (\textbf{A}) shows that the sequential single qubit rotations can be handled by changing subsequent applied angles by the client, while propagating Pauli-type corrections that can be applied in post-processing. (\textbf{B}) shows how sequential CZ gates can be handled and that the corrections are also Pauli-type. (\textbf{C}) Top,  $X, Y$ operations can propagate through $S$, while a global phase, -1, is applied for the propagation of $Y$. This global phase does not affect the operation. Down, the Pauli-type feedback can propagate through $S$ gate. (\textbf{D}) shows the Pauli operation can propagate through $S_1 S_2 \text{CZ}$ and remains Pauli-type. (\textbf{E}) Top: by storing the type of Pauli-operator feedback the client needs to apply as well as adjusting the subsequent blind gates, the circuit can be applied deterministically. Bottom: blind quantum gate set.}
    \label{fig:adapt}
\end{figure}

\begin{figure}
    \centering
    \includegraphics[width=1\linewidth]{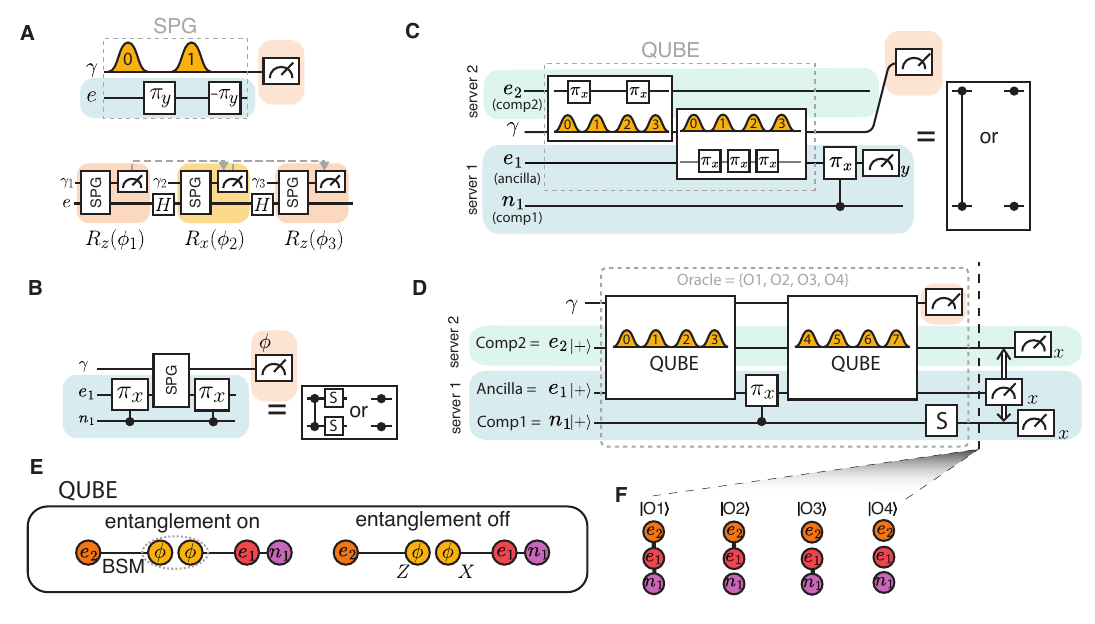}
    \caption{\textbf{Gate sequences for 1QBG, 2QBG, Deutsch-Jozsa-type algorithm.} (\textbf{A}) 1QBG; (\textbf{B}) Intra-node 2QBG. (\textbf{C}) Distributed 2QBG. (\textbf{D}) Deutsch-Jozsa-type algorithm. (\textbf{E})  Cluster states after applying the distributed 2QBG to $\ket{++}$ when entanglement is on or off. (\textbf{F}) Cluster states corresponding to  $\ket{O_{i,i=1\sim 4}}$}
    \label{fig:all_seq}
\end{figure}

\subsection{Adaptive Feedback \label{adap_feed}}

With Pauli-type correction feedback, the feedback operations can be realized by adaptively adjusting subsequent operations and propagating corrections that can be applied in data post-processing~\cite{anders2010ancilla}. We here consider the single-qubit Pauli-type feedback $\sigma_i^s$, which means we apply a feedback gate $\sigma_i$ (Identity) if the measurement outcome $s = 1 (0)$. We note that this assumption is consistent in our work as all shown blind gates are with Pauli-type feedback. To show that any gate can be implemented deterministically, we demonstrate that the feedback corrections can be propagated through all gates in our gate set:
\begin{align}
    \text{1QBG} &\xrightarrow{} U_{1q}\\
    \text{intra-node 2QBG} &\xrightarrow{} \{ I, S_1 S_2 \text{CZ} \}\\
    \text{distributed 2QBG} &\xrightarrow{} \{I, \text{CZ}\}
\end{align}, while keeping the form of Pauli operators after the propagation. The propagation does not affect the implementation of blind operations on the servers, and adjusting the client's  measurement angles is sufficient to complete the propagation. By propagating the feedback operations to the end of the computation, the client can apply them in post-processing by flipping or not flipping the final qubit measurement results. Therefore, the client only needs to store and keep track of the type of Pauli-operator feedback after each propagation step, as well as adjust the measurement angle of subsequent blind gates (Fig.~\ref{fig:adapt}E top). This shows an arbitrary gate sequence can be implemented deterministically and blindly with a one-way communication. The propagation of Pauli-type operations for all of our gate sets is explained below.

The equation in Fig.~\ref{fig:adapt}A shows that the feedback propagation over a single-qubit rotation can be handled by changing the client's applied measurement angles, while the feedback correction remains Pauli-type after a propagation step. Fig.~\ref{fig:adapt}B shows the same is true for CZ gates. Fig.~\ref{fig:adapt}C shows that a Pauli-type feedback correction can propagate through $S = \ket{0}\bra{0} + i \ket{1}\bra{1} = e^{i\pi/4}(1 - iZ)$ and remain Pauli-type. By using Fig.~\ref{fig:adapt}B, C, we show the same is true for gate $S_1 S_2 \text{CZ}$ (Fig.~\ref{fig:adapt}D). Fig.~\ref{fig:adapt}A, B, D correspond to our 1QBG, intra-node 2QBG and distributed 2QBG, all the blind operations in our work.

For example, in the 1QBG, the client performs the gates $R_z(\phi_1)$, $R_x(\phi_2)$, $R_z(\phi_3)$ in a row. Using the propagation rule in Fig.~\ref{fig:all_seq}A bottom, the client must measure the $i$-th photon in the $\ket{\pm_{\tilde{\phi}_i}}$ basis, with $\tilde{\phi}_i = (-1)^{s_{i-1}}\phi_i$ adjusted by the $(i-1)$-th measurement outcome.

\subsection{1QBG protocol} 
With the outcome of three photon measurements $s_1, s_2, s_3$, we propagate the feedback to the end of the computation (see section \ref{adap_feed}) and get the feedback operator $Z_1^{s_1 + s_3} X^{s_2}$. With the input quantum state $\rho_I$, the 1QBG (shown in Fig.~\ref{fig:all_seq}A) applies $R_z(\phi_3)R_x(\phi_2)R_z(\phi_1)$, resulting in output state $\rho_o$. Without any feedback corrections, the output as seen by the servers is a mixed state of all outcomes $s_1, s_2, s_3$,
\begin{align}
    \rho_o^{\text{server}} = \frac{1}{8}\sum_{s_1, s_2, s_3 = \{0,1 \}^3 } Z_1^{s_1 + s_3} X^{s_2} \rho_o X^{s_2} Z_1^{s_1 + s_3}  = \frac{1}{2}I,
\end{align}
which is effectively a maximally depolarizing channel returning a maximally mixed state, $\frac{1}{2}I$. The corresponding quantum channel is
\begin{align}
    \mathcal{N}_1(\rho) =  \frac{1}{2}I
\end{align}.

\subsection{Intra-node 2QBG protocol\label{sec:intra_2qbg}}
We define the projecting operators
\begin{align}
    P_0 := \ket{00}\bra{00} + \ket{11}\bra{11} \\
    P_1 := \ket{01}\bra{01} + \ket{10}\bra{10} 
\end{align}
The intra-2QBG operation (shown in Fig.~\ref{fig:all_seq}B) implements the gate $U_{\text{intra}}^{\phi}=P_0 + e^{i\phi} P_1$ with the client's measurement basis $\ket{\pm}_{\phi}$. Depending on the photon measurement outcome of the SPG ($s$), the feedback operation ($Z_s$) is propagated and spread to two qubits (see Sec.\ref{adap_feed}) and takes the form,
\begin{align}
    F_s = Z_1^s Z_2^s = P_0 - P_1
\end{align}

\subsubsection{Information leakage}
One can show that
\begin{align}
    \sum_{s=0,1} F_s &U_{\text{intra}}^{\phi}\rho_I (U_{\text{intra}}^{\phi})^\dagger  F_s^\dagger = P_0 \rho_I P_0 + P_1 \rho_I P_1
\end{align}

Therefore, for any arbitrary measurement angle $\phi$, the resulting quantum channel observed from the server is always
\begin{align}
    \mathcal{N}_2(\rho) = P_0 \rho_I P_0 + P_1 \rho_I P_1,
\end{align}
which contains no information about $\phi$. Note that $\phi=\pi/2$ and $0$ corresponds to $S_1 S_2 \text{CZ}$ and $\text{I}$ operations, so hiding information about $\phi$  hides whether an entangling or non-entangling gate was applied.


\subsection{Distributed 2QBG protocol\label{sec:distr_2qbg}}
The gate sequence for the distributed 2QBG is shown in Fig.~\ref{fig:all_seq}C. For simplicity, we first define the four parameters, $\{E, p, s_1, s_2\}$, explained below. The variable $E = \{ 0, 1\}$ represents the entanglement switch, which is secretly decided by the client's measurement choice. Specifically,
\begin{align*}
    E: 0 \xrightarrow{} \text{entanglement off}\\
      1 \xrightarrow{} \text{entanglement on},
\end{align*}
 The public variable $p$ denotes the ancilla ($e_1$) measurement outcome after the distributed 2QBG:
\begin{align*}
    &\text{p=0 := measure $e_1$ state to be $\ket{+i}_{e_1}$ }\\
    &\text{p=1 := measure $e_1$ state to be  $\ket{-i}_{e_1}$},
\end{align*}
and $s_1, s_2$ is the client's photon measurement outcome, shown in the Table~\ref{tab:E_s2_table}.

\subsubsection{Gate operation}
Assuming the two computational qubits $e_2, n_1$ are in an arbitrary initial and the ancilla $e_1$ is initialized in $\ket{+}$, we have the general initial state:
\begin{align*}
    (\alpha \ket{00} + \beta \ket{01} + \gamma \ket{10} + \delta \ket{11})_{e_2, n_1} (\ket{0} + \ket{1})_{e_1}
\end{align*}

After applying the QUBE gate sequence (see main text), which carves out the two electron qubits state with corresponding time bins, $\ket{0,1,2,3}_{\gamma} \xrightarrow{} \ket{01, 10, 11, 00}_{e_2, e_1} $, the state becomes,
\begin{align}
\begin{split}
        \ket{\psi}_{\gamma, e_2, n_1, e_1} =& ( \alpha \ket{3}\ket{000} + \alpha \ket{0}\ket{001} + \\
    &\beta \ket{3}\ket{010} + \beta \ket{0}\ket{011} + \\ &\gamma \ket{1}\ket{100} + \gamma \ket{2}\ket{101} + \\
    &\delta \ket{1}\ket{110} + \delta \ket{2}\ket{111})_{\gamma, e_2, n_1, e_1}
\end{split}
\label{eq:qudit_cav_carving}
\end{align}
 
Based on Eq.\ref{eq:qudit_cav_carving}, by selecting the photon time bins based on Table~\ref{tab:E_s2_table}, and following the sequence of Fig.~\ref{fig:all_seq}C, the resulting state can be represented as 
\begin{align}
    U^{E=1, s_1, s_2, p} = Z_1^{s_1+p+s_2} Z_2^{p+s_2} S_1 CZ\\
    U^{E=0, s_1, s_2, p} = Z_1^{s_1+s_2} Z_2^{p+s_2} S_1,
\end{align}.
where we ignore global phases.
By feeding back on qubits  $F_{s_1, s_2, p, E=1} := S_1 Z_1^{s_1+p+s_2} Z_2^{p+s_2}$ or $F_{s_1, s_2, p, E=0} := S_1 Z_1^{s_1+s_2} Z_2^{p+s_2} $ when the entanglement is on (off), we have
\begin{align}
    U^{E=1, s_1, s_2, p} = CZ\\
    U^{E=0, s_1, s_2, p} = I
\end{align}

\subsubsection{Information leakage}
Without information $s_1, s_2$ variables about the measurement outcomes, the servers effectively see a dephasing channel applied to the system, 
preventing the server from distinguishing $I$ from $CZ$ gates. Specifically, for any input state $\rho_I$,
\begin{align}
\begin{split}
     \sum_{s_1, s_2} F_{s_1, s_2, p, E=0} \rho_I F^{\dagger}_{s_1, s_2, p, E=0} = \sum_{s_1, s_2} F_{s_1, s_2, p, E=1} CZ \rho_I CZ F^{\dagger}_{s_1, s_2, p, E=1} \\
     = \frac{1}{4}(\rho + Z_1 \rho Z_1 + Z_2\rho Z_2 + Z_1 Z_2 \rho Z_1 Z_2 )
\end{split}
\end{align}

The server always observes the same quantum channel regardless of the client's choice to apply $I$ or $CZ$. Thus the gate operation is blind, and the corresponding quantum channel is
\begin{align}
    \mathcal{N}_2(\rho) = \frac{1}{4}(\rho + Z_1 \rho Z_1 + Z_2\rho Z_2 + Z_1 Z_2 \rho Z_1 Z_2 ).
\end{align}

\subsubsection{Cluster state representation}
As shown in Fig.~\ref{fig:all_seq}E, if the initial state of the computational qubits $e_2, n_1$ is $\ket{++}$, applying QUBE gate (section.\ref{sec:distr_2qbg}) results in the state
\begin{align}
    (\ket{0}\ket{01} + \ket{1}\ket{10} + \ket{2}\ket{11} + \ket{3}\ket{00})_{\gamma, e_2, e_1}.
    \label{eq:qube_qudit_carving}
\end{align}

Since a 4 time bin photonic qudit provides 4-dimensional Hilbert space, it can equivalently be thought of as two qubits as follows:
\begin{align}
    \ket{0}_{\gamma_2} \ket{0}_{\gamma_1} := \ket{0}_\gamma\\
    \ket{0}_{\gamma_2} \ket{1}_{\gamma_1} := \ket{3}_\gamma\\
    \ket{1}_{\gamma_2} \ket{0}_{\gamma_1} := \ket{2}_\gamma\\
    \ket{1}_{\gamma_2} \ket{1}_{\gamma_1} := \ket{1}_\gamma
\end{align}.
Eq.\ref{eq:qube_qudit_carving} then becomes
\begin{align}
        ( \ket{00}\ket{01} +  \ket{11}\ket{10} + \ket{10}\ket{11} + \ket{01}\ket{00})_{\gamma_2, \gamma_1, e_2, e_1},
        \label{eq:qube_cluster}
\end{align}
where the first (second) photon is only entangled with the $e_1$ ($e_2$). Including the local two-qubit gate on $e_1, n_1$ after the QUBE gate operation, the entangled state looks like Fig.~\ref{fig:all_seq}E, up to local single-qubit operation. 
Projecting this resource state onto the $\ket{0}_\gamma \pm \ket{1}_\gamma$ qudit basis is equivalent to the two photon Bell-state measurement (BSM) $\ket{0}_{\gamma_2}\ket{0}_{\gamma_1} 
\pm \ket{1}_{\gamma_2}\ket{1}_{\gamma_1}$. By performing this BSM, $e_1, e_2$ become entangled (see Fig.~\ref{fig:qube}C bottom left or Fig.~\ref{fig:all_seq}E left). The same applies to a projection onto the $\ket{2}_\gamma \pm \ket{3}_\gamma$ qudit basis.

Similarly, projecting this resource state onto the $\ket{0}_\gamma \pm \ket{2}_\gamma$ qudit basis or $\ket{1}_\gamma \pm \ket{3}_\gamma$ qudit basis is equivalent to projecting onto the two photon basis $(\ket{0} \pm \ket{1})_{\gamma_2}\ket{0}_{\gamma_1}$ or $(\ket{0} \pm \ket{1})_{\gamma_2}\ket{1}_{\gamma_1}$, respectively. This is a $X_{\gamma_2} Z_{\gamma_1}$ basis measurement on the two photons, which does not entangle $e_1, e_2$. Note that to correctly link to the cluster state picture, additional $H_{\gamma_2} H_{\gamma_1}$ gates should be applied to the photons to convert Eq.\ref{eq:qube_cluster} from two separate Bell states into two separate cluster states. These Hadamard gates converts the measurement basis from $X_{\gamma_2} Z_{\gamma_1}$ to $Z_{\gamma_2} X_{\gamma_1}$, as shown at the Fig.~\ref{fig:qube}C bottom right and Fig.~\ref{fig:all_seq}C right.

\subsection{Deutsch-Jozsa-type algorithm}

We use $e_2$, $n_1$ as our two computational qubits and $e_1$ as our query ancilla qubit.  In our experiment, the sequences (Fig.~\ref{fig:all_seq}D) use two QUBE gates with a local nuclear-controlled electron $\pi$ gate in the middle, and we only herald one-photon events over these two QUBE gates: we only select events that either one photon comes at the first QUBE gate and zero photon arrives at the second QUBE gate, or vice versa. 

\begin{figure}
    \centering
    \includegraphics[width=0.66\linewidth]{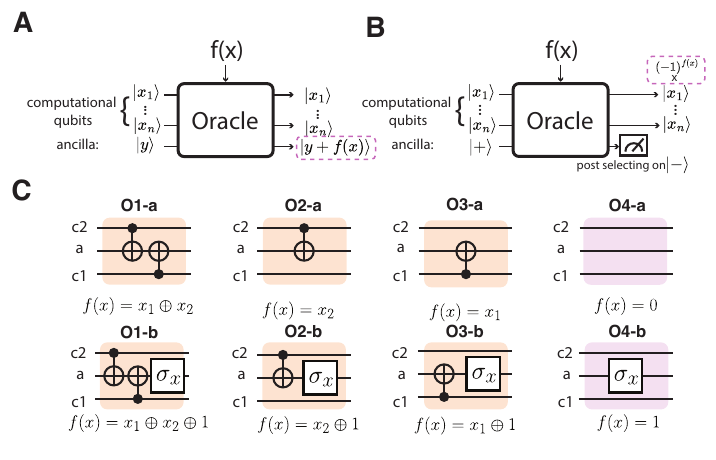}
    \caption{\textbf{Deutsch-Jozsa algorithm} (\textbf{A}) The Oracle in the standard Deutsch-Jozsa algorithm. The output function is encoded on the ancilla qubit. (\textbf{B}) The Oracle in our modified Deutsch-Jozsa-type algorithm. We fix the initial state of our ancilla qubit to be $\ket{+}$ and measure out at $\ket{-}$. The function output is encoded at the global phase. Here $x_1, x_2... x_n$ are either $0$ or $1$ in \textbf{A}  or \textbf{B}. That is to say, they are all on computational bases.
    (\textbf{C}) All eight oracles for $n=2$ Deutsch-Jozsa algorithm.}
    \label{fig:DJ_oracle}
\end{figure}

\begin{figure}
    \centering
    \includegraphics[width=0.66\linewidth]{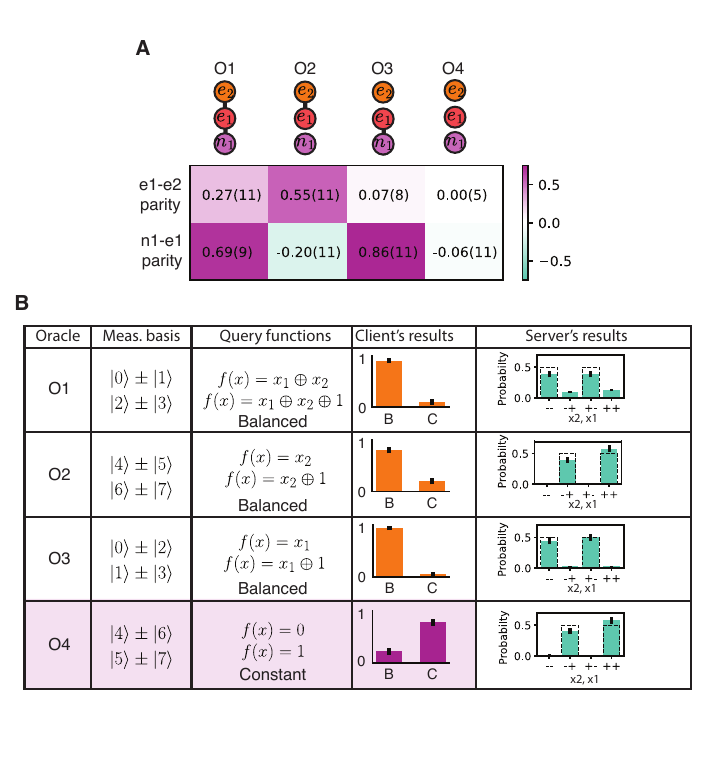}
    \caption{\textbf{Deutsch-Jozsa-type algorithm.} (\textbf{A}) Measured $e_1, e_2$ parity and $e_1, n_1$ parity for four cluster states generated by the protocol for our Deutsch-Jozsa-type algorithm. These four cluster states correspond to $\ket{O_{i, i=1\sim4}}$. (\textbf{B}) Experimental data for four oracles. $O_1, O_2, O_3$ are all three balanced oracles, while $O_4$ is the only constant oracle. We obtain the probability $0.90\pm0.05$, $0.80\pm0.06$, $0.94\pm0.04$, $0.78\pm0.06$ for determining the correct outcomes for $O_1, O_2, O_3, O_4$, respectively. The outcomes observed from the servers are shown at the rightmost column. The classical information leakage over $\{ O_1, O_3\}$ is $0.05^{+0.19}_{-0.05}$ bits, and the classical information leakage over  $\{ O_2, O_4\}$ is $0.07^{+0.14}_{-0.07}$ bits.}
    \label{fig:DJ_exp_data}
\end{figure}

In the standard Deutsch-Jozsa algorithm, to find out if a function $f$ is constant or balanced, an oracle (or black box) $O_f$ is implemented such that $O_f|x\rangle_c|y\rangle_a = |x\rangle_c|y + f(x)\rangle_a$, where $x=\{0,1\}^n$ here for computational qubits (denoted as $c$) and $y=\{0, 1\}$ here for the query ancilla qubit (denoted as $a$), as shown in Fig.~\ref{fig:DJ_oracle}A. To run the algorithm, the ancilla qubit starts at $\ket{-}_a$ and the computational qubits start at $\frac{1}{2^{n/2}}\sum_x \ket{x}_c$. After an oracle is applied, the computational qubit state is $\frac{1}{2^{n/2}} \sum_x (-1)^{f(x)} \ket{x}_c$ while the ancilla qubit remains in the state $\ket{-}_a$ and is decoupled from the other qubits~\cite{DJ1992}. In conclusion, with the ancilla qubit and the corresponding oracle, it transforms the computational qubit states as follows:
\begin{align}
    \frac{1}{2^{n/2}} \sum_x \ket{x}_c \xrightarrow{O_f}  \frac{1}{2^{n/2}}\sum_x \ket{x}_c (-1)^{f(x)}.
    \label{eq:dj}
\end{align}
If the function $f(x)$ is constant, the computational qubit state is $\frac{1}{2^{n/2}}\sum_x \ket{x}_c = \ket{+}_c^{\otimes n}$. Otherwise, if the function $f(x)$ is balanced, the probability of measuring the state $\ket{+}_c^{\otimes n}$ is
\begin{align}
    \frac{1}{2^{n}}\sum_x (-1)^{f(x)} = 0
\end{align}
.

Our gate sequence (Fig.~\ref{fig:all_seq}D), by selecting different photon measurement bases, generates the experimental operation listed in Table~\ref{tab:DJ_exp_table}. Due to the cavity carving nature of the operation, the operation is not unitary over the three matter qubits ($e_2, e_1, n_1$). We modify the oracle by first initializing the ancilla qubit ($e_1$) to $\ket{+}_a$. After applying our gate sequence, we post select on the resulting state of the query ancilla qubit $\ket{-}_a$. This encodes the output of functions on the global phase of computational qubits. Specifically, for $ x=\{0,1\}^n$, the gate sequence implements 
\begin{equation} 
\begin{split}
&|x\rangle_c|+\rangle_a\\
&\xrightarrow{\text{1. Servers apply QUBE1, local 2-qubit gate, QUBE2  (see Fig.~\ref{fig:all_seq}D)}}\\
&\xrightarrow{\text{2. Based on the targeted oracle, Client chooses the photon measurement basis in Table~\ref{tab:DJ_exp_table}}}\\
&\xrightarrow{\text{3. Server post selecting on } |-\rangle_a}(-1)^{f(x)}|x\rangle_c
\end{split}
\label{eq:dj_output_phase}
\end{equation}
(check Table~\ref{tab:DJ_exp_table}). This implements an oracle based on a specific function, with the output of the function imprinted on the phase of the computational qubits (see Eq.~\ref{eq:dj_output_phase} and Fig.~\ref{fig:DJ_oracle}B). By choosing the initial state of computational qubits to be a superposition state of all computational bases, $\frac{1}{2^{n/2}}\sum_x{\ket{x}_c}$, the state after applying this oracle on the computational qubits is the same as  Eq.\ref{eq:dj},
\begin{align}
    \frac{1}{2^{n/2}} \sum_x \ket{x}_c \xrightarrow{\text{Protocols in Fig.~\ref{fig:all_seq}D with } O_f \text{(see Eq.~\ref{eq:dj_output_phase})}}  \frac{1}{2^{n/2}}\sum_x \ket{x}_c (-1)^{f(x)}
    \label{eq:dj_modified}
\end{align}.
Therefore, our Deutsch-Jozsa-type algorithm performs the same actions on computational qubits as Deutsch-Jozsa algorithm, which suffices to answer the constant-or-balanced question with eight oracles in Fig.~\ref{fig:DJ_oracle}.

\begin{table}
    \centering
    \begin{tabular}{c|c|c|c}
        E & $s_1$ & $s_2$ & client's photon measurement basis \\\hline
        1 & 0 & 0 & $\ket{0} + \ket{1}$\\
        1 & 1 & 0 & $\ket{0} - \ket{1}$\\
        1 & 0 & 1 & $\ket{2} + \ket{3}$\\
        1 & 1&  1 & $\ket{2} - \ket{3}$\\
        0 & 0 & 0 & $\ket{0} + i\ket{2}$\\
        0 & 1 & 0 & $\ket{0} - i\ket{2}$\\
        0 & 0 & 1 & $\ket{1} + i\ket{3}$\\
        0 & 1 & 1 & $\ket{1} - i\ket{3}$\\
    \end{tabular}
    \caption{\textbf{The interference choice based on E, $s_1, s_2$ variables}}
    \label{tab:E_s2_table}
\end{table}

\begin{table}
    \centering
    \begin{tabular}{c|c|c|c}
         client's  & experimental & corresponding  \\
         measurement basis & operation &  oracle \\\hline
        $\ket{0} + \ket{1}$ & \ \ \ $\ket{+}_a\bra{+} I + \ket{-}_a\bra{-} Z_1 - \ket{+}_a\bra{-} Z_2 - \ket{-}_a\bra{+} Z_1Z_2 $ \ \ \ \ & O1-b\\
        $\ket{2} + \ket{3}$ & $\ket{+}_a\bra{+} I + \ket{-}_a\bra{-} Z_1 + \ket{+}_a\bra{-} Z_2 + \ket{-}_a\bra{+} Z_1Z_2 $ & O1-a\\
        $\ket{4} + \ket{5}$ & $\ket{+}_a\bra{+} I + \ket{-}_a\bra{-} Z_1 - \ket{+}_a\bra{-} Z_1 Z_2 - \ket{-}_a\bra{+} Z_2 $& O2-b\\
        $\ket{6} + \ket{7}$ & $\ket{+}_a\bra{+} I + \ket{-}_a\bra{-} Z_1 + \ket{+}_a\bra{-} Z_1 Z_2 + \ket{-}_a\bra{+} Z_2 $ & O2-a\\
        $\ket{0} + \ket{2}$ & $\ket{+}_a\bra{+} I + \ket{-}_a\bra{-} Z_1 - \ket{+}_a\bra{-} I - \ket{-}_a\bra{+} Z_1 $ & O3-b\\
        $\ket{1} + \ket{3}$ & $\ket{+}_a\bra{+} I + \ket{-}_a\bra{-} Z_1 - \ket{+}_a\bra{-} I + \ket{-}_a\bra{+} Z_1 $  & O3-a\\
        $\ket{4} + \ket{6}$ & $\ket{+}_a\bra{+} I + \ket{-}_a\bra{-} Z_1 - \ket{+}_a\bra{-} Z_1 - \ket{-}_a\bra{+} I $  & O4-b\\
        $\ket{5} + \ket{7}$ & $\ket{+}_a\bra{+} I + \ket{-}_a\bra{-} Z_1 + \ket{+}_a\bra{-} Z_1 + \ket{-}_a\bra{+} I $ & O4-a\\
    \end{tabular}
    \caption{\textbf{The client's photon measurement choice and the corresponding experimental operation on two computational qubits as well as an ancilla.} The detailed protocol is shown in Fig.~\ref{fig:all_seq}D. Here, the table lists general operations  for any arbitrary  initial states of two computational qubits and one ancilla qubit. To implement our Deutsch-Jozsa-type algorithm, we always start the ancilla qubit from $\ket{+}_a$ and post select on  $\ket{-}_a$ at the end. Then, this operation on the computational qubits corresponds to the oracles (Fig.~\ref{fig:DJ_oracle}) for the standard Deutsch-Jozsa algorithm.}
    \label{tab:DJ_exp_table}
\end{table}

\begin{table}
\centering
\begin{tabular}{| c | c | c | c | c |} 
\hline
  Initial state & Oracle and  & State after & Corresponding & Result  \\
  $\ket{e_2}\ket{e_1}\ket{n_1}$ &  and TDI  & photon measurement   & functions &   $\ket{e_2}\ket{n_1}$ \\
  & measurement basis & $\ket{e_2}\ket{e_1}\ket{n_1}$ &  & \\[0.5ex] 
 \hline & & & & \\[-2ex]
 \multirow{16}{4.5em}{$\ket{+}\ket{+}\ket{+}$} & $O_1\text{-b}: \ket{0}\pm\ket{1}$ & $(\ket{\uparrow}\ket{\downarrow}+\ket{\downarrow}\ket{\uparrow})\ket{\uparrow}$ & $f(x) = x_1 + x_2 + 1$ &  $\ket{-}\ket{-}$\\
 & & $+(\ket{\uparrow}\ket{\uparrow}+\ket{\downarrow}\ket{\downarrow})\ket{\downarrow}$ &   $ $ & (balanced)\\[0.5ex] 
 \cline{2-5}  & & & & \\[-2ex]
 & $O_1\text{-a}: \ket{2}\pm\ket{3}$ & $(\ket{\uparrow}\ket{\uparrow}+\ket{\downarrow}\ket{\downarrow})\ket{\uparrow}$ &  $f(x)=x_1 + x_2$ &  $\ket{-}\ket{-}$\\
 & & $+(\ket{\uparrow}\ket{\downarrow}+\ket{\downarrow}\ket{\uparrow})\ket{\downarrow}$ &   $ $ & (balanced)\\[0.5ex] 
 \cline{2-5} & & & & \\[-2ex]
 & $O_2\text{-b}: \ket{4}\pm\ket{5}$ & $(\ket{\uparrow}\ket{\uparrow}+\ket{\downarrow}\ket{\downarrow})\ket{+}$ & $f(x)=x_2 + 1$ &  $\ket{-}\ket{+}$\\
 & &  &  $ $  & (balanced)\\[0.5ex] 
 \cline{2-5} & & & & \\[-2ex]
 & $O_2\text{-a}: \ket{6}\pm\ket{7}$ & $(\ket{\uparrow}\ket{\downarrow}+\ket{\downarrow}\ket{\uparrow})\ket{+}$ &  $f(x)=x_2$  &  $\ket{-}\ket{+}$\\
 & &  &   $ $  & (balanced)\\[0.5ex] 
 \cline{2-5} & & & & \\[-2ex]
 & $O_3\text{-b}: \ket{0}\pm\ket{2}$ & $\ket{+}(\ket{\downarrow}\ket{\uparrow}+\ket{\uparrow}\ket{\downarrow})$ & $f(x)=x_1 + 1$  &  $\ket{+}\ket{-}$\\
 & & &  $ $ & (balanced)\\[0.5ex] 
 \cline{2-5} & & & & \\[-2ex]
 & $O_3\text{-a}: \ket{1}\pm\ket{3}$ & $\ket{+}(\ket{\uparrow}\ket{\uparrow}+\ket{\downarrow}\ket{\downarrow})$ &  $f(x)=x_1$  &  $\ket{+}\ket{-}$\\
 & & &  $ $  & (balanced)\\[0.5ex] 
 \cline{2-5} & & & & \\[-2ex]
 & $O_4\text{-b}: \ket{4}\pm\ket{6}$ & $\ket{+}\ket{\uparrow}\ket{+}$ &  $f(x)=1$ &  $\ket{+}\ket{+}$\\
 & &  &  $ $ & (constant)\\[0.5ex] 
 \cline{2-5} & & & & \\[-2ex]
 & $O_4\text{-a}: \ket{5}\pm\ket{7}$ & $\ket{+}\ket{\downarrow}\ket{+}$ & $f(x)=0$ &  $\ket{+}\ket{+}$\\
 & &  &   $ $& (constant)\\[0.5ex] 
 \hline
\end{tabular}
\caption{\textbf{Deutsch-Jozsa-type algorithm oracle implementation table.} By post selectong on query ancilla qubit $\ket{-}_{a}$, we get the correction operation on the computational qubits ($e_2, n_1$).}
\label{table:DJ-table}
\end{table}

We here take the oracle $\text{O1-b}$ in Fig.~\ref{fig:DJ_oracle} as an example. To run the standard Deutsch-Jozsa algorithm, we need to initialize the query ancilla qubit in $\ket{-}_a$. After the oracle $\text{O1-b}$ is implemented, the ancilla qubit remains in $\ket{-}_a$. The operation on the computational qubit is $\bra{-}_a \text{O1-b} \ket{-}_a = -Z_1 Z_2$. This operation  is the same as the experimental operation we have by initializing the ancilla qubit in $\ket{+}_a$ and post selecting on $\ket{-}_a$ (sandwiching $\bra{-}_a$ and $\ket{+}_a$ on the first row of the experimental operations in Table~\ref{tab:DJ_exp_table} also yields $-Z_1 Z_2$.) The more detailed information about the quantum states and corresponding protocol is in Table~\ref{table:DJ-table}.

Note that the oracle $\text{O1-a}$ and $\text{O1-b}$ differ only by a flip on the ancilla qubit, which does not affect whether a function is constant or balanced.
The same principle applies for $\{\text{O2-a, O2-b}\}$, $\{\text{O3-a, O3-b}$\}, $\{\text{O4-a, O4-b}\}$. 
For simplicity, we here group them as the same oracle. We define,
\begin{align*}
    O_1 := \{ \text{O1-a, O1-b}\} \\
    O_2 := \{ \text{O2-a, O2-b}\} \\
    O_3 := \{ \text{O3-a, O3-b}\} \\
    O_4 := \{ \text{O4-a, O4-b}\} \\
\end{align*}

We also note that the implementation of this algorithm utilizes the QUBE gate to hide the inter-node entanglement, which is the origin of the blindness in this protocol. Specifically, an additional  $Z_2^s$ operation is applied to the system, while $s$ is the measurement  outcome from the TDI and only known by the client. The client uses $s$ to decode the measurement results, while the server only observes the mixture of measurement results with $s=0$ and $s=1$. This makes the server unable to distinguish between the states with $e_1, e_2$ entanglement from the states without $e_1, e_2$ entanglement (see Table~\ref{tab:DJ_exp_table} with $\ket{-}_a\bra{+}$ term). As a result, $O_1$ and $O_3$ are indistinguishable from the servers, since the only difference of the resulting states between these two oracles is with or without $e_1, e_2$ entanglement (check Fig.~\ref{fig:DJ_exp_data}A), which is hidden by the protocol. The sample principle also applies for $O_2$ and $O_4$.

\subsubsection{Post selection}
There are several post selection components in our experiment, which makes our success probability much less than $50\%$ and therefore cancels the quantum speedup associated with the standard Deutsch-Jozsa algorithm.

First, our quantum network photonic link has an efficiency of $\sim 10^{-5}$ efficiency. This can in principle be solved by having a deterministic entangling operation, which requires a quantum memory that remains robust against decoherence during entanglement attempts~\cite{main2024distributed, drmota2024verifiable}. In our work, this can be solved by using weakly coupled $^{13}C$ spins in the diamond lattice~\cite{PhysRevX.9.031045}.

Second, as mentioned in the previous paragraph, we only post select on $\ket{e_1} = \ket{-}$ after a protocol is applied, which is with $50\%$ efficiency. An extra matter qubit, for example, a $^{13}C$ nuclear spin in the first node would directly allow us to fully implement the standard Deutsch-Jozsa algorithm with blind oracles. Specifically, it can be done by using $^{13}C$ at the first node as a query ancilla qubit for the  Deutsch-Jozsa algorithm and $e_1$ as an ancilla qubit for the QUBE gate operation.

Third, during the photon measurement, the client can only deterministically choose two measurement configurations: using a short TDI to measure oracles $\{ O_1, O_2 \}$, or using a long TDI to measure $\{ O_3, O_4 \}$. With a short TDI configuration, the client cannot deterministically choose to apply $O_1$ or $O_2$. However, the client can, for example, implement $O_1 (O_2)$ through post-selection by discarding the $O_2 (O_1)$ events. The same principle applies to $O_3, O_4$ with a long TDI measurement configuration. This post selection is also with an additional $50\%$ efficiency. This post selection can in principle be eliminated by executing a sequence with two one-photon gates over two nodes. For example, instead of only selecting one-photon events with two QUBE gates, we deterministically execute the first QUBE gate with one photon heralded and the second QUBE gate with the second photon heralded. This requires a deterministic entangling operation as mentioned in the above paragraph.

\subsubsection{Experimental data}

To characterize our gate performance, we first check the parity of the three spin qubits after applying our protocol. After applying our gate sequence to the qubit initialized in $\ket{e_2, e_1, n_1} = \ket{+++}$, we get
\begin{align}
    \ket{O_i} =  \frac{1}{2^{n/2+1}} \sum_{x =  \{0,1\}^{n} } \ket{x}_{\text{c}} \ket{+}_{\text{a}} + (-1)^{f(x)} \ket{x}_{\text{c}} \ket{-}_{\text{a}}
\end{align}.
Here $n=2$ means the number of computational qubits. Post selecting on $\ket{-}_a$ yields the quantum state $\frac{1}{2^{n/2}}(-1)^{f(x)} \ket{x}_{\text{c}}$ for computational qubits. This suffices to answer if a function is constant or balanced (Eq.\ref{eq:dj}) and consistent with the argument mentioned in the previous paragraph. In our experiment, $\ket{O_{i,i=1\sim4}}$ includes four different cluster states shown in Fig.~\ref{fig:all_seq}F. Its measured parity is shown in Fig.~\ref{fig:DJ_exp_data}A.

We later probe the population of $\ket{++}_c$ and execute our Deutsch-Jozsa-type algorithm. All oracles data are shown in Fig.~\ref{fig:DJ_exp_data}B.  
We obtain the probability
$0.90\pm0.05$, $0.80\pm0.06$, $0.94\pm0.04$, $0.78\pm0.06$ to detect the correct outcome if a function is constant (C) or
balanced (B), for  $O_1$, $O_2$, $O_3$, $O_4$ oracles, respectively. $O_4$  implements constant functions, while
$O_1$, $O_2$, $O_3$ are for balanced functions. 
The classical information leakage over $\{ O_1, O_3\}$ is $0.05^{+0.19}_{-0.05}$ bits, and the classical information leakage over $\{ O_2, O_4\}$ is $0.07^{+0.14}_{-0.07}$ bits.

\subsection{Photon saving from matter-based architecture}
One challenge for BQC comes from the low success rate of gate operations. This is due to the probabilistic nature of photon heralding efficiency $\eta$. Benefited from the capability to locally manipulate quantum information via matter qubits, we already have some photon savings.

\begin{figure}
    \centering
\includegraphics[width=0.66\linewidth]{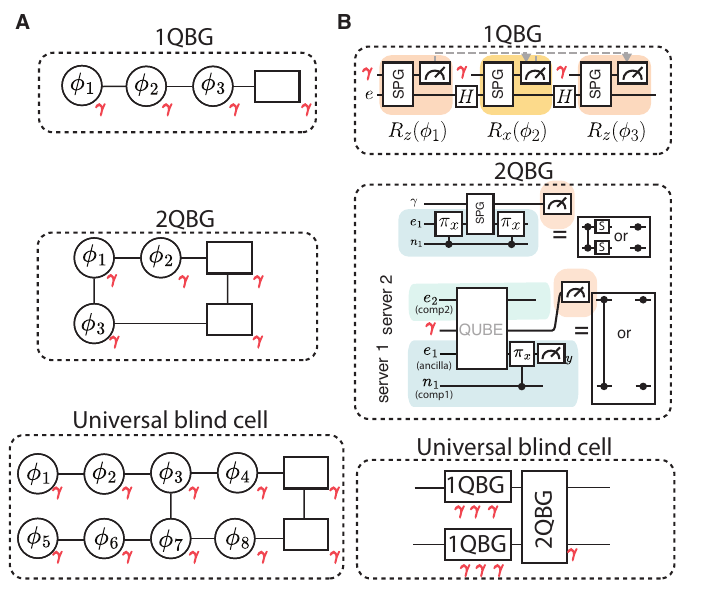}
    \caption{\textbf{Summarized comparison of BQC implementation between the all-photonic platform and the matter-based platform.} (\textbf{A}) For the all-photonic platform, 1QBG (2QBG) needs $3$ ($3$) photons for the gate operations, and additional $1$ ($2$) photons for output qubits. Therefore, for a full implementation, single 1QBG needs at least 4 photons, and single 2QBG needs at least 5 photons. An universal blind cell needs $8$ photons for the gate operations and  $2$ additional photons at the end as the output qubits. Full implementation needs at least 10 photons for a single universal blind cell. (\textbf{B}) In the demonstrated matter-based platform, 1QBG (2QBG) needs $3$ ($1$) photons for the full implementation. An universal blind cell can concatenate the above elements together, and $7$ photons needed in total.}
    \label{fig:si_photon_saving}
\end{figure}

Fig.~\ref{fig:si_photon_saving} shows the comparison of BQC implementation between all-photonic platforms and demonstrated matter-based platforms. For 1QBG, both all-photonic and matter-based platforms need 3 photons for the gate operation, while the all-photonic platforms need an additional one photon at the end as a output photon. For 2QBG, all-photon platforms need 3 photons for the gate operations and 2 extra photons at the end as output photons. Therefore, overall 5 photons are needed for the complete implementation of single 2QBG using the all-photonic scheme. Our demonstrated matter-based scheme requires 1 photon for the full implementation of 2QBG.

For the universal blind cell, all-photonic platforms can integrate 1QBG and 2QBG together, which requires 8 photons for blind operations within each universal blind cell. An additional $2$ photons are still required at the end of the circuits as output photons. Overall, $10$ photons are needed for a full implementation of a single universal blind cell. Matter-based BQC platforms can concatenate two 1QBG and one 2QBG together, which in total needs $7$ photons to fully implement one universal blind cell.

\subsection{Path towards deterministic operation}
Matter-photon entanglement is probabilistic due to photon loss. To enable deterministic operation, it is necessary to have robust quantum memories to hold quantum information while the ancilla qubits (or communication qubits) keep trying to establish successful matter-photon entanglement. Our demonstrated scheme can be directly integrated with an architecture with robust quantum memories.

\subsubsection{Blind rotation}
\begin{figure}
    \centering
    \includegraphics[width=1\linewidth]{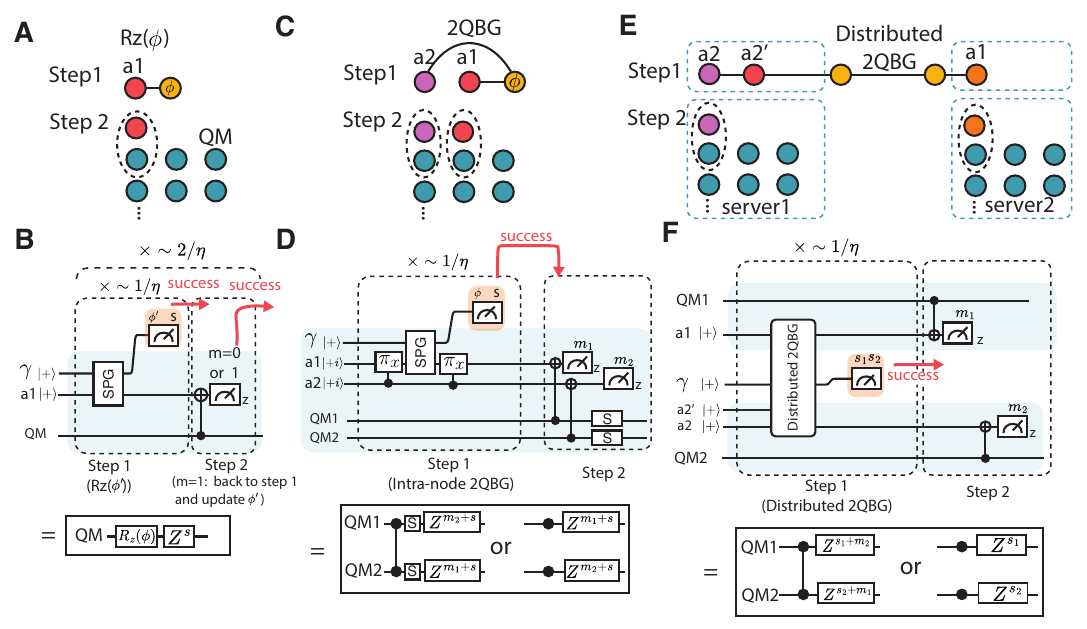}
    \caption{\textbf{Proposed  operation to integrate our demonstrated scheme into a quantum memory architecture.}(\textbf{A})
    High-level schematic diagram for blind $R_z(\phi)$ rotations with a quantum-memory architecture. $a_i$ denotes the $i^{th}$ ancilla qubit in the graph, and QM denotes the quantum-memory qubit. (\textbf{B}) Detailed circuit-level procedure for blind $R_z(\phi)$ rotations with a quantum-memory architecture.
    (\textbf{C})
    High-level schematic diagram for intra-node 2QBG with a quantum-memory architecture.  (\textbf{D}) Detailed circuit-level procedure for intra-node 2QBG with a quantum-memory architecture.       (\textbf{E})
    High-level schematic diagram for distributed 2QBG with quantum-memory architecture.  (\textbf{F}) Detailed circuit-level procedure for distributed 2QBG a with quantum-memory architecture.   
}
    \label{fig:SI_det_scheme}
\end{figure}

A deterministic blind z rotation, as already demonstrated in \cite{drmota2024verifiable}, can be achieved by first implementing the $R_z(\phi)$ blind rotation on the ancilla qubits (initialized at $\ket{+}$) as we described in the main text. In our proposed method, this can be followed by a teleportation to map the operation back to the quantum memory. The entire process is summarized in Figs.\ref{fig:SI_det_scheme}A,B. The first step consists of the server repeatedly trying to establish entanglement between an ancilla qubit and a photon sent to the client by repetitively running a SPG sequence. Once the entanglement is heralded by a photon detection at the client, the state of the ancilla qubit is $\ket{\uparrow}_a + (-1)^s e^{i\phi}\ket{\downarrow}_a$, where $s$ is the client's  photon measurement outcome. The second step then consists of a teleportation between the ancilla qubit and the quantum memory, as shown in the circuit Fig.~\ref{fig:SI_det_scheme}B. Assuming the initial quantum memory is in the state $\ket{\psi}_{\text{QM}} = a \ket{\uparrow}_{\text{QM}} + b \ket{\downarrow}_{\text{QM}}$, a local CNOT operation results in the state $a\ket{\uparrow}_{\text{QM}} (\ket{\uparrow}_a + (-1)^s e^{i\phi}\ket{\downarrow}_a) + b \ket{\downarrow}_{\text{QM}}(\ket{\downarrow}_a + (-1)^se^{i\phi}\ket{\uparrow}_a)$. Measuring the ancilla qubit at the $z$ basis returns the outcome $m$ ($m=0, 1$ corresponds to $\ket{\uparrow, \downarrow}_a$, respectively),
\begin{align}
    \ket{\psi}_{\text{QM}}  = a\ket{\uparrow} + (-1)^s e^{i\phi \times (-1)^m} b\ket{\downarrow}
\end{align}

For outcome $m=0$, the quantum memory is in the state of $Z^s R_z(\phi) \ket{\psi}_{\text{QM}}$, fulfilling a blind z rotation on the quantum memory qubit. If $m=1$, we start over from step one again, but this time updating the targeted measurement angle to $2\times \phi$. If the second teleportation succeeds (i.e., $m=0$), the quantum memory is rotated by a total angle $-\phi + 2\phi = \phi$ from the first and second trials. If $m=1$ again, we repeat the same steps recursively, doubling the targeted rotation angle each time the teleportation fails ($m=1$). The failure probability is thus exponentially suppressed by the teleportation trial number $n_{\text{tele}}$ as $2^{-n_{\text{tele}}}$. Once the teleportation succeeds, the desired rotation angle $2^{n_{\text{tele}}-1} \phi - \sum_{i=0}^{n_{\text{tele}}-2} 2^i\phi= \phi$ is applied on the memory qubit.  Assuming the SPG success probability to be $\eta$, and the number of ancilla qubit resets to be $N_{Rz}$, each trial (step 1 and step 2) has a success probability of $\eta \times 1/2$, which we can repeat until success without affecting the rest of the circuit. It therefore allows us to deterministically apply $R_z(\phi)$. To have a reasonably low failure rate and therefore deterministic successful $R_z(\phi)$, we have
\begin{align}
    N_{Rz} \sim \frac{2}{\eta}
\end{align}. The specific procedure is as follows:
\begin{enumerate}
    \item $\phi' \gets \phi$
    \item $success \gets 0$
    \item \textbf{While} $success \neq 1$
    \begin{enumerate}
        \item[] (Step1: SPG)
        \item[] $PhotonCount \gets 0$
        \item[] \textbf{While} $PhotonCount \neq 1$:
        \begin{enumerate}
            \item[] Initialize the  ancilla
            \item[] Execute Intra-node 2QBG with angle $\phi'$ and get the photon count
        \end{enumerate}
        \item[] (Step2: Teleportation) 
        \item[] Server executes teleportation and the announces the measurement results $m$
        \item[] \textbf{if} $m$ = 0: $success \gets 1$
        \textbf{else}: $\phi'\gets 2\phi'$ 
    \end{enumerate}
    \item  Client applies feedback based on s by changing the subsequent blind operation
\end{enumerate}
    

A deterministic 1QBG can straightforwardly be implemented by concatenating three deterministic blind $R_z(\phi)$ rotations interleaved with Hadamard gate on the quantum memory qubit.

\subsubsection{Intra-node 2QBG}
For a deterministic intra-node 2QBG, we similarly first try to establish entanglement between two ancilla qubits ($a_1, a_2)$ and the photon ($\gamma$), as shown in the first step of Fig.~\ref{fig:SI_det_scheme}D. Each time we initialize the ancilla qubits in the state $\ket{+i, +i}_{a_1, a_2}$, followed by applying an inter-node 2QBG on the aniclla qubits. Once it succeeds, we again move to the second step and teleport the information on the quantum memories. The measurement outcomes $m_1, m_2$ of each ancilla qubit are publicly announced by the server. As opposed to for the deterministic $R_z(\phi)$ blind rotation, here the feedback correction depending on the ancilla qubits measurements ($m_1, m_2$) after teleportation takes the form of Pauli operators ($Z_1$ and $Z_2$, see Fig.~\ref{fig:SI_det_scheme}D bottom), which can straightforwardly propagated to the end of the computation (see Section \ref{adap_feed}). The summarized operation of the circuit is shown at the bottom of Fig.~\ref{fig:SI_det_scheme}D. Importantly, after the teleportation, the feedback operator on the two quantum memories is $Z_1^sZ_2^s$, while entangling off and on operation still takes the form of $I$ and $CZ S_1 S_2$. Therefore, as discussed in the previous section, the $Z_1^sZ_2^s$ feedback correction form hides the operation between $I$ and $CZ S_1 S_2$. The teleportation therefore preserves the blindness of the intra-node 2QBG. 

The specific procedure is as follows:
\begin{enumerate}
\item[] (Step1: Intra-node 2QBG)
    \item $PhotonCount \gets 0$
    \item \textbf{While}{$PhotonCount \neq 1$}:
    \begin{enumerate}
        \item[] Initialize 2 ancilla qubits $\ket{+i, +i}_{a_1, a_2}$
        \item []  Execute Intra-node 2QBG with rotation basis $\phi$, get the photon count 
    \end{enumerate}
    \item [] (Step2: Teleportation)
    \item Server executes teleportation and the server announces the measurement results $m_1, m_2$
    \item Client applies feedback based on $s, m_1, m_2$ by changing the subsequent blind operation
\end{enumerate}


The required reset clock cycles ($N_{\text{I-2QBG}}$) are determined by the probability of successful intra-node 2QBG events using ancilla qubits $\eta$.
\begin{align}
    N_{\text{I-2QBG}} \sim 1/\eta
\end{align}

\subsubsection{Distributed 2QBG}
Similar to the intra-node 2QBG, we start by applying a distributed 2QBG to ancilla qubits and then teleport onto the memory qubits, as shown in Figs.~\ref{fig:SI_det_scheme}E,F. The operation after the teleportation is summarized at the bottom of Fig.~\ref{fig:SI_det_scheme}F. The feedback operator $Z_1^{s_1} Z_2^{s_2}$ and the operation $\{ I / CZ\}$ remain the same after teleportation, preserving the blindness of distributed 2QBG on the memory qubits. 

The specific procedure is as follows:
\begin{enumerate}
    \item [] (Step1: Distributed 2QBG)
    \item \textbf{While} $PhotonCount \neq 1$
    \begin{enumerate}
        \item[] Initialize 3 ancilla qubits $\ket{+, +, +}_{a_1, a_2, a_2'}$
        \item [] Execute distributed 2QBG  and get the photon count and outcome $s_1, s_2$
    \end{enumerate}
    \item [] (Step2: Teleportation)
    \item Server executes teleportation and the server announces the measurement results $m_1, m_2$
    \item Client applies feedback based on $s_1, s_2, m_1, m_2$ by changing the subsequent blind operation
\end{enumerate}

The required reset clock cycles ($N_{\text{D-2QBG}}$) are determined by the probability of successful distributed 2QBG events $\eta_D$.
\begin{align}
    N_{\text{D-2QBG}} \sim 1/\eta_D
\end{align}


Similarly, the client's choice of implementing $\{ CZ, I\}$ as well as the imprinted noise type onto the server qubits $ Z^{s_1}Z^{s_2}$ is exactly identical when compared to non-deterministic operation without teleportation. As already argued previously, this combination enables a fully blind 2QBG.

\section{Experimental Hardware}

\subsection{Cavity-QED parameters}

The SiV defect center strongly coupled to the nanophotonic cavity provides an efficient optical interface for the spin qubits. Due to this unique interface, we are able to perform multiple remote blind gates in our network in a row. The efficiency comes from the strong interaction between the emitter and sub-wavelength mode of the optical cavity, which results in high cooperativity $C = \frac{4g^2}{\kappa_{\text{tot}}\gamma} > 1$. where $g$ is the single-photon Rabi frequency, $\kappa_{\text{tot}}$ the total cavity decay rate, and $\gamma$ is the bare SiV linewidth. Decay rates and line widths are expressed as full widths at half max (FWHM). In addition, we are able to optimize the efficiency of retrieval of the photons after the coherent interaction with high probability by fabricating overcoupled one-sided cavities and optimizing the mode matching between the cavity mode and the tapered input fiber. The cavity reflection amplitude at frequency $\omega$ can be expressed as:

\begin{figure}
    \centering
    \includegraphics[width=1\linewidth]{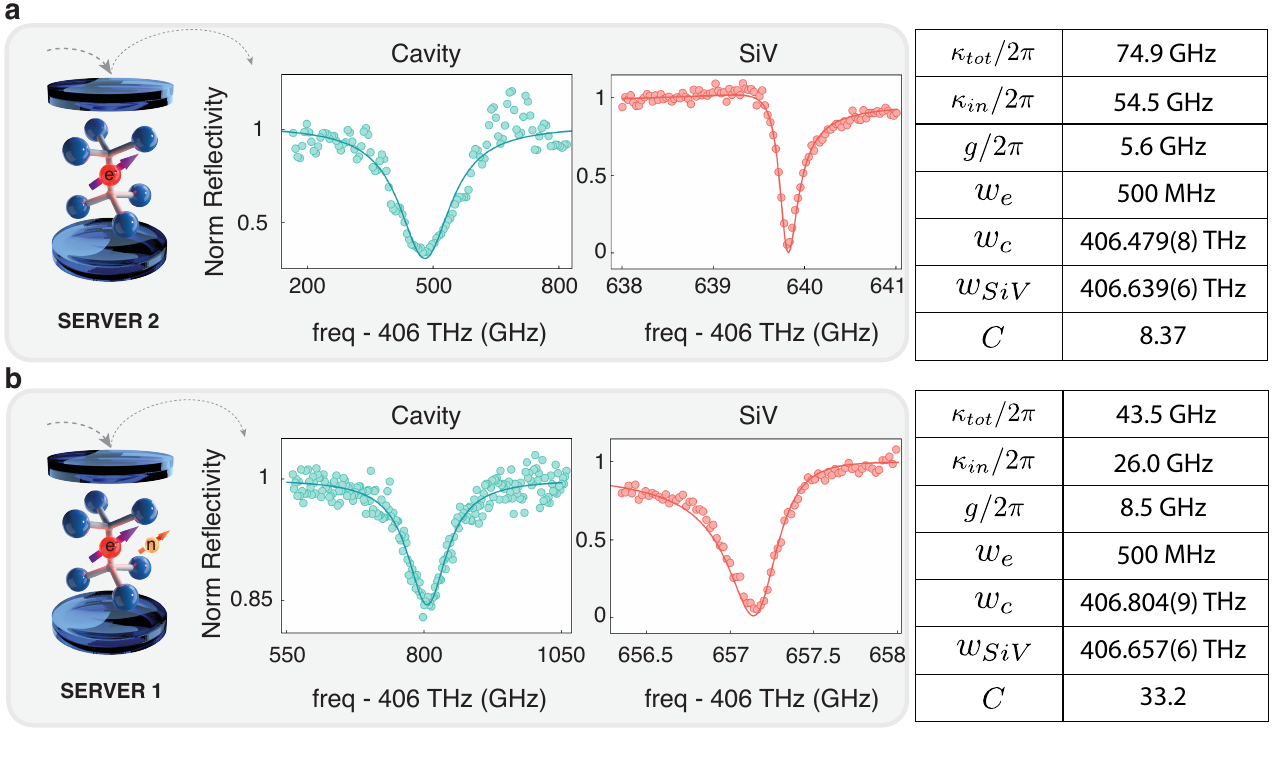}
    \caption{\textbf{Cavity-QED parameters of the two servers.} (\textbf{A}) Cavity reflectivity for the SiV at Server 2 with the corresponding fit parameters shown in the table to the right. (\textbf{B}) Cavity reflectivity fit for the SiV at Server 1 with the corresponding fit parameters shown in the table to the right. }
    \label{fig:SicavQED}
\end{figure}

\begin{figure}
    \centering
    \includegraphics[width=1\linewidth]{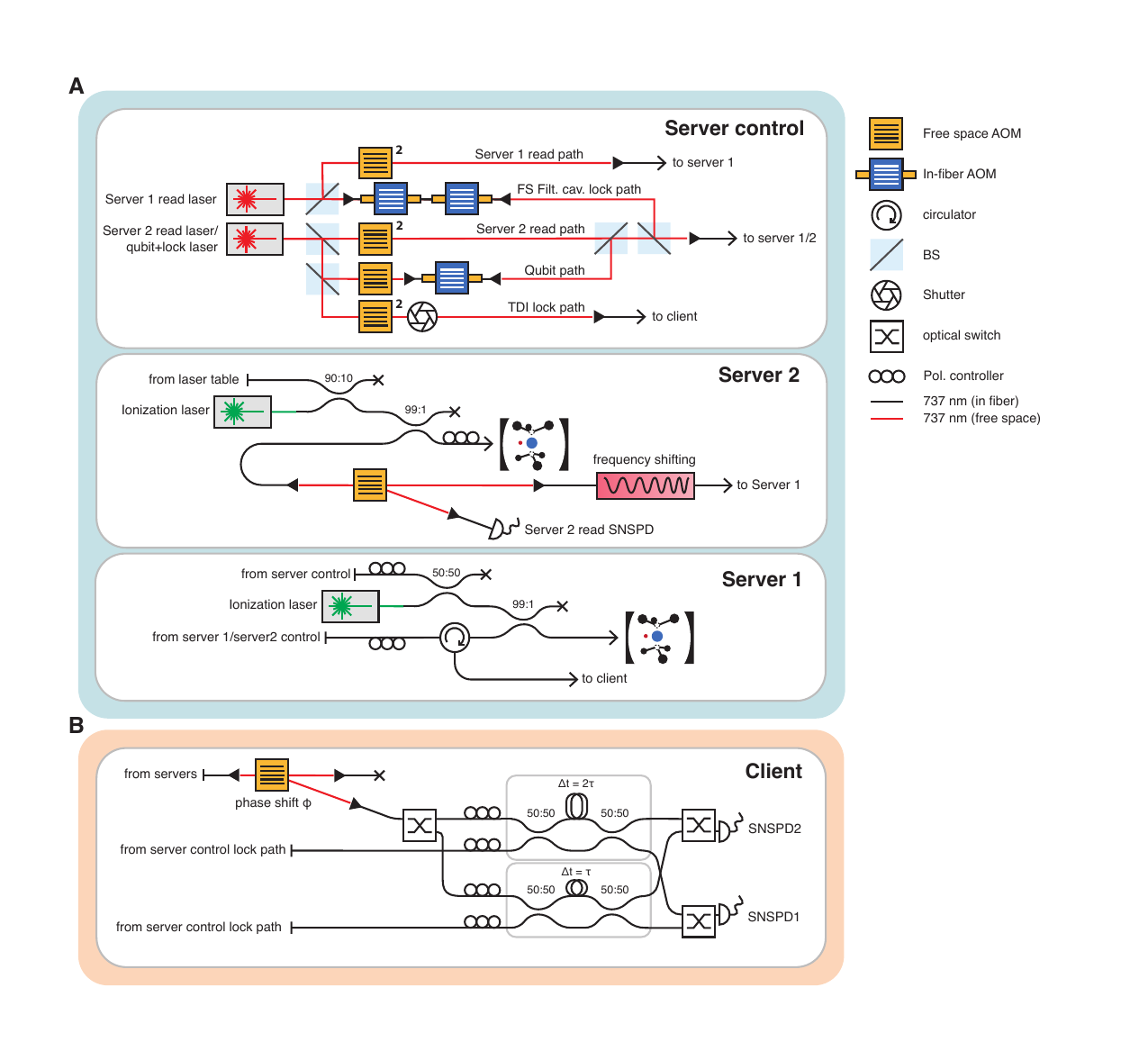}
    \caption{\textbf{Schematic drawing of the server and client experimental setups.} (\textbf{A}) Server setup, including server control with optical fields preparation for server 1 and 2 readout, frequency shifting and time-delay interferometer (TDI) locking, and the photonic qubit/qudit preparation, as well as servers 1 and 2 containing the quantum network nodes. Server 2 also contains a frequency shifting setup to bridge the optical frequency gap between the SiVs at Servers 1 and 2. (\textbf{B}) Client setup with AOM for photonic qubit phase shifting and optical switch to choose between two TDIs with differing delay lengths.}
    \label{fig:SI_exp_setup}
\end{figure}

\begin{figure}
    \centering
    \includegraphics[width=1\linewidth]{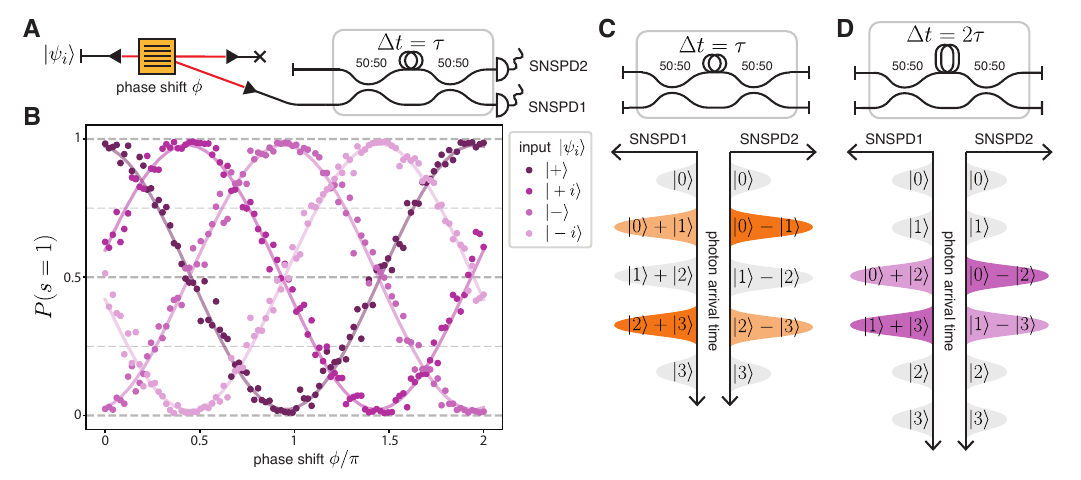}
    \caption{\textbf{Tunable photonic qubit and qudit readout.} (\textbf{A}) Photonic qubit measurement setup, including an AOM to shift the qubit phase and a TDI for superposition measurement. (\textbf{B}) Plot of the probability of measuring a click on SNSPD 1 (corresponding to the measurement of state $(\ket{0} + e^{i\phi}\ket{1})/\sqrt{2}$) while sweeping the  measurement phase $\phi$ applied by the AOM for different input states $\ket{\psi_i}$. (\textbf{C}) photonic qudit measurement outcomes depending on the photon arrival time at the SNSPDs and whether the click is recorded on SNSPD1 or 2 for a TDI with delay length $\Delta = \tau$. (\textbf{D}) Same as (\textbf{C}) but for TDI with delay length $\Delta t = 2\tau$.}
    \label{fig:SI_readout}
\end{figure}

\begin{figure}
    \centering
    \includegraphics[width=1\linewidth]{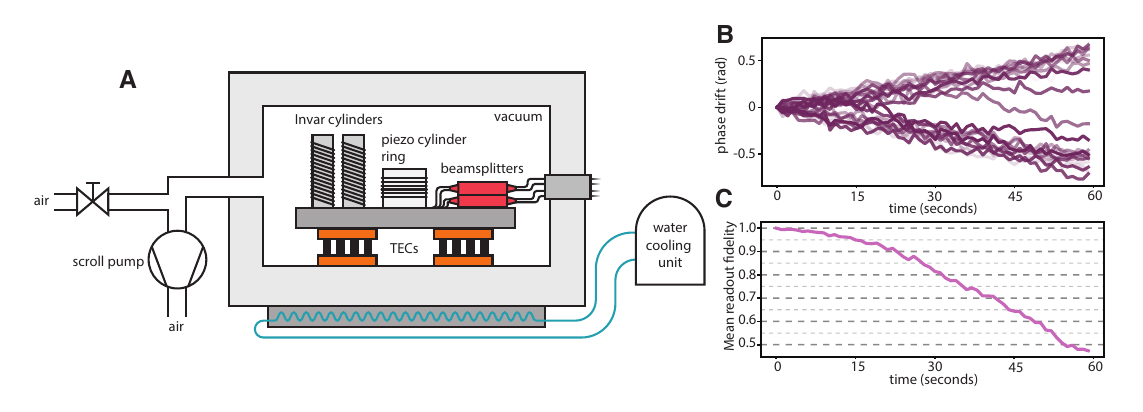}
    \caption{\textbf{Design and performance of \SI{290}{\nano \second} time delay interferometer.} (\textbf{A}) Components of temperature stabilized TDI under vacuum. The delay line of \SI{290}{\nano \second} ($\approx$ 58 m) is principally wrapped around Invar cylinders. (\textbf{B}) Traces of TDI differential phase drift over 60 seconds, repeatedly measured over a period of 30 minutes. (\textbf{C}) TDI mean readout fidelity over 60 seconds calculated from the average absolute phase drift in (\textbf{B}).}
    \label{fig:SI_TDI_design}
\end{figure}

\begin{equation}
        \text{Reflection}(\omega)  = 1 - \frac{2\kappa_{\text{in}}}{i (\omega - \omega_c) + \kappa_{\text{tot}} + g^2 /(i(\omega-\omega_{SiV}) + \gamma)} 
\label{eq:cavityQED}
\end{equation}

where $\kappa_{\text{in}}$ is the coupling rate of the cavity into the in-coupling port, and $\omega_{SiV (c)}$ is the resonance frequency of the SiV (cavity). The cavity-QED parameters and reflection data for the SiVs used in Server 1 and Server 2 are shown in Fig.~\ref{fig:SicavQED}. The design and fabrication of the nanophotonic cavity are described in \cite{Knall_2022}.



   
      
    
  
   

\subsection{Experimental setup}

The experimental setup consists of two servers (Fig.~ \ref{fig:SI_exp_setup}A) and a client (Fig.~ \ref{fig:SI_exp_setup}B). The server's side of the experimental setup can be separated in three main parts: the server control, Server 1 and Server 2. The server control prepares the necessary optical paths for optical readout and control of the qubits at Server 1 and 2, the locking of the frequency conversion setup, as well as the preparation of the optical photonic qubits and qudits. The server control also sends a light path to the client directly as a reference for the locking of time delay interferometers for photonic qubit/qudit readout. Server 2 contains an SiV in the nanophotonic cavity, a free space acousto-optic modulator (AOM) to switch between local readout and routing of the photonic qubit/qudit to Server 1, as well as a frequency shifting setup to bridge the optical frequency gap between the SiVs at Server 1 and Server 2. The frequency shifting method relies on optical frequency sideband generation with an electro-optic modulator (EOspace) and is described in detail in \cite{twofridge2024}. Server 1 contains a second SiV in the nanophotonic cavity. 
The client possesses a free space AOM to shift the phase of incoming photonic qubits/qudits and an optical switch to switch between TDIs of different delay lengths ($\Delta t = \tau$, $2\tau$, where $\tau$ is the photonic timebin separation) for qubit/qudit superposition basis readout with Superconducting Nanowires Single Photon Detectors (SNSPDs, Photon Spot). For experiments showing single server operation, photonic qubits are prepared in the server control, sent directly to Server 1 and subsequently to the client. For experiments involving two server operations, photonic qudits are again prepared in the server control, sent to Server 2, then Server 1, and finally to the client. Each SiV at Servers 1 and 2 is located in a separate dilution refrigeration unit (BlueFors BF-LD250) with a base temperature below \SI{200}{\milli \kelvin}. The SiV–cavity systems are optically accessed through a hydrofluoric acid-etched tapered fiber. Lasers at \SI{532}{\nano \meter} (ionization lasers in \ref{fig:SI_exp_setup}A) are used to stabilize the charge state of the SiVs. Counts from the SNSPDs are recorded on a time-tagger (Swabian Instruments Time Tagger Ultra). A Zurich Instrument HDAWG8 2.4 GSa/s arbitrary waveform generator (ZIHDAWG) is used for sequence logic, control of the AOMs (Gooch\&Housego, AA Opto-electronic), as well as microwave (MW) control. The MW and RF control chains for Servers 1 and 2 are described in detail in \cite{robustnode2022}.

\subsection{Photonic qubit and qudit readout}
The client's implementation of single qubit and intra-node two-qubit blind gates requires them to measure incoming photonic qubits in the basis $\{(\ket{0} + e^{i\phi}\ket{1})/\sqrt{2}, (\ket{0} - e^{i\phi}\ket{1})/\sqrt{2}\}$ with adjustable phase $\phi$. The client achieves this by making the photonic qubit pass through a continuously driven AOM. By carefully timing the phase of the RF signal driving the AOM and changing it by $\phi$ in between the arrival of the early and late time bins (without changing the frequency or amplitude of the RF signal), the client can imprint a phase $\phi$ on the late time bin only, effectively implementing the phase gate 
\begin{equation}
    P(\phi) = \begin{bmatrix}
    1 & 0 \\ 
    0 & e^{i\phi}
    \end{bmatrix}
\end{equation}
on the photonic qubit. The client then sends the photon to a TDI, which is a Mach-Zehnder interferometer with a delay line on one of the interferometer paths. Since the length of the delay line corresponds to the time spacing between early and late time bins ($\Delta t = \tau$), this superposes the early and late time bins and allows for the measurement of states $(\ket{0} + \ket{1})/\sqrt{2} =\ket{+}$ (corresponding to a click on SNSPD 1) and $(\ket{0} - \ket{1})/\sqrt{2}=\ket{-}$ (corresponding to a click on SNSPD 2), as long as the TDI is locked for a differential phase accumulation of $0$. This complete setup (Fig.~\ref{fig:SI_readout}A) effectively results in the POVM $\{P(\phi)\ket{\pm}\bra{\pm}P(-\phi)\}$ corresponding to the desired POVM $\{(\ket{0} \pm e^{i\phi}\ket{1})(\bra{0} \pm e^{i\phi}\bra{1})/2\}$ as shown in Fig.~\ref{fig:SI_readout}B.

For inter-node experiments, the server side uses 4-dimensional qudits for gate operations and sends them to the client for measurement. Similar to the case with photonic qubits, the client can use the AOM to apply the 4D qudit phase gate

\begin{equation}
    P_{4d}(\phi_1, \phi_2, \phi_3) = \begin{bmatrix}
    1 & 0 & 0 & 0 \\ 
    0 & e^{i\phi_1} & 0 & 0 \\
    0 & 0 & e^{i\phi_2} & 0 \\
    0 & 0 & 0 & e^{i\phi_3}
    \end{bmatrix}
\label{eq:4d_phasegate}
\end{equation}.

The client can then choose to send the photonic qudit to one of two TDIs: one with short delay length $\tau$ (Fig.~\ref{fig:SI_readout}C), and one with long delay length $2\tau$ (Fig.~\ref{fig:SI_readout}D). The short delay length TDI superposes the nearest time bins, resulting in the POVM $\{ \ket{0}\bra{0}, (\ket{0}\pm\ket{1})(\bra{0}\pm\bra{1})/2, (\ket{1}\pm\ket{2})(\bra{1}\pm\bra{2})/2, (\ket{2}\pm\ket{3})(\bra{2}\pm\bra{3})/2, \ket{3}\bra{3}\}$. The phase between the interfered time bins is, as in the qubit case, determined by whether a click was recorded on SNSPD1 or SNSPD2. We can distinguish which specific time bins are interfered by looking at the photon arrival time, or the timestamp of the recorded photon click at SNSPD1/2. By thresholding the arrival time of measured photons, we can herald detection of states, $(\ket{0}\pm\ket{1})/\sqrt{2}$, $(\ket{2}\pm\ket{3})/\sqrt{2}$ resulting in the final POVM $\{ (\ket{0}\pm e^{i\phi_1}\ket{1})(\bra{0}\pm e^{i\phi_1}\bra{1})/2, (\ket{2}\pm e^{i(\phi_3-\phi_2)}\ket{3})(\bra{2}\pm e^{i(\phi_3-\phi_2)}\bra{3})/2\}$. The long delay length TDI superposes next-nearest time bins instead. Similarly, by heralding the detection of desired states, the client implements the POVM $\{ (\ket{0}\pm e^{i\phi_2}\ket{2})(\bra{0}\pm e^{i\phi_2}\bra{2})/2, (\ket{1}\pm e^{i(\phi_3-\phi_1)}\ket{3})(\bra{1}\pm e^{i(\phi_3-\phi_1)}\bra{3})/2\}$. By choosing the phases $\phi_1, \phi_2, \phi_3$ and which TDI is used, the client controls which POVM is implemented.

\subsection{Robust time-delay interferometer design}
In order for the TDI to consistently measure photons in the $\{(\ket{i}\pm\ket{j})/\sqrt{2}\}$, the differential phase accumulation between the short and long paths of the TDI must be kept to 0. In between experimental trials, the TDI phase is locked with a CW laser by equalizing counts at the output SNSPDs through adjusting the length of the long path with a spool of fiber wrapped around a piezo cylinder ring acting as a fiber stretcher (Fig.~\ref{fig:SI_TDI_design}A). Then, during the experimental trials, the TDI is left unlocked. It is therefore crucial for the unlocked differential phase to remain stable for the duration of an experimental trial. Concretely, to keep the measurement error below 5\%, this corresponds to keeping the delay length stable within $arccos(0.9)\lambda/2\pi \approx $\SI{53}{\nano \meter}. Particularly for the long delay TDI with a delay length of \SI{290}{\nano \second} $\approx$ 58 m, this means keeping the delay length constant within a factor of $9.14 \cdot 10^{-10}$. (We also note that the qubit/qudit laser frequency is required to be stable within $arccos(0.9)/(2\pi \Delta t) \approx 248$~kHz.) To achieve high delay length stability, we place the TDI inside a vacuum box with temperature stabilized by a water cooling system. The TDI components are positioned on a plate further temperature stabilized with Thermo-Electric Couplers (TECs) and the bulk of the fiber is wrapped around low thermal expansion coefficient Invar cylinders (Fig.~\ref{fig:SI_TDI_design}A). We measure the average readout fidelity of the TDI while unlocked to stay above 95\% for $\sim15$ seconds (Fig.~\ref{fig:SI_TDI_design}B,C).

\subsection{Control sequence}

\begin{figure}
    \centering
    \includegraphics[width=1\linewidth]{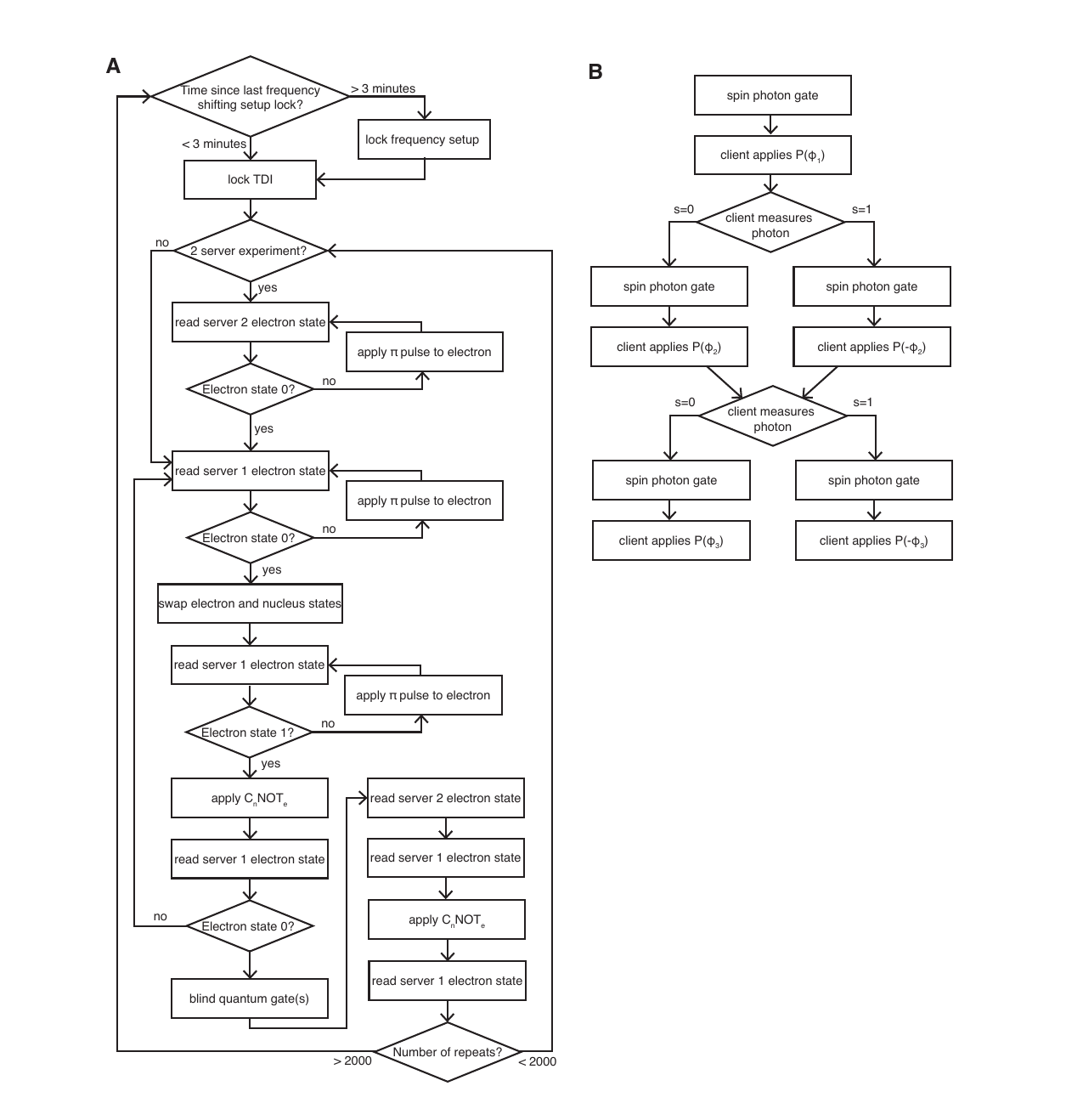}
    \caption{\textbf{Experimental sequence flowcharts.} (\textbf{A}) General control sequence for all blind quantum computing experiments shown in this work. (\textbf{B}) Fast feed forward flowchart for the client's implementation of the blind universal gate.}
    \label{fig:SI_exp_flowchart}
\end{figure}

The control sequence for all blind quantum computing experiments is shown in Fig.~\ref{fig:SI_exp_flowchart}A. Three cores across two ZIHDAWGs synchronize sequence logic, as well as MW- and RF-pulse generation. The first two cores (with one controller and one worker) act as the server and control the SiVs of servers A and B, as well as the AOMs and shutters in the server control setup (Fig.~\ref{fig:SI_exp_setup}A). The third core acts as the client, controlling the elements of Fig.~\ref{fig:SI_exp_setup}B. Fig.~\ref{fig:SI_exp_flowchart}B shows the client's feedforward logic for the implementation of the blind universal gate with three successive spin photon gates.

\begin{figure}
    \centering
    \includegraphics[width=1\linewidth]{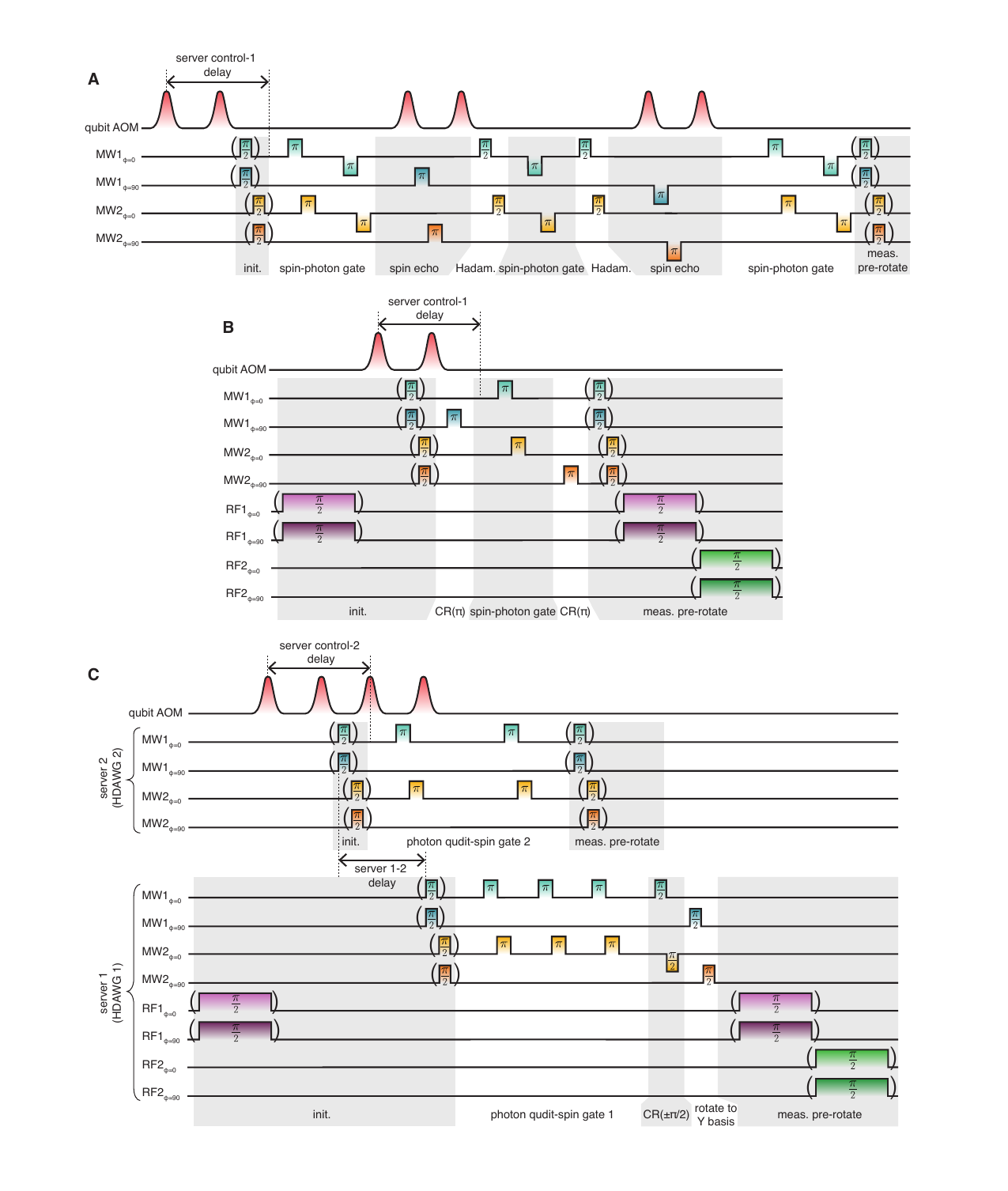}
    \caption{\textbf{Experimental MW and RF pulse sequences.} (\textbf{A}) Single qubit universal blind gate(\textbf{B}), intra-node two-qubit blind gate (\textbf{C}), and inter-node two-qubit blind gate.}
    \label{fig:SI_MWRF_pulse_seq}
\end{figure}

\begin{figure}
    \centering
    \includegraphics[width=1\linewidth]{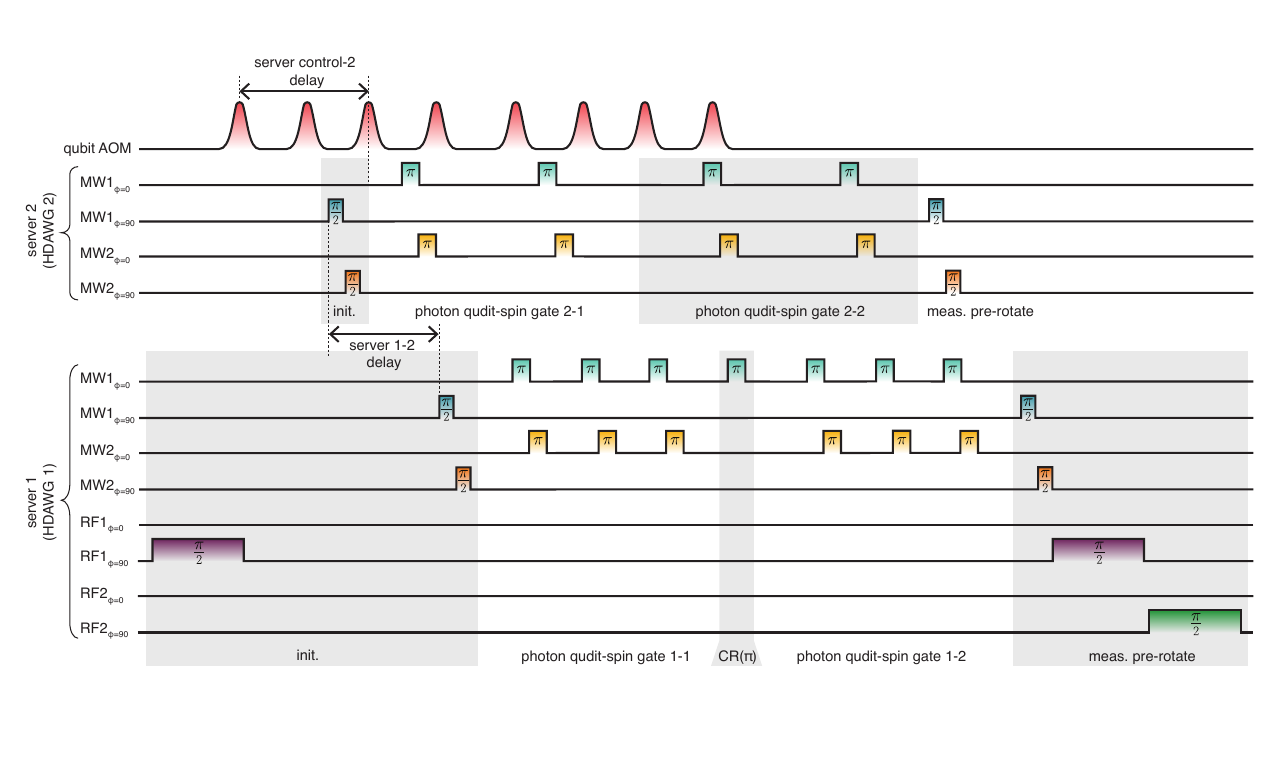}
    \caption{\textbf{Experimental MW and RF pulse sequences for the implementation of the Deutsch-Jozsa-type algorithm.}}
    \label{fig:SI_DJ_pulse_seq}
\end{figure}

\subsection{Gate and protocol efficiencies}

\begin{table}
\centering
\begin{tabular}{c c c c} 
  & One-qubit blind gate & Intra-node two-qubit & Inter-node two-qubit \\
  &  (1QBG) & blind gate & blind gate \\[0.5ex] 
 \hline \\ 
 WCS $\mu$ & 0.2$^3$ & 0.05 & 0.2 \\ 
 Server 2 coupling & - & - & 0.6 \\
 Server 2 cavity average reflectivity & - & - & 0.35 \\
 Frequency shifting & - & - & 0.05 \\
 Circulator & - & - & 0.64 \\
 Server 1 coupling & 0.6$^3$ & 0.6 & 0.6$^2$ \\
 Server 1 cavity average reflectivity & 0.35$^3$ & \multicolumn{2}{c}{0.35} \\
 Client phase shifting & 0.3$^3$ & \multicolumn{2}{c}{0.3} \\
 TDI measurement & 0.35$^3$ &\multicolumn{2}{c}{0.35} \\ 
 SNSPD efficiency & 0.9$^3$ & \multicolumn{2}{c}{0.9}  \\ [0.5ex] 
 \hline \\ 
 Expected total efficiency & $6\cdot10^{-8}$ & $1\cdot10^{-3}$ & $1.5\cdot10^{-5}$ \\ [0.5ex] 
 Measured total efficiency & $4.1\cdot10^{-8}$ & $8\cdot10^{-4}$ & $1.2\cdot10^{-5}$\\
\end{tabular}
\caption{\textbf{Efficiency breakdown for the one-qubit blind gate (1QBG), intra-node two-qubit blind gate and inter-node two-qubit blind gate.} Note that all efficiencies are cubed for the single qubit universal blind gate since the gate requires three successive spin-photon entanglement events.}
\label{table:efficiency_breakdown}
\end{table}

\begin{table}
\centering
\begin{tabular}{c c c c} 
  & Single qubit & Intra-node two-qubit & Inter-node two-qubit \\
  & universal blind gate & blind gate & blind gate \\[0.5ex] 
 \hline \\ 
 WCS $\mu$ & 0.9$^3$ & 0.975 & 0.9 \\ 
 contrast error server 2 & - & - & 0.95 \\
 MW pulse error server 2 & - & - & 0.95 \\
 Electron readout error server 2 & - & - & 0.99 \\
 contrast error server 1 & 0.97$^3$ & 0.97 & 0.97 \\
 MW pulse error server 1 & 0.99$^3$ & 0.97 & 0.95 \\
 RF pulse error server 1 & - & 0.99 & 0.99 \\
 Electron readout error server 1 & 0.99 & 0.99 & 0.99 \\
 Nucleus readout error server 1 & - & 0.98 & 0.98 \\
 TDI measurement error & 0.98$^3$ & 0.98 & 0.98 \\ [0.5ex] 
 \hline \\ 
 Expected total fidelity & 0.67 & 0.87 & 0.71 \\ [0.5ex] 
 Measured total fidelity & $0.73\pm0.04$ & $0.85\pm0.02$ & $0.71\pm0.02$ \\
\end{tabular}
\caption{\textbf{Fidelity breakdown for the single qubit universal blind gate, intra-node two-qubit blind gate and inter-node two-qubit blind gate.} Note that all fidelities are cubed for the single qubit universal blind gate since the gate requires three successive spin-photon entanglement events.}
\label{table:fidelity_breakdown}
\end{table}

The optical losses for our system are broken down in table~\ref{table:efficiency_breakdown} for different applied gates. 
Although all gates are designed to be used with single photons, we use much cheaper and readily available weak coherent states (WCS) at the cost of a reduction in gate efficiency with the WCS's average photon number $\mu$. 
The gate efficiencies are reduced by server coupling efficiency (due to inefficiencies in the tapered fiber-nanocavity interface), the client phase shifting setup efficiency (including a free space AOM first order shifting efficiency and free space to fiber coupling efficiency), and SNSPD detection efficiency. 
Two further sources of efficiency reduction come from the server cavity average reflection and TDI measurement efficiency. 
The server cavity average reflection is limited to a maximum of 50\% due to the reflection amplitude-based gates we use to generate spin-photon entanglement. This corresponds to the perfect contrast case of one spin state having a reflectivity of 1, and the other of 0. In practice, this efficiency number gets further reduced by the non-unity reflectivity of the reflective state due to the SiV optical transition frequency being located in the shoulder of the cavity. 
The TDI measurement efficiency is also limited to 50\% since only half of the photonic qubit state is shifted into a superposition basis (and only half is shifted into the correct superposition basis in the case of qudits, see Fig.~\ref{fig:SI_readout}C,D), with a further reduction due to fiber losses. 
While the single qubit universal blind gate and intra-node two-qubit blind gate are both single server operations, the inter-node two-qubit blind gate is also affected by frequency shifting losses due to the optical frequency difference between the SiVs of Servers 1 and 2. All the efficiencies for the 1QBG are cubed because the gate requires 3 successful photon events in a row.

Several steps can be taken to improve the overall efficiencies. 
The WCS can be replaced with a single photon source for an in principle arbitrarily high efficiency, although we note that it is important for the source to match both the SiV frequency and linewidth.
The server coupling efficiencies could be increased to $>0.95$ with careful nanocavity design and tapered fiber etching~\cite{Bhaskar2020}, and the client phase shifting setup can be improved by using a low-loss fiber AOM (with spliceable single mode fiber) or phase shifter to avoid free space to fiber coupling losses. 
We could increase server cavity average reflection to close-to-unity values by replacing the photon reflection amplitude-based gates by reflection phase-based gates. A key requirement for this is highly overcoupled, high cooperativity nanocavities, which could potentially be achieved with recent advances in thin-film diamond nanofabrication~\cite{Ding2024thinfilm}.
We could bring the TDI qubit/qudit measurement efficiency to higher values by replacing the first beamsplitter of the TDI with a fast switch (the switch has to be faster than the inter-pulse spacing of the photonic qubit/qudits) that can deterministically route photonic time bins to the short or long path.
Finally, active strain-tuning of SiVs to shift their optical resonance frequencies would allow us to bypass the requirement (and losses) of frequency shifting, further increasing inter-node operations.

\subsection{MW and RF pulse sequences}
Fig.~\ref{fig:SI_MWRF_pulse_seq} and Fig.~\ref{fig:SI_DJ_pulse_seq} show the MW and RF pulses applied for the different blind gates and protocols. MW $\pi$-pulses and $\pi/2$-pulses are \SI{30}{\nano \second} and \SI{15}{\nano \second}, respectively. RF $\pi$-pulses and $\pi/2$-pulses are $\sim$\SI{38}{\micro \second} and $\sim$\SI{19}{\micro \second}, respectively. The qubit/qudit AOM time bin time width is \SI{25}{\nano \second}. The initialization and measurement pre-rotation pulses (in parentheses in Fig.~\ref{fig:SI_MWRF_pulse_seq}) are applied depending on the desired initial state and final measurement basis, respectively. At the beginning of each sequence, all participating spins are initialized in the $\ket{\downarrow}$ state.

\subsection{Gate and protocol error sources}
Several error sources limit the fidelities of gates demonstrated in this work as shown in Table~\ref{table:fidelity_breakdown}. 
Multi-photon components of the photonic qubit/qudit WCS introduce errors scaling roughly with $\sim\mu/2$, where $\mu$ is the average photon number. 
Finite contrast errors are caused by the nonzero reflectivity of the non-reflective spin state and scale as $\sim 1/(C+1)$, where $C$ is the optical contrast, i.e., the reflective state reflectivity divided by the non-reflective state reflectivity.
TDI measurement errors are caused by a combination of imperfect 50:50 beamsplitters (where the ratio is not exactly 50:50), drift of the short-long path phase accumulation difference during experiments, and nonzero linewidth of the qubit/qudit laser (as compared to the TDI delay length).
While MW pulse sequences are designed in a way to have rephasing properties (see Fig.~\ref{fig:SI_MWRF_pulse_seq}), heating caused by the pulses and drift of the MW and RF resonance frequencies introduce errors, particularly for longer sequences involving more RF and MW gates.
Readout of the electron spin state is limited by the distinguishability of the reflective and non-reflective states, as well as the probability of the spin state flipping during readout. Since the nuclear spin state is read out by measuring the electron after applying a C$_n$NOT$_e$ gate introducing more errors, the fidelity is further reduced.

\section{Simulations}

In this section, we numerically explore the fidelities of each blind gate as a function of most likely errors in the server's systems. In addition, we evaluate the information leakage as a function of different errors to the server from the quantum state left after the gate.  It is important to note that only if the server knows that the gate has succeeded, there could be an information leakage to the server. Otherwise, there is no information transfer from the client to the server. But since we would begin the next gate after the first one succeeded, we assume that that server knows that the gate under question has succeeded.

\begin{figure}[t]
    \centering
    \includegraphics[width=1\linewidth]{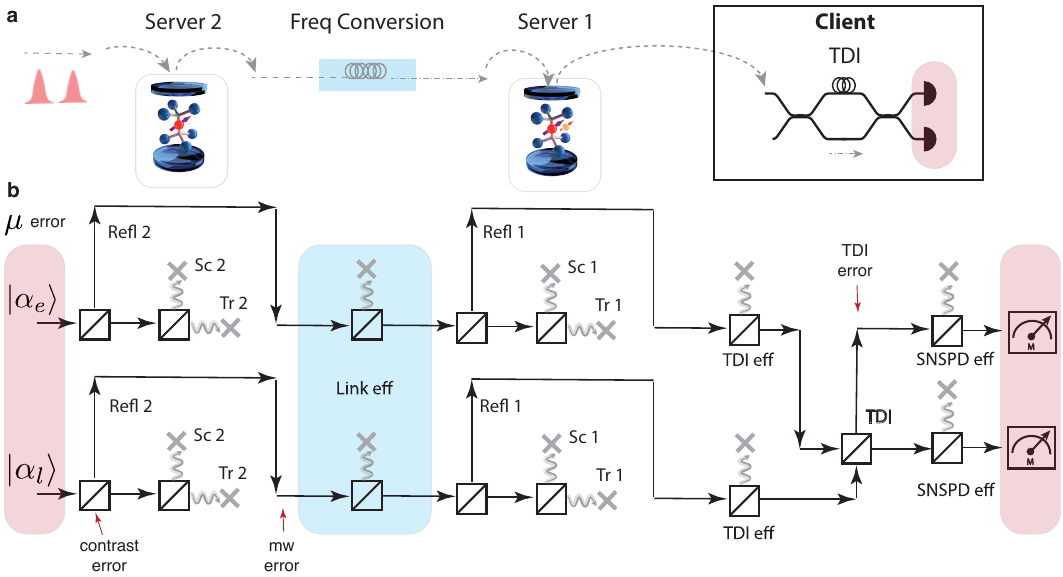}
    \caption{\textbf{Network Schematic:} An example schematic of a chain of beam splitters for a distributed two-qubit gate with a time-bin encoding. Each new port of a beam splitter corresponds to a new loss channel or new mode for the photons, e.g., from the input mode to the reflected mode.}
    \label{fig:SimScheme}
\end{figure}

The numerical simulations are built as a simple chain of beam splitter operators, shown in Fig.~\ref{fig:SimScheme}. The code can be found at \cite{github_repo}. The chain accepts a photonic qubit or qudit state in time-bin encoding represented as:

\begin{align}
\ket{\psi_{phot}} &= \frac{1}{norm_{\alpha}}(\ket{\alpha_e}\ket{0} + \ket{0}\ket{\alpha_l}) \\
\ket{\psi_{phot}} &= \frac{1}{norm_{\alpha}}(\ket{\alpha_0}\ket{0}\ket{0}\ket{0} + \ket{0}\ket{\alpha_1}\ket{0}\ket{0} + \ket{0}\ket{0}\ket{\alpha_2}\ket{0} + \ket{0}\ket{0}\ket{0}\ket{\alpha_3})
\end{align}

shown here in X basis. This multi-rail representation is especially useful for our implementation since we use the average photon number $\mu = |\alpha|^2 < 1$, which is a weak coherent state. The goal is to herald the event where the photonic input state had a single photon per qubit. Since it's a weak coherent state, most of the time, the number of photons is 0, which translates to loss, and sometimes the number of photons $> 1$, which translates to infidelity.  

The chain of beam splitters, shown in Fig.~\ref{fig:SimScheme}, corresponds to steps that the photonic modes go through on the way to detection. It starts with the interaction with the cavity, which separates the reflected mode, which goes on to the next step from transmitted and scattered modes, which are traced out as loss. Next, the reflected photonic modes pass through the link beam splitter, which attenuates the signal proportionally to the efficiency of the link, including multiple losses such as the frequency conversion efficiency. The attenuated mode then interacts with the cavity in the second node. The reflected mode then travels to the client with some link losses. At the client, the photon passes through the beam splitter operator, which corresponds to the action of the time delay interferometer (TDI). TDI interferes with specific rails of the photonic state to convert into a chosen basis $\phi$. After losses proportional to the TDI efficiency and the detection efficiency, the photon gets measured, which projects the state of the matter qubits at the server into the final state, which depends on $\phi$ and the detector that got fired. A single photon detection based on a click from + or - detectors heralds the success of the gate, which is immediately known to the client but unknown to the server unless the client chooses to communicate with the server later. 

The error sources we are considering here are:
\begin{enumerate}
    \item \textbf{$\mu$ errors} - the infidelity which occurs due to the occurrence of n = 2, 3, ... photon events. Since our photonic qubit is created using weak coherent sources for $\mu < 1$, there is a $\approx \frac{\mu^2}{2}$ chance of getting two photons in the same mode. When two-photon events occur, there is a chance of both photons interacting with the SiV and one of them getting subsequently lost. This loss decoheres the matter qubits but still gives us the heralding signal of a successful gate - which reduces the fidelity of the operation. A higher number of photons per qubit $\mu$ results in higher success rates of the gate but lower fidelities. We simulate this error by varying the average number of photons in the initial photonic state and evaluate the reduction in the overall fidelity of the gate after the heralding click.  
    
    \item \textbf{the contrast errors} - the infidelity that comes from residual reflectivity of the cavity when the electric is in a non-reflective state. This deviation from $\frac{Ref(\uparrow)}{Ref(\downarrow)} \to \infty$ can come from non-ideal cavity-QED parameters, the spectral diffusion of the SiV and also due to stray reflections on the path. The effect of the contrast error is an asymmetry to our spin-photon entangled resource before the client's measurement.
    For the matter and photon qubits starting in $\ket{+}$ states, we can approximately think of the spin-photon state which can be detectable to be:

    \begin{align}
    \approx R (- \ket{+}_{phot}\ket{x-}_{spin} + \ket{-}_{phot}\ket{+}_{spin}) + ( 1- R) (\ket{+}_{phot}\ket{-}_{spin} + \ket{-}_{phot}\ket{+}_{spin}) \\ 
    \end{align}

    So when $R < 1$, there is an infidelity in the gate performance.  We simulate this error by varying the complex reflectivity in the beam splitter operator, which corresponds to the cavity-QED-photon interaction. The varying complex reflectivity is calculated using a full cavity-QED fit of the experimental data. Note that in our simulations, we also included time-dependent contrast drifts, which are likely to occur in the solid-state platforms. 
    
    \item \textbf{microwave gate errors} - the infidelity that comes from imperfect single qubit rotations driven by microwave or RF signals. We include the infidelities as a coherent under- or over-rotation of the electron or nuclear spins in each experimental shot between beam splitter operations. The deviation in the angle is picked randomly from a Gaussian distribution. This results in an incoherent error in microwave gates average over many experimental runs.
    
    \item \textbf{TDI errors} - the infidelity that comes from imperfect interference of the photonic modes in the time delay interferometer. We simulate the phase offset of the interference operator due to imperfect locking of the TDI, which results in the deviation of the basis $\phi$ chosen by the client and also an amplitude offset of the interference operation, which can result in asymmetry of the probability of a click between detectors. Note that in our simulations, we also included time-dependent TDI drifts, which are likely to occur in our system. 
    
\end{enumerate}

\subsection{Reproducing the experimental results}

\begin{figure}[t]
    \centering
    \includegraphics[width=1\linewidth]{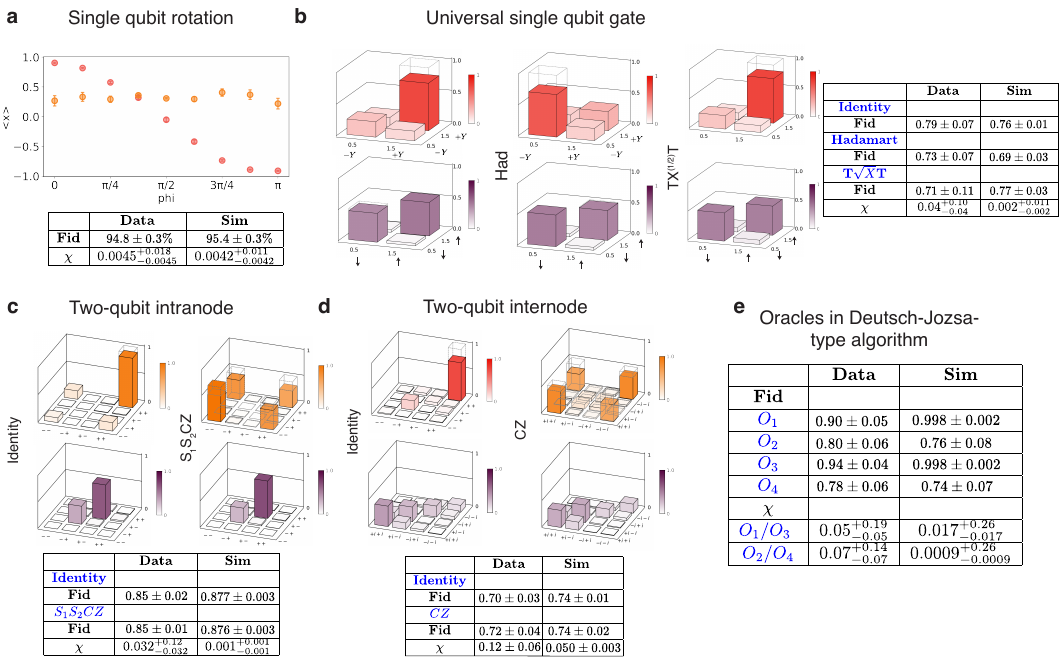}
    \caption{\textbf{Simulation results for experimental blind gates} (a) single qubit rotation as a function of $\phi$, (b) universal single gates, (c) intranode 2 qubit gates, (d) internode two-qubit gate, (e) the Deutsch-Jozsa-type algorithm performance}
    \label{fig:datamatch}
\end{figure}

First, to demonstrate the predictive ability of the numerics, we reproduce the single and two-qubit blind gates described in the main text. Fig.~\ref{fig:datamatch} shows the results. For the single qubit rotations, we use $\mu = 0.05$, contrast is $ = 28$, microwave gate fidelity $=99\%$, and it is incoherent, TDI phase error is $1.6\%$. For the universal single qubit gates, the noise sources are $\mu = 0.2$, contrast is $ = 28$, microwave gate fidelity $=99\%$, and it is incoherent, TDI phase error is $6.4\%$. The larger $\mu$ was chosen to reflect the experiment, where we increased the number of photons per event to improve rates while sacrificing a bit on fidelity. Similarly, the increase in TDI error reflects the fact that 3-photon event experiments are slower, so TDI has a longer time to drift. 
For a two-qubit internode gate, we use $\mu = 0.25$, contrast is $ = 28$, microwave gate fidelity $=99\%$, and it is incoherent, the TDI phase error is $6.4\%$. For these single-node experiment simulations, the detection efficiency from the SiV to the photon detector is $0.057$.

In the internode two-qubit gate simulations, the noise sources are $\mu = 0.07$, contrast is $ = 28$, microwave gate fidelity $=99\%$, and it is incoherent, the TDI phase error is $6.4\%$. In the Deutsch-Jozsa-type algorithm, we use $\mu = 0.25$, contrast is $ = 28$, microwave gate fidelity $=99\%$, and it is incoherent, the TDI phase error is $6.4\%$. Here, Fig.~\ref{fig:datamatch} shows a comparison between the probability of getting the correct answer - Balanced or Constant. As a reference, the two-qubit state fidelities of the final computational qubits of $O_1$,  $O_2$,  $O_3$, and $O_4$ are $0.74(7)$, $0.76(8)$, $0.70(7)$,  and $0.74(5)$, respectively. Note that the correctness of the Balanced vs Constant in simulation for $O_1$ and $O_3$ is a bit higher than in the experiment. We believe it is due to the fact that the simulation does not have the extra dephasing of the nuclear qubit due to the photon-electron interactions. This could be easily added to the simulation package.
For these two node experiment simulations, the detection efficiency from the SiV of server 1 to the photon detector is $0.057$, and the link efficiency between the SiVs is $0.012$.

\subsection{Full gate set tomography of two-qubit gates } 

In the experiment, it was costly to run a full gate-set tomography to calculate gate fidelities for two-qubit gates. Therefore, the fidelities quoted are of final states after the action of the gate on the critical input states -- those which require quantum coherence and entanglement generating ones. Using numerics, we can simulate a full gate tomography to estimate expected gate fidelities. 

\begin{figure}[t]
    \centering
    \includegraphics[width=1\linewidth]{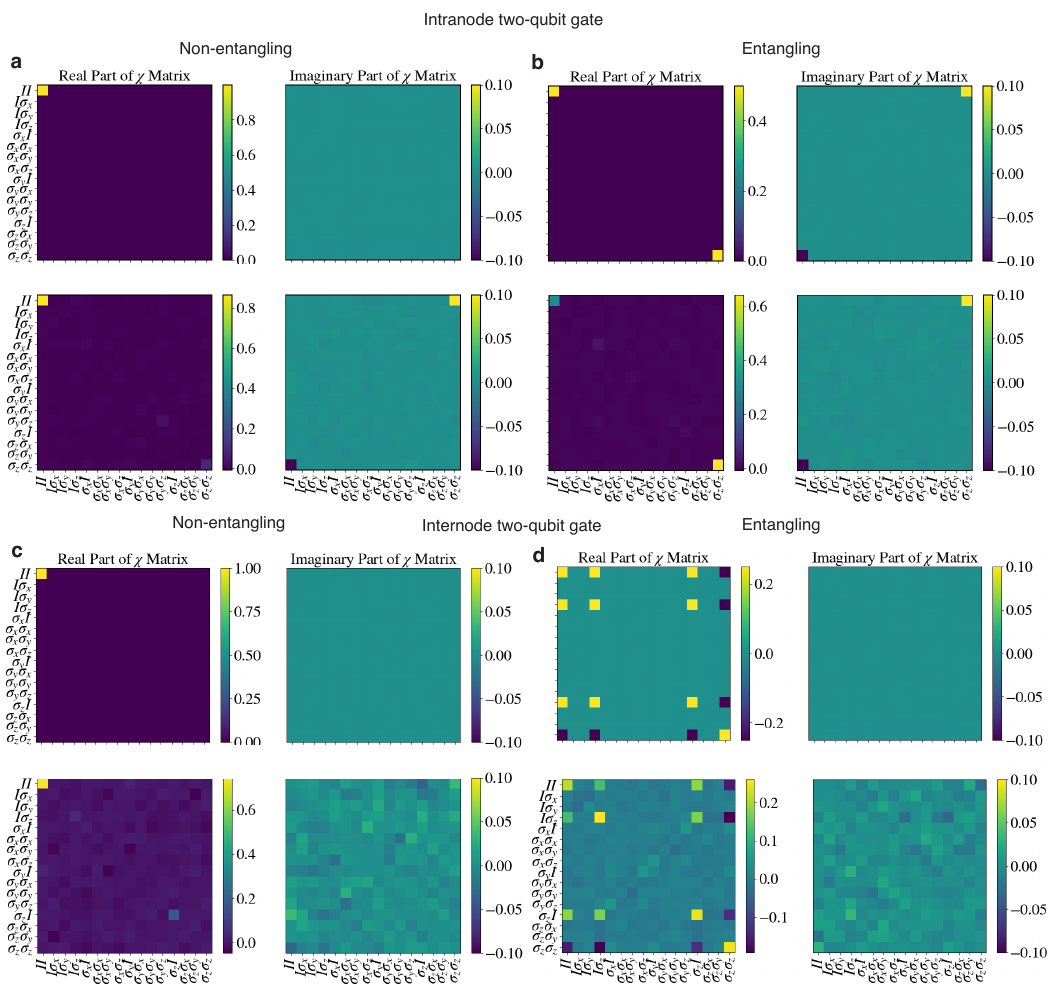}
    \caption{\textbf{Simulation results for  $\chi$ tables: average $\chi$ matrices from gate set tomography for two-qubit gates with and without noise sources} (a) for intranode non-entangling gate (Identity) $\phi = 0$ without(top) and with(bottom) error sources,  (b) for intranode entangling gate ($S_1 S_2 CZ$) $\phi = \pi/2$ without(top) and with(bottom) error sources, (c) for internode non-entangling gate (Identity) without(top) and with(bottom) error sources, (a) for internode entangling gate (CZ) without(top) and with(bottom) error sources}
    \label{fig:intranode_chitables}
\end{figure}

Fig.~\ref{fig:intranode_chitables} shows the $\chi$ matrices of two-qubit intranode and internode blind gates. The  $\chi$ matrix elements are calculated using the process matrix for gate A:

\begin{equation}
A = \sum^{d^2}_{i,j= 1} \chi_{ij}P_i \rho P_j 
\end{equation}

where $\rho$ is arbitrary input state, $P_i$ are Pauli operators acting on the state and $d$ is the dimension of the Hilbert space.

The top row is for noiseless operation of the experiment, and the bottom row is with error sources including $\mu = 0.25$, contrast is $ = 28$, microwave gate fidelity $=99\%$, and it is incoherent, and the TDI phase error is $6.4\%$. 
The simulated process fidelities, defined as $F_p = Tr[\chi_{ideal}*\chi_{sim}]$ where $\chi_{ideal}$ is the $\chi$ matrix of the ideal process and $\chi_{sim}$ is the $\chi$ matrix of the simulated actual process, for the intranode two-qubit gate are $0.87$ for the non-entangling Identity gate and $0.84$ for the entangling $S_1 S_2 CZ$ gate, respectively. The average gate fidelities, defined as $F_{ave} = \frac{d F_p +1}{d+1}$, for the intranode two-qubit gate are $0.91$ for the non-entangling Identity gate and $0.89$ for the entangling $S_1 S_2 CZ$ gate, respectively.  For the internode two-qubit gate the simulated process fidelities are $0.74(1)$ for the non-entangling Identity gate and $0.72(1)$ for the entangling $CZ$ gate, respectively. The average gate fidelities for the internode two-qubit gate are $0.83(1)$ for the non-entangling Identity gate and $0.81(1)$ for the entangling $S_1 S_2 CZ$ gate, respectively The fidelity is calculated by taking the overlap between the simulated chi matrix of the gate operator and the ideal chi matrix. 

\subsection{Single-qubit rotations}

\begin{figure}[t]
    \centering
    \includegraphics[width=1\linewidth]{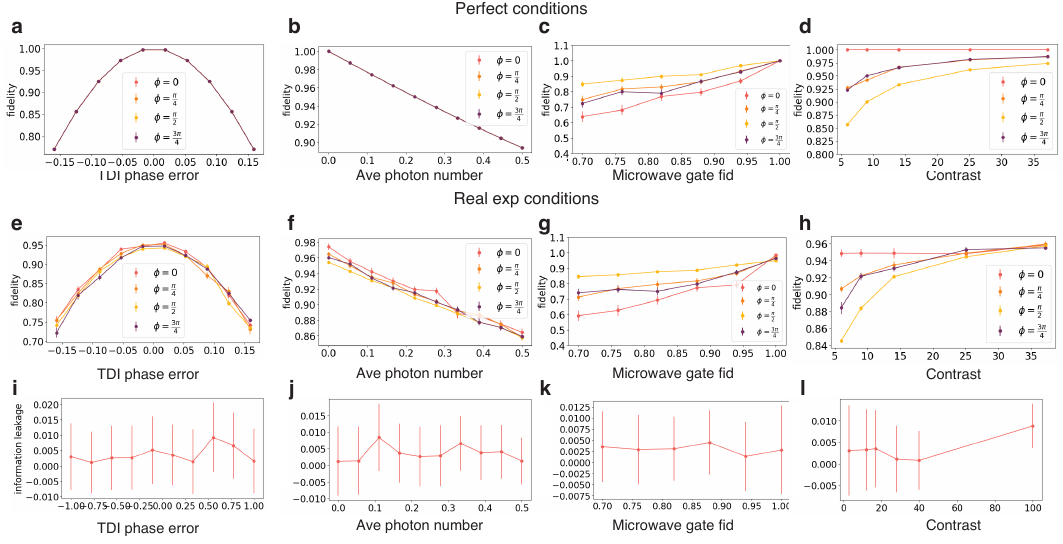}
    \caption{\textbf{Simulation results for fidelities and information leakage of blind single qubit gates as a function of specific errors} (a-d) TDI phase errors, Ave photon number errors, microwave fidelity, contrast dependence of fidelity of the gates with no other errors included, (e-h) TDI phase errors, Ave photon number errors, microwave fidelity, contrast dependence of fidelity of the gates with all realistic experimental errors included. Here the experimental errors which are not beiing scannes are set to $\mu = 0.05$, contrast $= 25$, mw fidelity $ = 99$, and TDI phase error $= - 0.1$ rad or $1.6 \%$, (i-l) TDI phase errors, Ave photon number errors, microwave fidelity, contrast dependence of the information leakage to the server while client performs gates over all four values of $\phi$ with all realistic experimental errors included.}
    \label{fig:perf_vs_exp}
\end{figure}

Since it's the simplest case, we use single qubit rotations to explore the propagation of errors and the effect of different types of noise on the fidelity of the computation and information leakage. Fig.~\ref{fig:perf_vs_exp} shows the output of the simulation for a single qubit rotation by $\phi = 0, \pi/4, \pi/2, 3\pi/4$ dependent on four main types of major sources of errors in our system. On top row we only track behavior of the fidelity as a function of the specific error source and no other errors and at the bottom two rows we plot fidelities and information leakage with all other errors in the experiment included. We can see that TDI phase errors and photon number errors affect all $\phi$ gates similarly, while microwave and contrast errors change the behavior for different angles. 

It is worthy to note some of the important features of contrast errors for example. If the client measures in the X basis, there is no reduction in fidelity from contrast. In other words, we expect the fidelity of single rotations by $\phi = 0$ or $\pi$ to be independent of contrast. Away from the X basis, the contrast becomes an issue. We can see this effect clearly in Fig.~\ref{fig:perf_vs_exp}d and h. Another interesting feature of contrast errors is their effect on the probability of different detectors, in our case SNSPDs, firing depending on the client rotation angle. For perfect contrast, we expect the probabilities of + and - detectors firing to be equal but, in the case of low contrast, to have significant asymmetries in rates of firing depending on the client's angle. If we have low contrast and we tell the server when we succeed, the contrast error can cause information leakage from this asymmetry.

On the other hand, the microwave errors similarly not only affect the overall fidelities of the blind gates but also affect relative fidelities and likelihoods of $\phi +$ and $\phi -$ detectors firing, which can be seen in different variability of fidelities at different $\phi$ angles. In contrast to contrast errors, microwave infidelity affects the $\phi = 0$ angles the most and the $\phi = \pi/2$ the least. However, similar to contrast errors, high infidelity in microwave errors could become a source of information leakage if the server knows when the client succeeded in the gate performance. 

Each point in the plots include 100 experimental rounds and the error bars are evaluate from the standard deviation of the output. The ambient errors outside of the source that is being scanned are set to $\mu = 0.05$, contrast $= 25$, mw error $ = 99$, and TDI$= - 0.1$ rad or $1.6\%$ error. 

The simulated blindness or leakage of information for single qubit gates are smaller than the error bars from the random nature of noise encoded in the simulations to reprosduce the experiment.

\section{Blind gates data analysis}

\subsection{Information leakage calculation}

The server does not have access to the photonic measurement outcomes, but he or she does hold the physical qubits and can measure them to learn information about the computation.
To show blindness for a given gate $U(\Phi)$, where $\Phi = (\phi_1,...,\phi_n)$ is the client's chosen parameter(s) to apply a specific desired unitary, we need to show that the information the server has access to, i.e., the information contained in the final density matrix of the server's qubit(s), is independent of the choice of $\Phi$ by the client.

To quantify the information leakage in our experiment, we measure the Holevo information of the server's final density matrix for a given set of parameter(s) choices $\Upsilon = \{\Phi_1,...,\Phi_k\}=\{(\phi_1^{(1)},...,\phi_n^{(1)}),...(\phi_1^{(k)},...,\phi_n^{(k)})\}$:
\begin{align}
    \chi(\rho) = S(\rho) - \frac{1}{N_{\Phi}}\sum_{\Phi_i\in\Upsilon} S(\rho_{\Phi_i}),
\end{align}
Where $S(\rho)$ is the von Neumann entropy defined as:

\begin{align}
    S(\rho) = - Tr[\rho \ln{\rho}].
\end{align}

$N_{\Phi}$ is the number of parameter(s) choices in $\Upsilon$, $\rho_{\Phi_i}$ is the final server qubit(s) density matrix after implementation of gate $U(\Phi_i)$, and $\rho$ is the average: 

\begin{align}
    \rho = \frac{1}{N_{\Phi}}\sum_{\Phi_i\in\Upsilon} \rho_{\Phi_i}.
\label{eq:general_blindness}
\end{align}

$\chi = 0$ for perfect blindness, meaning the server does not receive any information about which parameter(s) $\Phi_i$ the client chooses within the set $\Upsilon$, and thus which gate the client applies within the set $U(\Phi_1), ..., U(\Phi_k)$.

\subsubsection{Single qubit blind rotation}

The single qubit blind rotation about $z$ axis ($R_z$) is characterized by a single parameter $\Phi_i = \phi_i$ where $\phi_i$ is the rotation angle of the applied gate $R_z(\phi_i)$. We sweep $\Phi_i$ over 4 values: $\Upsilon = \{0, \frac{\pi}{4},  \frac{\pi}{2}, \frac{3\pi}{4}\}$, so that the information leakage can be calculated with equation \ref{eq:general_blindness} and $N_{\Phi}=4$. Since the server does not have access to the photonic measurement result $s$, he or she sees the density matrix as:

\begin{align}
    \rho_{\Phi_i}  = p_{\Phi_i}^{(s=0)}\rho_{\Phi_i}^{(s=0)} + p_{\Phi_i}^{(s=1)}\rho_{\Phi_i}^{(s=1)},
\end{align}
where $p_{\Phi_i}^{M}$ and $\rho_{\Phi_i}^{M}$ are the probability of the client's photonic measurement outcome $M$ and the server qubit's density matrix after the client's photonic measurement outcome $M$ when the client chooses parameter $\Phi_i$, respectively. Concretely, $s=0$ corresponds to the measurement of state $(\ket{0}+e^{i\phi_i}\ket{1})/\sqrt{2}$ and $s=1$ of state $(\ket{0}-e^{i\phi_i}\ket{1})/\sqrt{2}$.
We can explicitly write the density matrix as a function of the expectation values:
\begin{align}
    \rho_{\Phi_i} = \frac{1}{2}(I + \Vec{\sigma}\cdot \Vec{n}_{\Phi_i}) = \frac{1}{2}(I + \sigma_x \braket{\sigma_x}_{\Phi_i} + \sigma_y \braket{\sigma_y}_{\Phi_i} + \sigma_z \braket{\sigma_z}_{\Phi_i}),
\end{align}
with:
\begin{align}
    \braket{\sigma_j}_{\Phi_i} = p_{\Phi_i}^{(s=0)}\braket{\sigma_j}_{\Phi_i}^{(s=0)} + p_{\Phi_i}^{(s=1)}\braket{\sigma_j}_{\Phi_i}^{(s=1)}
\end{align}
We can then write the eigenvalues:
\begin{align}
    p_{\Phi_i}^{\pm} = \frac{1}{2} \left ( 1 \pm \sqrt{ \braket{\sigma_x}_{\Phi_i}^2 + \braket{\sigma_y}_{\Phi_i}^2 + \braket{\sigma_z}_{\Phi_i}^2 } \right ).
\label{eq:eigenvalues_sq}
\end{align}
Resulting in the entropy:
\begin{align}
    S(\rho_{\Phi_i}) = -p_{\Phi_i}^{+} \ln{p_{\Phi_i}^{+}} -p_{\Phi_i}^{-} \ln{p_{\Phi_i}^{-}} 
\end{align}
We express the average final density matrix as:
\begin{align}
    \frac{1}{2} (I + \frac{1}{4}\sum_{\Phi_i\in\Upsilon}( \sigma_x \braket{\sigma_{x}}_{\Phi_i} + \sigma_y \braket{\sigma_{y}}_{\Phi_i} + \sigma_z \braket{\sigma_{z}}_{\Phi_i} )),
\end{align}
with eigenvalues
\begin{align}
    p^{\pm} = \frac{1}{2} \left ( 1 \pm \frac{1}{4}\sqrt{  (\sum_{i=1}^4 \braket{\sigma_{x}}_{\Phi_i})^2 +  (\sum_{i=1}^4 \braket{\sigma_{y}}_{\Phi_i})^2 + (\sum_{i=1}^4 \braket{\sigma_{z}}_{\Phi_i})^2 } \right ).
\label{eq:avg_eigenvalues_sq}
\end{align}
We finally can explicitly write the expression for information leakage (equation \ref{eq:general_blindness}) as:
\begin{align}
    \chi = -p^{+} \ln{p^{+}} -p^{-} \ln{p^{-}} - \frac{1}{4} \sum_{i=1}^4 (-p_{\Phi_i}^{+} \ln{p_{\Phi_i}^{+}} -p_{\Phi_i}^{-} \ln{p_{\Phi_i}^{-}}).
\label{eq:explicit_blindness_sq}
\end{align}

\begin{figure}
    \centering
    \includegraphics[]{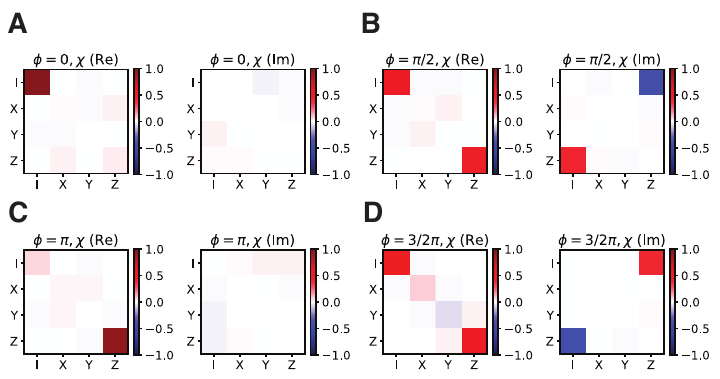}
    \caption{\textbf{Quantum process tomography on $R_z$}. Using experimental data in Table~\ref{table:sq_blind_rotation_x_cl}, Table~\ref{table:sq_blind_rotation_x_sr} Table~\ref{table:sq_blind_rotation_y_cl}, Table~\ref{table:sq_blind_rotation_y_sr},
    Table~\ref{table:sq_blind_rotation_z_cl}, and Table~\ref{table:sq_blind_rotation_z_sr} , we do quantum process tomography on $R_z$ operation with $\phi=0, \pi/2, \pi, 3\pi/2$, plotted in \textbf{A}, \textbf{B}, \textbf{C}, \textbf{D}, respectively. $\chi$ here denotes the full state tomography coefficients matrices.}
\label{fig:SI_rz_qpt}
\end{figure}

From the Table~\ref{table:sq_blind_rotation_x_cl}, ~\ref{table:sq_blind_rotation_x_sr}, ~\ref{table:sq_blind_rotation_y_cl}, ~\ref{table:sq_blind_rotation_y_sr}, ~\ref{table:sq_blind_rotation_z_cl}, and ~\ref{table:sq_blind_rotation_z_sr}, we can perform quantum process tomography~\cite{main2024distributed} shown in Fig.~\ref{fig:SI_rz_qpt} and get the gate fidelity $94.8 \pm0.3 \%$.

\subsubsection{One-qubit blind gate (1QBG)}

The 1QBG consists of three consecutive blind rotations and is characterized by three parameters $\Phi_i = (\phi_i^{(0)}, \phi_i^{(1)}, \phi_i^{(2)})$ defining the applied gate $R_z(\phi_i^{(2)})R_x(\phi_i^{(1)})R_z(\phi_i^{(0)})$. We choose three different values of $\Phi_i$: $\Upsilon = \{(0,0,0), (\frac{\pi}{2}, \frac{\pi}{2}, \frac{\pi}{2}),  (\frac{\pi}{4}, \frac{\pi}{2}, \frac{\pi}{4})\}$, corresponding to the Identity, Hadamard, and $T\sqrt{X}T$ gates, respectively. We again calculate the information leakage with equation \ref{eq:general_blindness}, but here with $N_{\Phi}=3$. With 3 blind rotations and thus 3 photonic measurement outcomes, the server sees the density matrix:

\begin{align}
    \rho_{\Phi_i}  = \sum_{k=\{0,1\}}\sum_{l=\{0,1\}}\sum_{m=\{0,1\}}p_{\Phi_i}^{(s_1=k, s_2=l,s_3=m)}\rho_{\Phi_i}^{(s_1=k, s_2=l,s_3=m)}.
\end{align}
Similarly, the Pauli operator expectation values can be calculated as:

\begin{align}
    \braket{\sigma_j}_{\Phi_i} = \sum_{k=\{0,1\}}\sum_{l=\{0,1\}}\sum_{m=\{0,1\}}p_{\Phi_i}^{(s_1=k, s_2=l,s_3=m)}\braket{\sigma_j}_{\Phi_i}^{(s_1=k, s_2=l,s_3=m)}
\end{align}
We can then use equations \ref{eq:eigenvalues_sq}, \ref{eq:avg_eigenvalues_sq}, and \ref{eq:explicit_blindness_sq} (with $N_{\Phi}=3$) to calculate the information leakage.

\subsubsection{Intra-node two-qubit blind gate}

The intra-node 2QBG, as with the single qubit blind rotation, is characterized by a single parameter $\Phi_i = \phi_i$, but here $\phi_i$ is the rotation angle of the SPG in the intra-node 2QBG sequence (Fig.\ref{fig:all_seq}B)
. We sweep $\Phi_i$ over 4 values: $\Upsilon = \{0, \frac{\pi}{4},  \frac{\pi}{2}, \frac{3\pi}{4}\}$ ($N_{\Phi}=4$). We can explicitly calculate the density matrices with 2-qubit Pauli operator measurements:

\begin{align}
    \rho_{\Phi_i} = \frac{1}{4}(II + \sigma_x I \braket{\sigma_x I}_{\Phi_i} + \sigma_y I \braket{\sigma_y I}_{\Phi_i} + \sigma_z I \braket{\sigma_z I}_{\Phi_i} \nonumber \\
    + \sigma_x \sigma_x \braket{\sigma_x \sigma_x}_{\Phi_i} + \sigma_y \sigma_x \braket{\sigma_y \sigma_x}_{\Phi_i} + \sigma_z \sigma_x \braket{\sigma_z \sigma_x}_{\Phi_i} + I \sigma_x \braket{I \sigma_x}_{\Phi_i} \nonumber \\
    + \sigma_x \sigma_y \braket{\sigma_x \sigma_y}_{\Phi_i} + \sigma_y \sigma_y \braket{\sigma_y \sigma_y}_{\Phi_i} + \sigma_z \sigma_y \braket{\sigma_z \sigma_y}_{\Phi_i} + I \sigma_x \braket{I \sigma_y}_{\Phi_i} \nonumber \\
    + \sigma_x \sigma_z \braket{\sigma_x \sigma_z}_{\Phi_i} + \sigma_y \sigma_z \braket{\sigma_y \sigma_z}_{\Phi_i} + \sigma_z \sigma_z \braket{\sigma_z \sigma_z}_{\Phi_i} + I \sigma_z \braket{I \sigma_z}_{\Phi_i}),
\label{eq:2q_DM}
\end{align}
with 

\begin{align}
    \braket{\sigma_j\sigma_k}_{\Phi_i} = p_{\Phi_i}^{(s=0)}\braket{\sigma_j\sigma_k}_{\Phi_i}^{(s=0)} + p_{\Phi_i}^{(s=1)}\braket{\sigma_j\sigma_k}_{\Phi_i}^{(s=1)}.
\end{align}
We then find the eigenvalues of the calculated density matrices and use equation \ref{eq:general_blindness} with $N_{\Phi}=4$ to calculate information leakage.

\subsubsection{Inter-node two-qubit blind gate}
For the application of the inter-node 2QBG, the client can choose the three phases of the 4D qudit phase gate (equation \ref{eq:4d_phasegate}), as well as the use of the short TDI (to measure in the $\{\ket{0}\pm\ket{1},\ket{2}\pm\ket{3}\}$ basis for an entangling gate) or the long TDI (to measure in the $\{\ket{0}\pm\ket{2},\ket{1}\pm\ket{3}\}$ basis for a non-entangling gate), see Fig.~\ref{fig:SI_readout}C,D. This gives the client four parameters: $\Phi_i = (\phi_i^{(0)}, \phi_i^{(1)}, \phi_i^{(2)}, k_i)$, where $\phi_i^{(j)}$ are the phases of the qudit phase gate, and $k_i$ is the choice of TDI for the measurement basis $\{\ket{0}\pm\ket{1+k_i},\ket{2-k_i}\pm\ket{3}\}$ ($k_i=0$ for the short TDI, and $k_i=1$ for the long TDI). We choose 2 different values of $\Phi_i$: $\Upsilon = \{(0,0,0,0), (0, \frac{\pi}{2}, \frac{\pi}{2}, 1)\}$, corresponding to an entangling gate and non-entangling gate, respectively.

Here, the client's measurement results in four different outcome states. We characterize this with two different numbers $s_1, s_2 \in \{0,1\}$, where $s_1$ distinguishes between states $\ket{i}+e^{i\phi}\ket{j}$ ($s_1=0$) and $\ket{i}-e^{i\phi}\ket{j}$ ($s_1=1$) and $s_2$ distinguishes between states $\ket{0}\pm e^{i\phi}\ket{1+k}$ ($s_2=0$) and $\ket{2-k}\pm e^{i\phi}\ket{3}$ ($s_2=1$). We calculate the combined density matrix of the electron of Server 2 and the nucleus of Server 1 with \ref{eq:2q_DM}, modifying the expectation value expression to:

\begin{align}
    \braket{\sigma_j\sigma_k}_{\Phi_i} = \sum_{l=\{0,1\}}\sum_{m=\{0,1\}}p_{\Phi_i}^{(s_1=l, s_2=m)}\braket{\sigma_j\sigma_k}_{\Phi_i}^{(s_1=k, s_2=m)}
\end{align}
We once again find the eigenvalues of the calculated density matrices and use equation \ref{eq:general_blindness} with $N_{\Phi}=2$ to calculate information leakage. We note that the server sees two different (Server 2 electron-Server 1 nucleus) density matrices depending on the measurement outcome of the Server 1 electron. We therefore separately calculate the density matrices for each Server 1 electron measurement outcome, separately compute the information leakage, and finally take an average of the two for the final total information leakage of the inter-node two-qubit blind gate.

\subsection{Fidelity calculation}

The fidelity of a given gate is calculated as the overlap of the resulting density matrix $\rho_{\mathrm{exp}}$ and the ideal target state$\ket{\psi_{\mathrm{target}}}$: $F = \bra{\psi_{\mathrm{target}}}\rho_{\mathrm{exp}}\ket{\psi_{\mathrm{target}}}$. Explicitly writing out this expression in terms of Pauli operator measurements, for 1-qubit gates:

\begin{align}
    F = \frac{1}{2}(1 + \braket{\sigma_x}_{\mathrm{target}} \braket{\sigma_x}_{\mathrm{exp}} + \braket{\sigma_y}_{\mathrm{target}} \braket{\sigma_y}_{\mathrm{exp}} + \braket{\sigma_z}_{\mathrm{target}} \braket{\sigma_z}_{\mathrm{exp}}),
\end{align}
where $\braket{O}_{\mathrm{target}}$ is the expectation value of operator $O$ of the ideal target state, and $\braket{O}_{\mathrm{exp}}$ the experimentally measured expectation value. For 2-qubit gates, we explicitly calculate the fidelity as:

\begin{align}
    F = \frac{1}{4}(1 + \braket{\sigma_x I}_{\mathrm{target}} \braket{\sigma_x I}_{\mathrm{exp}} + \braket{\sigma_y I}_{\mathrm{target}} \braket{\sigma_y I}_{\mathrm{exp}} + \braket{\sigma_z I}_{\mathrm{target}} \braket{\sigma_z I}_{\mathrm{exp}} \nonumber \\
    + \braket{\sigma_x \sigma_x}_{\mathrm{target}} \braket{\sigma_x \sigma_x}_{\mathrm{exp}} + \braket{\sigma_y \sigma_x}_{\mathrm{target}} \braket{\sigma_y \sigma_x}_{\mathrm{exp}} + \braket{\sigma_z \sigma_x}_{\mathrm{target}} \braket{\sigma_z \sigma_x}_{\mathrm{exp}} + \braket{I \sigma_x}_{\mathrm{target}} \braket{I \sigma_x}_{\mathrm{exp}} \nonumber \\
    + \braket{\sigma_x \sigma_y}_{\mathrm{target}} \braket{\sigma_x \sigma_y}_{\mathrm{exp}} + \braket{\sigma_y \sigma_y}_{\mathrm{target}} \braket{\sigma_y \sigma_y}_{\mathrm{exp}} + \braket{\sigma_z \sigma_y}_{\mathrm{target}} \braket{\sigma_z \sigma_y}_{\mathrm{exp}} + \braket{I \sigma_y}_{\mathrm{target}} \braket{I \sigma_y}_{\mathrm{exp}} \nonumber \\
    + \braket{\sigma_x \sigma_z}_{\mathrm{target}} \braket{\sigma_x \sigma_z}_{\mathrm{exp}} + \braket{\sigma_y \sigma_z}_{\mathrm{target}} \braket{\sigma_y \sigma_z}_{\mathrm{exp}} + \braket{\sigma_z \sigma_z}_{\mathrm{target}} \braket{\sigma_z \sigma_z}_{\mathrm{exp}} + \braket{I \sigma_z}_{\mathrm{target}} \braket{I \sigma_z}_{\mathrm{exp}}).
\end{align}
In all cases error bars are calculated assuming a binomial distribution for the operator expectation value measurements.

\subsection{Information leakage and fidelity calculation additional data}
\begin{figure}
    \centering
    \includegraphics[width=1\textwidth]{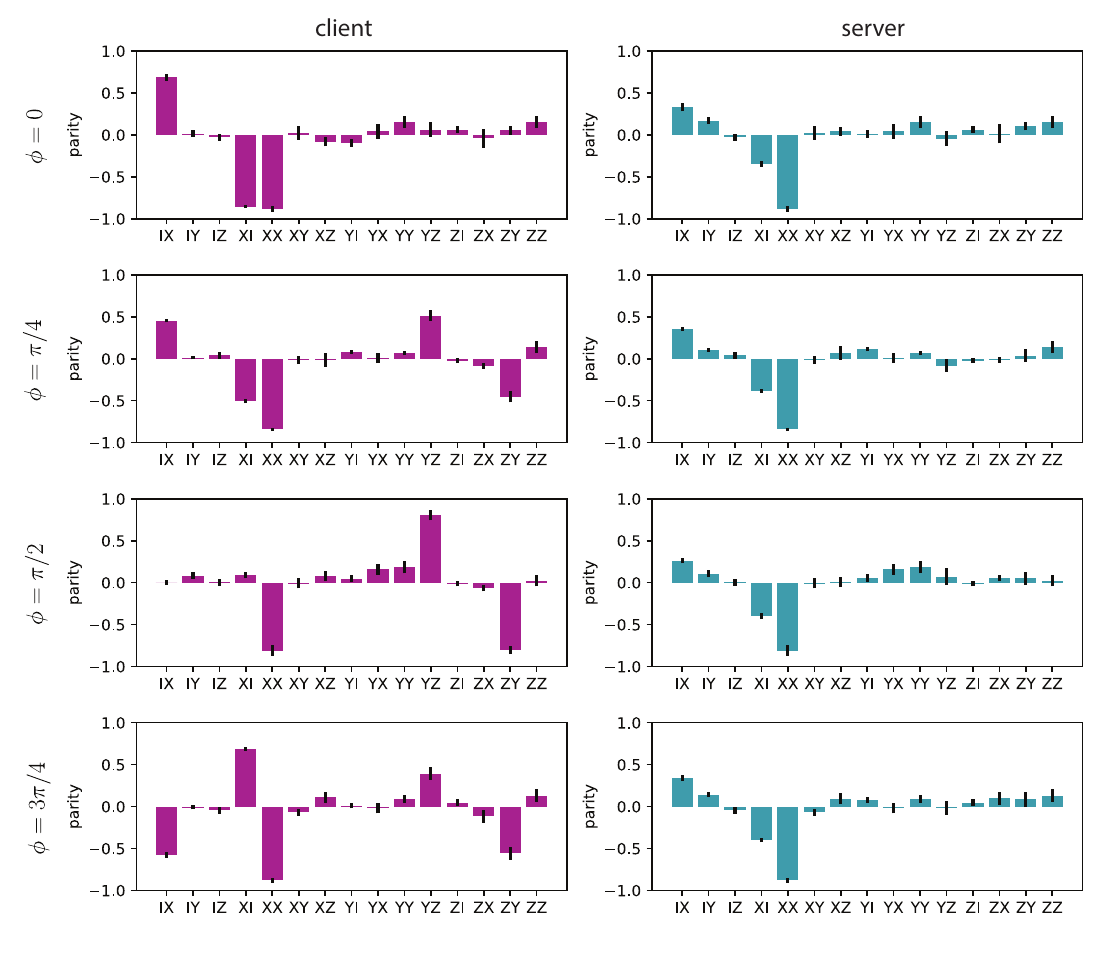}
    \caption{\textbf{2-qubit operator expectation values of the server and client for the intra-node 2-qubit gate shown in Fig.~\ref{fig:intranode}.}}
    \label{fig:SI_intranode_exp_val}
\end{figure}

\begin{figure}
    \centering
    \includegraphics[]{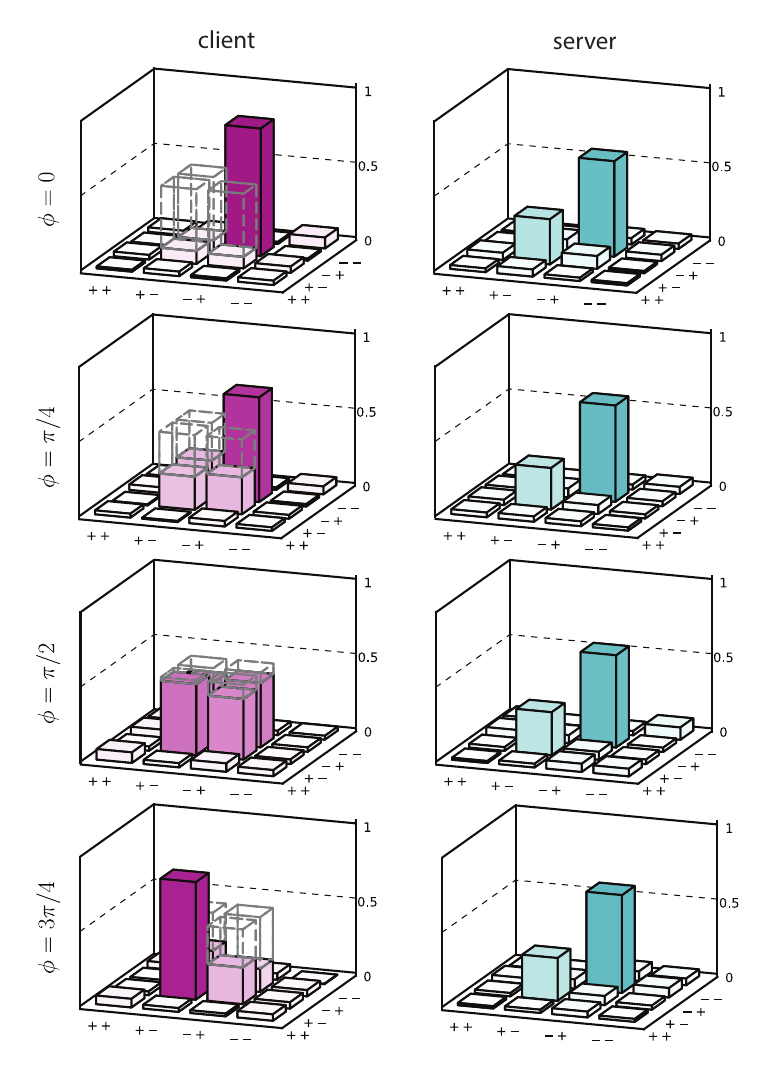}
    \caption{\textbf{Reconstructed density matrices of the server and client for the intra-node 2-qubit gate shown in Fig.~\ref{fig:intranode}.}}
    \label{fig:SI_intranode_DMs}
\end{figure}

\begin{figure}
    \centering
    \includegraphics[]{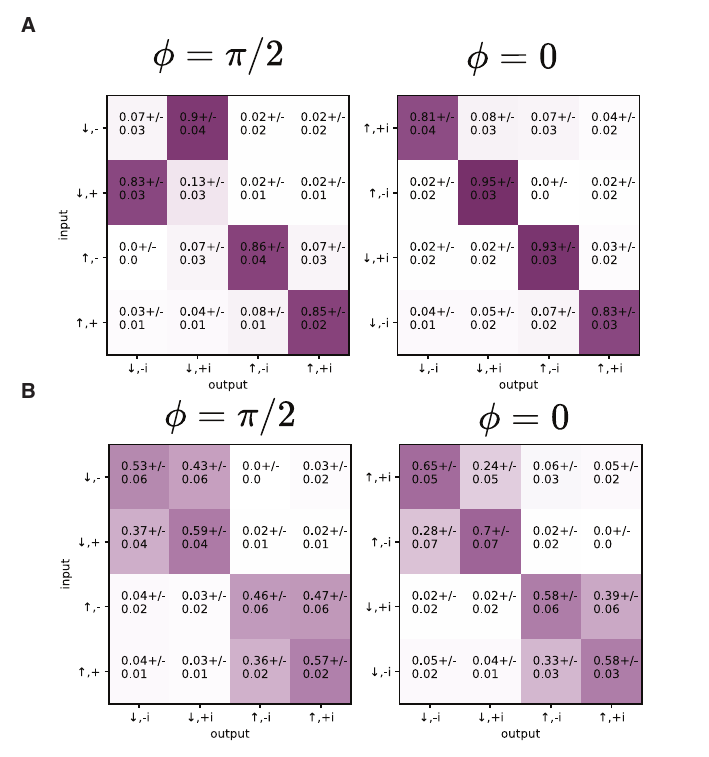}
    \caption{\textbf{Gate truth table for intra-node 2QBG operation shown in Fig.~\ref{fig:intranode}} (\textbf{A}) Gate truth table observed from the client. (\textbf{B}) Gate truth table observed from the server. }
    \label{fig:SI_intranode_table}
\end{figure}

\begin{figure}
    \centering
    \includegraphics[width=1\textwidth]{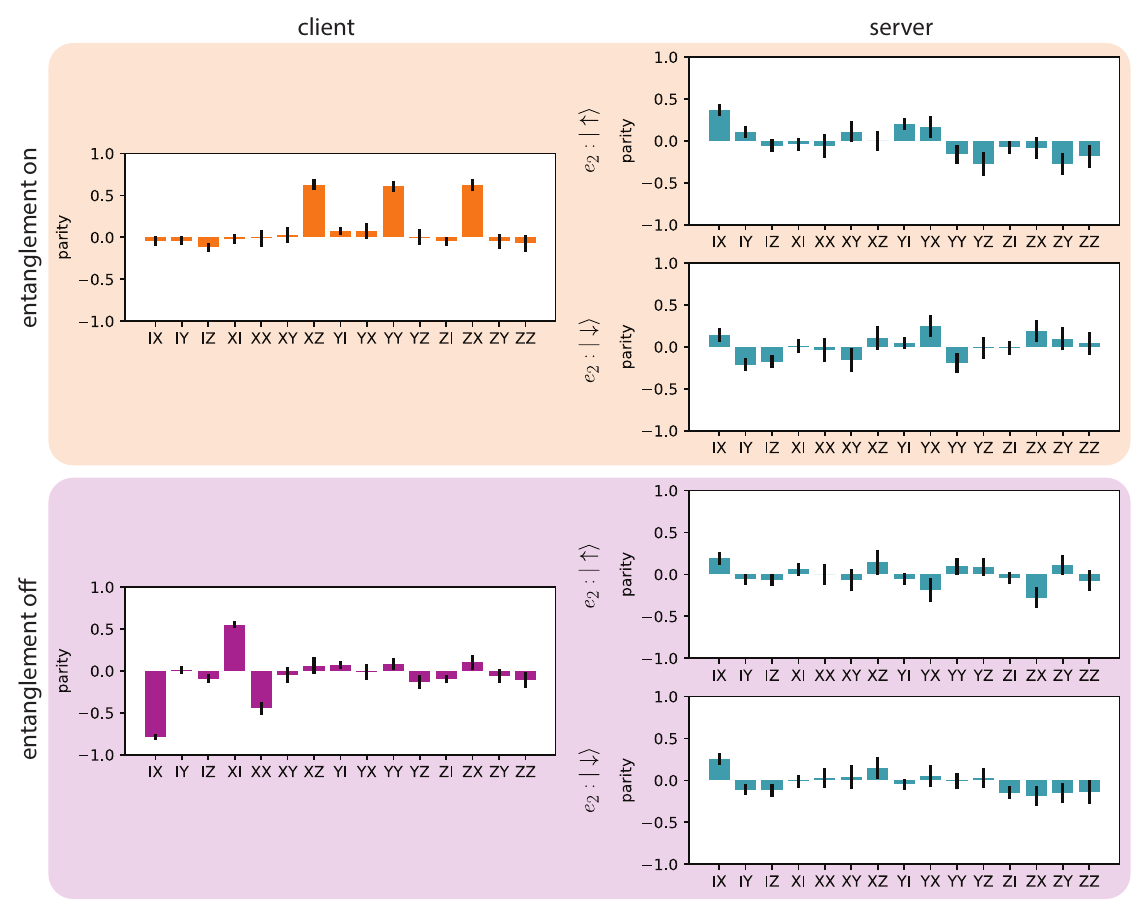}
    \caption{\textbf{2-qubit operator expectation values of the server and client for the distributed 2-qubit gate shown in Fig.~\ref{fig:qube}.}}
    \label{fig:SI_internode_DMs}
\end{figure}

\begin{figure}
    \centering
    \includegraphics[]{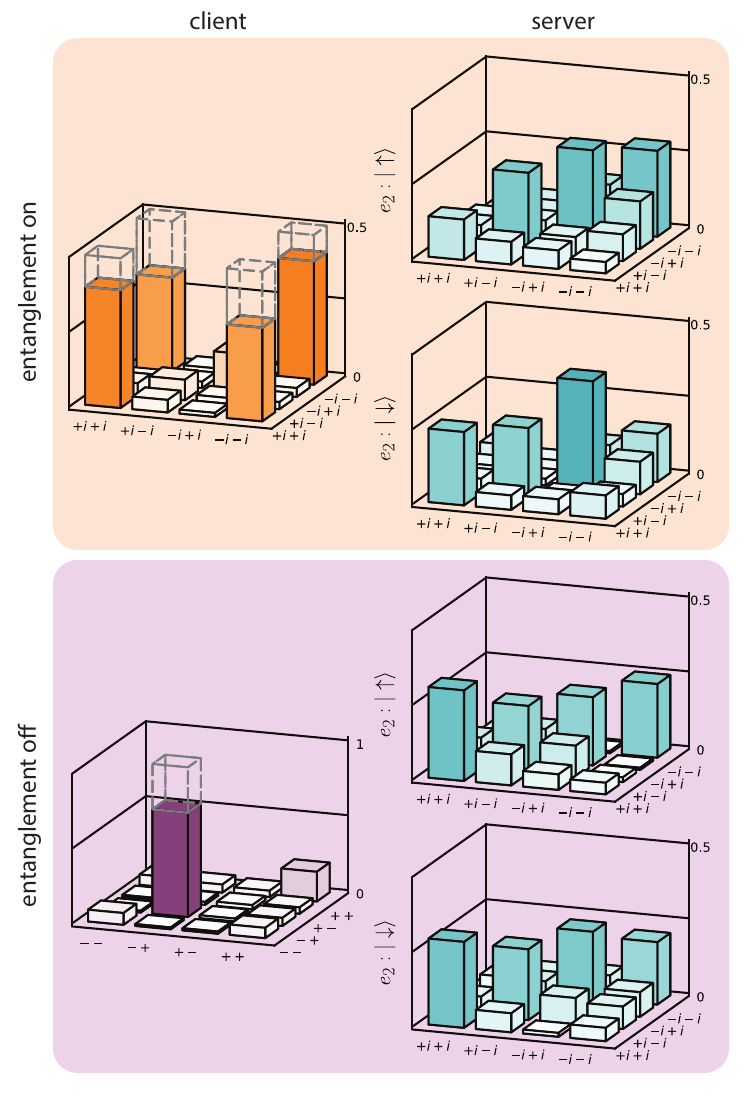}
    \caption{\textbf{Reconstructed density matrices of the server and client for the distributed 2-qubit gate shown in Fig.~\ref{fig:qube}.}}
    \label{fig:SI_internode_exp_val}
\end{figure}

This section contains additional data to calculate fidelity and information leakage of all the gates shown in the main text. Tables~\ref{table:sq_blind_rotation_x_cl}, ~\ref{table:sq_blind_rotation_x_sr}, ~\ref{table:sq_blind_rotation_y_cl}, ~\ref{table:sq_blind_rotation_y_sr}, ~\ref{table:sq_blind_rotation_z_cl}, ~\ref{table:sq_blind_rotation_z_sr}, ~\ref{table:sq_u_blind_gate_cl}, and ~\ref{table:sq_u_blind_gate_sr} contain expectation values, state fidelities, and state entropies for all single qubit blind operations. Fig.~\ref{fig:SI_intranode_exp_val} and Fig.~\ref{fig:SI_intranode_DMs}, as well as Table~\ref{table:intranode_entropy} show the complete data for fidelity and entropy calculation for the intra-node 2QBG.
Fig.~\ref{fig:SI_intranode_table} shows the complete gate truth tables of intra-node 2QBG observed from the client and server.
Fig.~\ref{fig:SI_internode_exp_val}, Fig.~\ref{fig:SI_internode_DMs}, and Table~\ref{table:internode_entropy} show the complete data for fidelity and entropy calculation for the distributed 2QBG.

\begin{table}
\centering
\begin{tabular}{| c || c | c | c | c |} 
\hline
\multicolumn{5}{|c|}{$\ket{\psi_i}=\ket{+}$} \\
\hline
   & \multicolumn{4}{c|}{Client}\\
   \hline
   $\phi_i$ & $\langle \sigma_x \rangle$ & $\langle \sigma_y \rangle$ & $\langle \sigma_z \rangle$ & Fidelity \\
   \hline
   $0$ & $0.91\pm0.02$ & $-0.10\pm0.03$ & $-0.12\pm0.04$ & $0.96\pm0.01$\\
   \hline
   $\pi/2$ & $-0.04\pm0.07$ & $-0.83\pm0.03$ & $0.02\pm0.06 $ & $0.92\pm0.02$ \\
   \hline
   $\pi$ & $-0.86\pm0.02$ &$ 0.07\pm0.06$ &$0.04\pm0.05$ & $0.93\pm0.01$ \\
   \hline
   $3\pi/2$ & $0.03\pm0.04$ & $0.88\pm0.03$ & $-0.03\pm0.06$ & $0.94\pm0.02$\\
   \hline
   \hline
   Server average state & - & - & - & - \\
   \hline
   \hline
   Information leakage & - & - & - & - \\
   \hline
   \hline
\multicolumn{5}{|c|}{$\ket{\psi_i}=\ket{-}$} \\
\hline
   & \multicolumn{4}{c|}{Client} \\
   \hline
   $\phi_i$ & $\langle \sigma_x \rangle$ & $\langle \sigma_y \rangle$ & $\langle \sigma_z \rangle$ & Fidelity \\
   \hline
   $0$ & $-0.88\pm0.02$ & $0.21\pm0.05$ & $-0.07\pm0.07$ & $0.94\pm0.01$\\
   \hline
   $\pi/2$ & $0.02\pm0.08$& $0.85\pm0.04$ & $-0.02\pm-0.03$ & $0.93\pm0.02$ \\
   \hline
   $\pi$ & $0.82\pm0.04$ & $-0.31\pm0.07$ & $-0.11\pm0.06$ & $0.91\pm0.02$\\
   \hline
   $3\pi/2$ & $-0.20\pm0.06$ & $-0.90\pm0.03$ & $-0.07\pm0.07$ & $0.95\pm0.02$ \\
   \hline
   \hline
   Server average state & - & - & - & - \\
   \hline
   \hline
   Information leakage & - & - & - & - \\
   \hline
\end{tabular}
\caption{\textbf{Additional data for the single qubit blind rotation $R_z(\phi)$ shown in Fig.~\ref{fig:single_blind}.} }
\label{table:sq_blind_rotation_x_cl}
\end{table}

\begin{table}
\centering
\begin{tabular}{| c || c | c | c | c |} 
\hline
\multicolumn{5}{|c|}{$\ket{\psi_i}=\ket{+}$} \\
\hline
   & \multicolumn{4}{c|}{Server} \\
   \hline
    $\phi_i$ & $\langle \sigma_x \rangle$ & $\langle \sigma_y \rangle$ & $\langle \sigma_z \rangle$ & Entropy \\
   \hline
   $0$ & $-0.23\pm0.05$ & $-0.06\pm0.03$ & $-0.12\pm0.04$ & $0.95\pm0.02$\\
   \hline
   $\pi/2$ & $-0.09\pm0.07$ & $-0.02\pm0.05$ & $0.02\pm0.06$ & $0.99\pm0.01$ \\
   \hline
   $\pi$ &$ -0.24\pm0.04$ & $0.01\pm0.06 $& $0.04\pm0.05$ & $0.96\pm0.01$\\
   \hline
   $3\pi/2$  & $-0.21\pm0.04$ & $0.09\pm0.07$ & $-0.03\pm0.06$ & $0.96\pm0.01$\\
   \hline
   \hline
   Server average state & $-0.19\pm0.03 $ & $0.01\pm0.03$ & $-0.02\pm0.03$ & $0.97\pm0.07$\\
   \hline
   \hline
   Information leakage  & - & - & - & $0.007^{+0.006}_{-0.007}$\\
   \hline
   \hline
\multicolumn{5}{|c|}{$\ket{\psi_i}=\ket{-}$} \\
\hline
 & \multicolumn{4}{c|}{Server} \\
   \hline
   $\phi_i$ & $\langle \sigma_x \rangle$ & $\langle \sigma_y \rangle$ & $\langle \sigma_z \rangle$ & Entropy \\
   \hline
   $0$  & $0.16\pm0.05$ & $0.16\pm0.07$ & $-0.07\pm0.07$ &$ 0.96\pm0.02$\\
   \hline
   $\pi/2$ & $0.26\pm0.08$ &$ 0.11\pm0.07$ & $-0.02\pm0.03$ &$ 0.94\pm0.02$\\
   \hline
   $\pi$ & $0.30\pm0.06$ & $-0.05\pm0.07$ & $-0.11\pm0.06$ & $0.91\pm0.03$\\
   \hline
   $3\pi/2$ & $0.35\pm0.06$ & $0.02\pm0.07$ & $-0.07\pm0.07$ & $0.91\pm0.02$ \\
   \hline
   \hline
   Server average state  & $0.27\pm0.03$ & $0.06\pm0.04$ & $-0.05\pm0.03$ & $0.95\pm0.01$\\
   \hline
   \hline
   Information leakage  & - & - & - & $0.006^{+0.010}_{-0.006}$\\
   \hline
\end{tabular}
\caption{\textbf{Additional data for the single qubit blind rotation $R_z(\phi)$ shown in Fig.~\ref{fig:single_blind}.} }
\label{table:sq_blind_rotation_x_sr}
\end{table}

\begin{table}
\centering
\begin{tabular}{| c || c | c | c | c |} 
\hline
\multicolumn{5}{|c|}{$\ket{\psi_i}=\ket{+i}$} \\
\hline
   & \multicolumn{4}{c|}{Client} \\
   \hline
   $\phi_i$ & $\langle \sigma_x \rangle$ & $\langle \sigma_y \rangle$ & $\langle \sigma_z \rangle$ & Fidelity\\
   \hline
   $0$ & $0.01\pm0.05$ & $0.88\pm0.03$ & $0.03\pm0.06$ & $0.94\pm0.02$\\
   \hline
   $\pi/2$ & $0.83\pm0.04 $& $0.05\pm0.06$ & $0.04\pm0.05$ & $0.92\pm0.02$\\
   \hline
   $\pi$ & $-0.05\pm0.05$ &$ -0.92\pm0.03$ & $-0.01\pm0.05 $&$0.96\pm0.02$\\
   \hline
   $3\pi/2$ & $-0.90\pm0.03$ & $-0.15\pm0.06$ & $-0.05\pm0.06$ &$ 0.95\pm0.02$\\
   \hline
   \hline
   Server average state & - & - & - & - \\
   \hline
   \hline
   Information leakage & - & - & - & - \\
   \hline
   \hline
\multicolumn{5}{|c|}{$\ket{\psi_i}=\ket{-i}$} \\
\hline
   & \multicolumn{4}{c|}{Client}  \\
   \hline
   $\phi_i$ & $\langle \sigma_x \rangle$ & $\langle \sigma_y \rangle$ & $\langle \sigma_z \rangle$ & Fidelity \\
   \hline
   $0$ & $0.24\pm0.05$ & $-0.92\pm0.02 $ & $0.17\pm0.05$ & $0.96\pm0.01$\\
   \hline
   $\pi/2$ & $-0.79\pm0.03$ & $-0.35\pm0.06$ & $0.07\pm0.08$ & $0.89\pm0.02$\\
   \hline
   $\pi$ & $0.35\pm0.05$ & $0.94\pm0.03$ & $0.11\pm0.05$ & $0.97\pm0.02$\\
   \hline
   $3\pi/2$ & $0.88\pm0.04$ &$ 0.06\pm0.05$ & $-0.01\pm0.04$ & $0.94\pm0.02$\\
   \hline
   \hline
   Server average state & - & - & - & - \\
   \hline
   \hline
   Information leakage & - & - & - & -\\
   \hline
\end{tabular}
\caption{\textbf{Additional data for the single qubit blind rotation $R_z(\phi)$ shown in Fig.~\ref{fig:single_blind}}. }
\label{table:sq_blind_rotation_y_cl}
\end{table}

\begin{table}
\centering
\begin{tabular}{| c || c | c | c | c |} 
\hline
\multicolumn{5}{|c|}{$\ket{\psi_i}=\ket{+i}$} \\
\hline
   & \multicolumn{4}{c|}{Server} \\
   \hline
   $\phi_i$ & $\langle \sigma_x \rangle$ & $\langle \sigma_y \rangle$ & $\langle \sigma_z \rangle$ & Entropy \\
   \hline
   $0$ & $0.02\pm0.05$ & $-0.22\pm0.06$ & $0.03\pm0.06$ & $0.96\pm0.02$\\
   \hline
   $\pi/2$  & $0.08\pm0.07$ & $-0.34\pm0.06$ & $0.04\pm0.05$ & $0.92\pm0.03$\\
   \hline
   $\pi$ & $-0.01\pm0.05$ & $-0.21\pm0.06$ & $-0.01\pm0.05 $& $0.97\pm0.02$\\
   \hline
   $3\pi/2$ & $0.02\pm0.06$ & $-0.23\pm0.06$ & $-0.05\pm0.06$ &$ 0.96\pm0.02$\\
   \hline
   \hline
   Server average state  & $0.03\pm0.03$ & $-0.25\pm0.03$ & $0.00\pm0.03 $& $0.96\pm0.01$\\
   \hline
   \hline
   Information leakage & - & - & - & $0.005^{+0.01}_{-0.005}$\\
   \hline
   \hline
\multicolumn{5}{|c|}{$\ket{\psi_i}=\ket{-i}$} \\
\hline
   & \multicolumn{4}{c|}{Server} \\
   \hline
   $\phi_i$ & $\langle \sigma_x \rangle$ & $\langle \sigma_y \rangle$ & $\langle \sigma_z \rangle$ & Entropy \\
   \hline
   $0$ & $0.12\pm0.05$ & $0.12\pm0.06$ & $0.17\pm0.05$ & $0.96\pm0.02$\\
   \hline
   $\pi/2$ & $-0.04\pm0.05$ & $0.17\pm0.07$ & $0.07\pm0.08$ & $0.97\pm0.02$\\
   \hline
   $\pi$ & $0.08\pm0.05$ & $0.30\pm0.07 $& $0.11\pm0.05$ & $0.92\pm0.03$\\
   \hline
   $3\pi/2$& $0.02\pm0.07$ & $0.20\pm0.05$ & $-0.01\pm0.04$ & $0.96\pm0.01$\\
   \hline
   \hline
   Server average state &$ 0.05\pm0.03$ & $0.20\pm0.03$ & $0.09\pm0.03$ & $0.96\pm0.01$\\
   \hline
   \hline
   Information leakage  & - & - & - & $0.002^{+0.037}_{-0.002}$\\
   \hline
\end{tabular}
\caption{\textbf{Additional data for the single qubit blind rotation $R_z(\phi)$ shown in Fig.~\ref{fig:single_blind}}. }
\label{table:sq_blind_rotation_y_sr}
\end{table}

\begin{table}
\centering
\begin{tabular}{| c || c | c | c | c |} 
\hline
\multicolumn{5}{|c|}{$\ket{\psi_i}=\ket{\uparrow}$} \\
\hline
   & \multicolumn{4}{c|}{Client} \\
   \hline
   $\phi_i$ & $\langle \sigma_x \rangle$ & $\langle \sigma_y \rangle$ & $\langle \sigma_z \rangle$ & Fidelity  \\
   \hline
   $0$ & $0.02\pm0.06$ &$ 0.04\pm0.05$ & $0.97\pm0.02$ & $0.98\pm0.01$\\
   \hline
   $\pi/2$ &$ 0.02\pm0.05 $& $0.05\pm0.06$ & $0.98\pm0.01$ & $0.99\pm0.01$\\
   \hline
   $\pi$ &$-0.06\pm0.05$ & $0.10\pm0.05$ &$ 0.94\pm0.02$ & $0.97\pm0.01$ \\
   \hline
   $3\pi/2$ &$ 0.03\pm0.05$ & $-0.05\pm0.05$ & $0.99\pm0.01$ & $0.99\pm0.01$\\
   \hline
   \hline
   Server average state & - & - & - & - \\
   \hline
   \hline
   Information leakage & - & - & - & - \\
   \hline
   \hline
\multicolumn{5}{|c|}{$\ket{\psi_i}=\ket{\downarrow}$} \\
\hline
   & \multicolumn{4}{c|}{Client} \\
   \hline
   $\phi_i$ & $\langle \sigma_x \rangle$ & $\langle \sigma_y \rangle$ & $\langle \sigma_z \rangle$ & Fidelity  \\
   \hline
   $0$ & $0.00\pm0.05$ &$ 0.04\pm0.05$ & $-0.99\pm0.01 $& $0.99\pm0.01$\\
   \hline
   $\pi/2$ & $0.06\pm0.05$ & $0.01\pm0.05$ & $-0.99\pm0.01$ & $0.99\pm0.01$\\
   \hline
   $\pi$ & $-0.09\pm0.06$ & $0.03\pm0.06$ & $-0.97\pm0.01 $& $0.98\pm0.01$ \\
   \hline
   $3\pi/2$ & $0.07\pm0.05$ & $0.01\pm0.03$ & $-0.97\pm0.01 $& $0.98\pm0.01$ \\
   \hline
   \hline
   Server average state & - & - & - & -\\
   \hline
   \hline
   Information leakage & - & - & - & - \\
   \hline   
\end{tabular}
\caption{\textbf{Additional data for the single qubit blind rotation $R_z(\phi)$ shown in Fig.~\ref{fig:single_blind}.} }
\label{table:sq_blind_rotation_z_cl}
\end{table}

\begin{table}
\centering
\begin{tabular}{| c || c | c | c | c |} 
\hline
\multicolumn{5}{|c|}{$\ket{\psi_i}=\ket{\uparrow}$} \\
\hline
   &  \multicolumn{4}{c|}{Server} \\
   \hline
   $\phi_i$  & $\langle \sigma_x \rangle$ & $\langle \sigma_y \rangle$ & $\langle \sigma_z \rangle$ & Entropy \\
   \hline
   $0$  & $-0.03\pm0.06$ & $-0.02\pm0.05$ &$ 0.97\pm0.02$ & $0.11\pm0.03$\\
   \hline
   $\pi/2$  & $-0.07\pm0.05$ & $-0.09\pm0.06$ & $0.98\pm0.02$ & $0.06\pm0.05$\\
   \hline
   $\pi$  & $-0.03\pm0.05$ & $0.03\pm0.05$ &$ 0.94\pm0.02$ & $0.20\pm0.06$\\
   \hline
   $3\pi/2$  &$ 0.06\pm0.05$ & $-0.02\pm0.07 $& $0.99\pm0.01 $& $0.04\pm0.06$\\
   \hline
   \hline
   Server average state &$-0.02\pm0.03$ & $-0.03\pm0.03$ &$ 0.97\pm0.01$ & $0.11\pm0.02$\\
   \hline
   \hline
   Information leakage  & - & - & - & $0.009^{+0.02}_{-0.009}$\\
   \hline
   \hline
\multicolumn{5}{|c|}{$\ket{\psi_i}=\ket{\downarrow}$} \\
\hline
   &  \multicolumn{4}{c|}{Server} \\
   \hline
   $\phi_i$  & $\langle \sigma_x \rangle$ & $\langle \sigma_y \rangle$ & $\langle \sigma_z \rangle$ & Entropy \\
   \hline
   $0$  &$0.00\pm0.05$ & $0.05\pm0.05 $& $-0.99\pm0.01$ & $0.04\pm0.03$\\
   \hline
   $\pi/2$ & $-0.01\pm0.05$ & $0.25\pm0.05$ & $-0.99\pm0.01$ & $0.00\pm0.01$\\
   \hline
   $\pi$  & $-0.01\pm0.06$ & $0.16\pm0.06$ & $-0.97\pm0.01$ & $0.07\pm0.04$\\
   \hline
   $3\pi/2$  & $0.07\pm0.05$ &$ 0.03\pm0.03$ & $-0.97\pm0.01$ &$ 0.10\pm0.03$\\
   \hline
   \hline
   Server average state  & $0.02\pm0.02$ & $0.12\pm0.02$ & $-0.98\pm0.01$ & $0.06\pm0.06$\\
   \hline
   \hline
   Information leakage  & - & - & - & $0.005^{+0.044}_{-0.005}$\\
   \hline   
\end{tabular}
\caption{\textbf{Additional data for the single qubit blind rotation $R_z(\phi)$ shown in Fig.~\ref{fig:single_blind}.} }
\label{table:sq_blind_rotation_z_sr}
\end{table}

   

\begin{table}
\centering
\begin{tabular}{| c || c | c | c | c|} 
\hline
   & \multicolumn{4}{c|}{Client}\\
   \hline
   gate $(\phi_i^{(0)}, \phi_i^{(1)},\phi_i^{(2)})$ & $\langle \sigma_x \rangle$ & $\langle \sigma_y \rangle$ & $\langle \sigma_z \rangle$ & Fidelity \\
   \hline
   Identity $(0,0,0)$ & $0.26\pm0.16$ & $-0.58\pm0.13$ & $-0.17\pm0.18$ & $0.79\pm0.07$ \\
   \hline
   Hadamard $(\pi/2, \pi/2, \pi/2)$ & $0.05\pm0.16$ & $0.46\pm0.15$ & $-0.06\pm0.18$ & $0.73\pm0.07$\\
   \hline   T$\sqrt{\mathrm{X}}$T $(\pi/4, \pi/2, \pi/4)$ & $0.25\pm0.24$ & $0.38\pm0.23$ & $0.14\pm0.22$ & $0.71\pm0.11$\\
   \hline
   \hline
   Server average state & - & - & - & - \\
   \hline
   \hline
   Information leakage & - & - & - & - \\
   \hline
   
\end{tabular}
\caption{\textbf{Additional data for the universal blind gate shown in Fig.~\ref{fig:single_blind}.}}
\label{table:sq_u_blind_gate_cl}
\end{table}

\begin{table}
\centering
\begin{tabular}{| c || c | c | c | c |} 
\hline
   &  \multicolumn{4}{c|}{Server} \\
   \hline
   gate $(\phi_i^{(0)}, \phi_i^{(1)},\phi_i^{(2)})$  & $\langle \sigma_x \rangle$ & $\langle \sigma_y \rangle$ & $\langle \sigma_z \rangle$ & Entropy \\
   \hline
   Identity $(0,0,0)$ & $-0.31\pm0.16$ & $-0.11\pm0.16$ & $-0.17\pm14$ & $0.90\pm0.20$ \\
   \hline
   Hadamard $(\pi/2, \pi/2, \pi/2)$ & $-0.30\pm0.15$ & $0.08\pm0.16$ & $-0.19\pm0.17$ & $0.90\pm0.20$\\
   \hline
   T$\sqrt{\mathrm{X}}$T $(\pi/4, \pi/2, \pi/4)$ &  $0.00\pm0.25$ & $0.13\pm0.25$ & $0.14\pm0.22$ & $0.97\pm0.05$\\
   \hline
   \hline
   Server average state & $-0.20\pm0.11$ & $0.03\pm0.11$ & $-0.07\pm0.10$ & $0.96\pm0.04$\\
   \hline
   \hline
   Information leakage  & - & - & - & $0.04\pm0.10$\\
   \hline
   
\end{tabular}
\caption{\textbf{Additional data for the universal blind gate shown in Fig.~\ref{fig:single_blind}.}}
\label{table:sq_u_blind_gate_sr}
\end{table}


\begin{table}
\centering
\begin{tabular}{| c || c |} 
   \hline
   $\phi_i$ & Server entropy \\
   \hline
   $0$ & $1.15\pm0.09$\\
   \hline
   $\pi/4$ & $1.21\pm0.08$\\
   \hline
   $\pi/2$ & $1.27\pm0.18$\\
   \hline
   $3\pi/4$ & $1.11\pm0.21$\\
   \hline
   \hline
   Server average state & $1.22\pm0.05$\\
   \hline
   \hline
   Information leakage & $0.03\pm0.09$\\
   \hline
   
\end{tabular}
\caption{\textbf{Server entropy and information leakage for the intra-node 2-qubit gate shown in Fig.~\ref{fig:intranode}.}}
\label{table:intranode_entropy}
\end{table}

\begin{table}
\centering
\begin{tabular}{| c || c | c |} 
   \hline
   entanglement ON/OFF & Server entropy ($e_2:\ket{\uparrow})$& Server entropy ($e_2:\ket{\downarrow})$\\
   \hline
   ON & $1.64\pm0.13$ & $1.77\pm0.11$ \\
   \hline
   OFF & $1.82\pm0.08$ & $1.83\pm0.07$\\
   \hline
   \hline
   Server average state & $1.88\pm0.03$ & $1.89\pm0.04$\\
   \hline
   \hline
   Information leakage & $0.15\pm0.08$ & $0.09\pm0.08$\\
   \hline
   
\end{tabular}
\caption{\textbf{Server entropy and information leakage for the distributed 2-qubit gate shown in Fig.~\ref{fig:qube}.}}
\label{table:internode_entropy}
\end{table}

\subsection{Data Analysis and Thresholding}
This section outlines the steps used to analyze the data produced in the experiment and the methods applied to filter it for enhanced performance.

In the experiment, multiple parameters were available for data filtering:
\begin{itemize}
    \item SiV optical contrast
    \item $\pi$ pulse fidelity
    \item Initialization fidelity
    \item Laser drift
\end{itemize}

These parameters are independently tracked over time during the duration of the experiment. For each parameter, we select threshold values to filter or analyze the data accordingly. For each experimental set, we systematically explore all combinations of these parameters to optimize the thresholds based on fidelity. The chosen thresholds are then used to calculate the information leakage of the gate implemented in each experiment. It is important to note that no more than 50\% of the data for each initial state and measurement basis combination is filtered at any point.

The following subsections detail the thresholds chosen for each experimental setup:

\subsubsection{Blind  rotation ($R_z$)}
For each combination of thresholds, phase $\phi$, initial state, and measurement basis, we calculate the total number of shots before and after filtering, as well as the parity values for both the client and server. We compute the fidelity for each phase $\phi$, average over all phases, and select the parameter set yielding the highest fidelity to calculate the information leakage. The information leakage is determined for each initial state separately, examining a table with rows corresponding to $\phi \in \{0, \pi/2, \pi, 3\pi/2\}$ and columns corresponding to measurement bases in $\{X, Y, Z\}$.

The thresholds are optimized over the following ranges:
\begin{align*}
\text{SiV residual reflectivity of non-reflective state ($1/\text{contrast}$) :} & \quad 0.06 \text{ to } 0.2 \text{ in steps of } 0.02, \\
\pi \text{ pulse fidelity:} & \quad 0.93 \text{ to } 0.96 \text{ in steps of } 0.01, \\
\text{Initialization fidelity:} & \quad 0.93 \text{ to } 0.97 \text{ in steps of } 0.01, \\
\text{Laser drift:} & \quad 0.2 \text{ to } 0.5 \text{ in steps of } 0.1.
\end{align*}

The optimal parameters found were: 
\[
\text{Residual reflectivity threshold: } 0.1, \; \pi \text{ fidelity: } 0.95
\]
\[
\text{Initialization fidelity: } 0.94, \; \text{Laser drift: } 0.3.
\]

\subsubsection{One-qubit blind gate (1QBG)}

For the 1QBG), which is composed of three spin-photon gates, data are analyzed for each gate type \(\{TX^{1/2}T, I, H\}\), initial state $\ket{+i}$, and measurement bases \(\{X, Y, Z\}\). For each set of thresholds, gates, and measurement bases, we calculate the total number of shots before and after filtering, as well as the parity for both client and server.

For each parameter set, a table of client fidelities is generated, with columns representing the gates and rows the measurement bases. The average fidelity for each gate type \(\{I, TX^{1/2}T, H\}\) is computed and averaged over all gates, to determine the final fidelity for the considered parameter set. The parameter set with the highest fidelity is selected, and the information leakage is calculated for this set only.

The thresholds are optimized over the following parameter ranges:
\begin{align*}
\text{SiV residual reflectivity of non-reflective state ($1/\text{ contrast}$) :} & \quad 0.1 \text{ to } 0.15 \text{ in steps of } 0.01, \\
\pi \text{ pulse fidelity:} & \quad 0.7 \text{ to } 0.9 \text{ in steps of } 0.02, \\
\text{Initialization fidelity:} & \quad 0.91 \text{ to } 0.99 \text{ in steps of } 0.01, \\
\text{Laser drift:} & \quad 0.3 \text{ to } 0.8 \text{ in steps of } 0.1.
\end{align*}

The optimal parameters found are:
\[
\text{Residual reflectivity threshold: } 0.13, \quad \pi \text{ pulse fidelity: } 0.7
\]
\[
\text{Initialization fidelity: } 0.91, \quad \text{Laser drift: } 0.3.
\]

\end{document}